\documentclass[modern]{aastex63}
\usepackage{amsmath}
\usepackage{amssymb}
\usepackage{amsfonts}
\usepackage{amstext}
\usepackage{pstricks}
\usepackage{graphicx}
\usepackage{fancybox}
\usepackage{bm}
\usepackage{soul, color}
\usepackage{latexsym}
\usepackage{pifont}
\usepackage{shorttoc}
\usepackage{subfigure}
\usepackage{textcomp}
\usepackage{float}
\usepackage{enumitem}
\usepackage{longtable}
\usepackage{calrsfs}
\usepackage{bm}
\usepackage{lipsum}
\usepackage{threeparttablex}
\usepackage{hyperref}
\hypersetup{urlcolor=blue}
\def\aap{A\&A}

\def\apjs{ApJS}
\def\apjl{ApJL}

\newcommand{\uvec}[1]{\boldsymbol{\hat{\textbf{#1}}}}

\newcommand{\lya}{Ly$\alpha$\ }
\newcommand{\lyb}{Ly$\beta$\ }
\DeclareMathAlphabet{\pazocal}{OMS}{zplm}{m}{n}
\newcommand{\po}{\pazocal{P}}

\newcommand{\no}[1]{}
\renewcommand{\exp}[1]{\mathrm{exp}\left(#1\right)}

\newcommand{\myemail}{lluis.mas-ribas@jpl.nasa.gov}
\def\lsim{~\rlap{$<$}{\lower 1.0ex\hbox{$\sim$}}}

\def\gsim{~\rlap{$>$}{\lower 1.0ex\hbox{$\sim$}}}

\newcommand{\ra}[1]{\renewcommand{\arraystretch}{#1}}

\shorttitle{Lyman-{$\alpha$} Polarization Intensity Mapping}
\shortauthors{Mas-Ribas \& Chang}

\begin{document}

\title{\Large Lyman-{$\alpha$} Polarization Intensity Mapping}

\email{\myemail}

\author{Llu\'is Mas-Ribas} 
\author{Tzu-Ching Chang}
\affiliation{Jet Propulsion Laboratory, California Institute of Technology, 4800 Oak Grove Drive, 
Pasadena, CA 91109, U.S.A.}
\affiliation{California Institute of Technology, 1200 E. California Blvd, Pasadena, CA 91125, U.S.A. \\ 
\copyright 2020. California Institute of Technology. Government sponsorship acknowledged.}

\begin{abstract}  

      We present a formalism that incorporates hydrogen Lyman-alpha (Ly$\alpha$) polarization arising 
from the scattering of radiation in galaxy halos into the intensity mapping approach.  Using the halo 
model, and \lya emission profiles based on simulations and observations, we calcualte auto and cross 
power spectra at redshifts $3\leq z \leq13$ for  the \lya total intensity, $I$, polarized intensity, $\po$, 
degree of polarization, $\Pi=\po/I$, and two new quantities, the astrophysical $E$ and $B$ modes of \lya 
polarization. The one-halo terms of the $\Pi$ power spectra show a turnover that signals the average 
extent of the polarization signal, and thus the extent of the scattering medium. The position of this feature 
depends on redshift, as well as on the specific emission profile shape and extent, in our formalism. 
Therefore, the comparison of various \lya polarization quantities and redshifts can break degeneracies 
between competing effects, and it can reveal the true shape of the emission profiles, which, in turn, are 
associated to the physical properties of the cool gas in galaxy halos. Furthermore, measurements of \lya 
$E$ and $B$ modes may be used as probes of galaxy evolution, because they are related to the average 
degree of anisotropy in the emission and in the halo gas distribution across redshifts. The 
detection of the polarization signal at $z \sim 3-5$ requires improvements in the sensitivity of current 
ground-based experiments by a factor of $\sim 10$,  and of $\sim 100$ for space-based instruments 
targeting the redshifts $z\sim 9-10$, the exact values depending on the specific redshift and experiment. 
Interloper contamination in polarization is expected to be small, because the interlopers need to also 
be polarized. Overall, \lya polarization boosts the amount of physical information retrievable on 
galaxies and their surroundings, most of it not achievable with total emission alone.

\vspace{2.5cm}

\end{abstract}

\section{ Introduction}

   Intensity mapping (IM) is  a novel method to study the formation and 
evolution of galaxies, by statistically analyzing the collective emission present in large areas of the 
sky, and at different epochs, regardless of the number of bright (individually detectable) sources in them 
\citep[][see the recent review by \citealt{Kovetz2017}]{Madau1997,Suginohara1999,Visbal2010}. The 
IM methodology takes into account the emission from the entire galaxy population, and thus, 
contrary to more traditional galaxy studies, it is not limited to the sources above observational 
detection thresholds. 

   IM considers a broad range of frequencies and emission lines, such as those of [C{\sc ii}] at 158 
$\mu$m \citep[e.g.,][]{Gong2012,Silva2015,Yue2015}, the CO molecule 
\citep[e.g.,][]{Righi2008,Gong2011,Lidz2011,Pullen2013,Chung2019}, the hydrogen 21cm 
spin-flip transition \citep[e.g.,][]{Scott1990,Madau1997,Chang2008,Chang2010,Switzer2013}, or 
X-rays, as recently proposed by \cite{Caputo2019}. 

In addition to the aforementioned frequencies, 
the hydrogen Lyman-alpha (Ly$\alpha$) radiation is one of the main targets of IM. \lya emission is 
especially useful for studies of cosmic reionization at $z\gtrsim 5$ \citep[e.g.,][]{Silva2012,Pullen2014}, 
but also for studies up to the pre-reionization epoch at $z\sim 20-30$ \citep{Loeb1999}, and down to 
the peak of cosmic star formation at $z\sim2-3$ \citep[e.g.,][]{Hogan1987,Gould1996,Croft2018}. In 
these cases, \lya is mostly produced by young (blue) stars, and it is the brightest emission line from 
star formation \citep{Partridge1967}. 

    A particular characteristic of \lya radiation compared to other emission lines, is its resonant nature. 
Because \lya is the only radiative channel allowed by quantum mechanics between the hydrogen ground 
and first excited atomic states, the absorption of a \lya photon by a  hydrogen atom (H{\sc i}) typically 
results in the immediate emission of another \lya photon. 
This is the well-known \lya scattering process, which enables the 
\lya photons to transfer (diffuse) through a neutral hydrogen medium, until the photons escape the 
medium or they become destroyed by dust \citep[see][for a review]{Dijkstra2014}. Scattering, together 
with other potential mechanisms \citep[see][for a discussion of the various processes]{Masribas2017b}, 
contributes to the diffuse and extended \lya emission  currently detectable with instruments such as 
MUSE \citep{Bacon2014} or KCWI \citep{Morrissey2018}, down to surface brightness levels of 
$\sim 10^{-19}\,{\rm erg\,s^{-1}\,cm^{-2}\,arcsec^{-2}}$, and out to several tens of physical kpc from the 
center of most individual $2\lesssim z \lesssim 7$ star-forming galaxies and quasars, constituting the 
so-called \lya halos  \citep[e.g.,][]{Borisova2016,Wisotzki2016,Leclercq2017,Wisotzki2018,
Arrigonibattaia2019,Farina2019}. The IM approach will enable studying this faint emission far from the 
sources for a large number of objects  statistically, as well as the \lya emission arising directly from the 
distant intergalactic medium, especially at redshifts $z\gtrsim 3 - 5$, where the fraction of cosmic neutral  
hydrogen gas is significant \citep[e.g.,][]{Gould1996,Loeb1999,Laursen2011,Silva2012,Pullen2014, 
Davies2016,Kakiichi2016,Visbal2018}. 

     In this paper, we focus on the polarization of \lya radiation around galaxies, which is another 
effect arising from the scattering of   \lya  photons, and that has not been  previously 
considered in the  intensity mapping formalism\footnote{Previous references to polarization in  
intensity mapping studies are for the case of  the 21cm radiation. \cite{Cooray2005} assessed the 
21cm polarization arising from Zeeman splitting due to magnetic fields, and \cite{Babich2005} 
discussed the polarizing effects of Thomson scattering during reionization on the pre-reionization 21cm 
emission. The latter effect will also be  suffered by any other frequency (IM or CMB radiation) 
due to the achromaticity of electron scattering. Finally, a more recent series of papers by 
\cite{Gluscevic2017,Hirata2017,Venumadhav2017,Mishra2018}, have revisited the 21cm polarization 
arising from Zeeman splitting, as well as from anisotropies in the CMB radiation, which enable the 
study of primordial magnetic fields and gravitational waves, respectively.}. Because in the scattering 
event the \lya photons become polarized \citep[e.g.,][]{Chandrasekhar1960}, and because scattering 
contributes to the extended \lya emission around sources, a net polarization 
fraction can appear in the diffuse \lya emission in galaxy halos. Indeed, 
\cite{Rybicki1999} first noted that high degrees of \lya polarization, up to $\sim 40 - 60 \,\%$, could 
occur around pre-reionization sources due to \lya radiation scattered by the neutral intergalactic gas. 
Later, \cite{Dijkstra2008} performed \lya radiative transfer simulations in idealized spherically-symmetric 
expanding H{\sc i} shells, resembling the environment around high-redshift galaxies, and showed that 
this polarization signal can also be found in the halo of $z \sim 5 - 6$ galaxies \citep[see also][]{
Dijkstra2012}. These calculations 
indicated that the degree of polarization increases with impact parameter, from a few per cent at the 
center to a few tens of per cent at large impact parameters, and  that the angle of polarization forms 
concentric rings projected on the sky around the radiation source. Observations confirming these  
theoretically-predicted trends  were presented by \cite{Hayes2011}, and supported more recently 
by \cite{Beck2016} and \cite{Herenz2020}, for a bright and extended \lya nebula at $z\approx3$,   LAB1;
\cite{Steidel2000}. Unlike \lya halos around individual galaxies, \lya nebulae or blobs can be 
powered by bright and/or multiple sources, such as quasars or bright galaxies, and they typically extend 
to distances on the order of $\gtrsim 100$ physical kpc, larger than typical \lya halos. \cite{Hayes2011} 
found a polarization fraction value increasing from a few per cent around the center of the brightest LAB1 
region, up to $\sim 18\,\%$ at $\sim 50$ pkpc, beyond which the signal-to-noise ratio did not enable 
precise measurements (see their Figure 3). The measured polarization pattern broadly agreed with the 
circular (tangential) directions predicted by the numerical models, but there are cases where the 
differences are significant \citep[see Figure S3 in][]{Hayes2011}. These differences are not surprising, 
because the actual environment around LAB1 is not spherically symmetric, as was the case in the 
numerical work of \cite{Dijkstra2008}, and because several sources could contribute to the \lya emission, 
as demonstrated by the observations and simulations of LAB1 by \citealt{Geach2016} \citep[although 
see][where their numerical simulations favor a gravitational cooling scenario driving the 
observations]{Trebitsch2016}. 
\cite{Prescott2011} performed observations of another \lya blob at $z\sim 2.6$, LABd05; \cite{Dey2005}, 
but their low spatial resolution enabled them to only set an upper limit of $\sim2.5\,\%$ for the polarization 
fraction in a single aperture with radius $\sim 33$ kpc.
More recently, \cite{Humphrey2013} and \cite{You2017} have also reported \lya polarization observations 
in  \lya nebulae at $z\approx 2.3$ and $z\approx 3$, respectively. Their results are in broad agreement 
with those found by \cite{Hayes2011}.

    Given the ubiquitous extended and diffuse \lya emission around high-redshift sources that overall 
covers large portions of the sky \citep[see Figure 1 in][]{Wisotzki2018}, it is plausible to expect also a 
global \lya polarization signal.  The exact value of the degree of polarization depends strongly 
on the physical properties of the scattering medium (i.e., its bulk and turbulent velocity, as well as 
its H{\sc i} density). The \lya polarization pattern (i.e., the angle of polarization) around the 
sources depends strongly on the isotropy and homogeneity of the emission and gas distribution 
\citep{Lee1998,Ahn2002,Chang2016,Eide2018}. 
These dependences make the polarization signal very sensitive to the specific conditions of the 
medium,  and, as we will show, this boosts the amount of information on galaxies and their 
environment retrievable from polarization, compared to that from emission alone. 
The goal of this work is to provide a first theoretical benchmark to assess the 
utility of such a polarized emission, and to investigate whether the expected signal is  within 
reach of current and future intensity mapping experiments.

        In \S~\ref{sec:formalism} below, we derive the mathematical formalism for characterizing 
the global polarized \lya signal. The physical origin and modeling of the \lya emission around 
sources is detailed in \S~\ref{sec:polalpha}. The results and estimates for the detectability are 
presented in \S~\ref{sec:results} and \S~\ref{sec:detection}, respectively. We discuss the case 
of \lya B modes in \S~\ref{sec:bmodes}, future work in 
\S~\ref{sec:future}, and conclude in \S~\ref{sec:conclusions}.

We assume a flat ($\Omega_{\rm k}=0$) $\Lambda$CDM cosmology with the parameter values from \cite{Planck2015}, and use comoving units throughout unless stated otherwise.

\section{A Halo Model Formalism for \lya Polarization}\label{sec:formalism}

    This section describes a simple formalism for parameterizing the \lya polarization signal. 
We use the halo model to assess the spatial distribution of \lya in \S~\ref{sec:2d}, and derive  
the formalism of $E$ and $B$ modes for the case of \lya polarization in \S~\ref{sec:eb}. We consider 
the case of cross-correlations between polarization quantities in \S~\ref{sec:cross}. 

     We characterize the polarization signal by considering the four Stokes parameters $I, Q, U, V$. 
The quantity $I$ is the total intensity of radiation, and the parameters $Q$ and $U$ relate to the 
polarized radiation along the coordinate axes, and along the directions at $\pi/4$ from them, 
respectively. This definition  implies that the values of $Q$ and $U$ depend on the choice of the 
coordinate system that defines them, while $I$ is simply a scalar quantity invariant under a change 
of coordinates (we address this coordinate system dependence in \S~\ref{sec:eb}). 
We ignore the parameter describing circular polarization, $V$, because the scattering of \lya 
radiation yields linear polarization alone when the incoming radiation is non-circularly polarized, 
which we assume to be the case here \citep{Chandrasekhar1960}. Derived quantities also useful 
for our work are the polarized intensity, $\po=\sqrt{Q^2 + U^2}$, and the degree of polarization (or 
polarization fraction), $\Pi=\po / I$. For completeness, we define the polarization angle to be 
$2\gamma = \tan^{-1}(U/Q)$. 

    To parameterize the spatial distribution of radiation, we adopt the halo model formalism 
\citep{Peacock2000,Seljak2000,Scoccimarro2001,Cooray2002}.  The halo model assumes that all the matter in the universe is contained 
in spherical halos, and that these halos do not overlap with each other. The signal from the halos is  
characterized by the one- and two-halo terms, which describe the contribution to the quantity of 
interest from regions within the same or different halos, respectively. For our work, this description 
implies that the total power spectra of any quantity can be simply calculated as the sum of the 
power spectra from the two terms,  $P = P^{\rm 1h} + P^{\rm 2h}$, as detailed in the following section.

    Below, we derive the two-dimensional (projected) halo-model formalism for the \lya polarization 
signal. In practice, the intensity of \lya radiation, $I$, could be modeled assuming spherical symmetry 
around the source, which allows one to compute three-dimensional quantities, such as the 3D 
power spectrum and correlation function. However, the other parameters characterizing the polarization 
signal are better defined as projected onto the plane of the sky, with a dependence on the impact 
parameter 
distance from the center of the emission source (instead of radial distance), and integrated along 
the line of sight within the source halo. 

     We assume the validity of the flat-sky approximation throughout, 
implying that our expressions are consistent with the full curved-sky calculation at 
multipole values $\ell \gg 1$.

\subsection{The 2D \lya Polarization Power Spectra}\label{sec:2d}

  We start by expressing the real-space projected signal of $I$ and $\po$ around 
a halo of mass $M$, and at redshift $z$, as the product of the total amplitude and the profile 
shape, $I(M)\,u_I(r_{\perp}|M,z)$ and $I(M) \,u_{\po}(r_{\perp}|M,z)$, respectively. 
Here, the amplitude of the intensity only depends on halo mass (see \S~\ref{sec:toy}), and 
$\int {\rm d}r_{\perp} 2\pi r_{\perp} u_{I}(r_{\perp}|M,z) = 1$, where $r_{\perp}$ 
is the comoving impact parameter from the center of the halo, and $u_I$ is the intensity profile shape. 
Because the polarization degree, $\Pi$, is not directly an additive quantity (one needs to count the 
intervening photons instead), we do not normalize the profile in this case and simply express the entire 
signal as $u_{\Pi}(r_{\perp}|M,z)$. 
With these definitions, we can then write $u_{\po}(r_{\perp}|M,z)  = u_I(r_{\perp}|M,z) \,
u_{\Pi}(r_{\perp}|M,z)$, and equivalently for the amplitude of the polarized intensity, $\po(M) =  I(M) \int {\rm d}r_{\perp} 2\pi r_{\perp} u_{\po}(r_{\perp}|M,z)$. 

   The projected Fourier transforms of these real-space profiles are 
\begin{align}\label{eq:fouri}
\tilde u_{\{I, \po,\Pi\}}(\ell|M,z) =  \int_0^\infty {{\rm d}\theta}\, 2 \pi \theta\, 
J_0(l\theta)\, u_{\{I, \po,\Pi\}}\left(\theta=\cfrac{r_{\perp}}{D_{\rm A}(z)}|M,z\right)   ~,  
\end{align}
where $\theta$ is the angular distance from the center of the halo, resulting from dividing 
$r_{\perp}$ by the comoving angular diameter distance, $D_{\rm A}(z)$, 
and the term $J_0(l\theta)$ is the Bessel function of the first kind and zeroth order. 

    Finally, the one- and two-halo terms of the projected power spectra for $I$, $\po$ and $\Pi$ are 
computed as \citep[see appendix A in][]{Hill2013}\footnote{ A simple way to view these 
expressions is considering the usual projection of the three-dimensional (3D) power spectra components 
along the line of sight \citep[e.g., equation 37 and appendix A in][for the case of near-infrared continuum 
radiation]{Fernandez2010}. Here, however, the 3D halo profiles in the 3D power calculation, $\tilde u(k)$, 
with $k$ denoting the 3D Fourier modes, are replaced by their two-dimensional (2D) counterparts, $\tilde u(\ell)$, 
both related under the Limber approximation \citep{Limber1953} as $\tilde u(\ell)\approx \tilde u(k) / 
\chi^2(z)$, where $\chi(z)$ is the comoving distance to redshift $z$.}
\begin{align}\label{eq:h1}
C_{\ell,\,\{I, \po, \Pi \}}^{\rm 1h} = \int {\rm d}z \frac{{\rm d}^2V}{{\rm d}z\,{\rm d}\Omega} \int  {\rm d}M \frac{{\rm d}n} {{\rm d}M} \, 
 w^2(M) \,|\tilde u_{\{I, \po,\Pi\}}(\ell|M,z)|^2 ~, 
\end{align}
and
\begin{align}\label{eq:h12}
C_{\ell,\,\{I, \po, \Pi \}}^{\rm 2h} = \int {\rm d}z \frac{{\rm d}^2V}{{\rm d}z\,{\rm d}\Omega} P_{\rm lin} \left( k=\frac{\ell + 1/2}{\chi(z)},z \right) 
\left[ \int  {\rm d}M \frac{{\rm d}n} {{\rm d}M} \, b_{\{I, \po, \Pi \}} \,   w(M) \,\tilde u_{\{I, \po, \Pi \}}(\ell|M,z) \right]^2 .   
\end{align}
The term ${{\rm d}^2V}/{{\rm d}z\,{\rm d}\Omega} = {c\, \chi^2(z)}/{H(z)}$ is the comoving 
volume element per steradian and redshift, where $H(z)$ is the Hubble parameter, the speed of light 
is denoted by 
$c$, and $\chi(z)$ is the comoving radial distance to redshift $z$. In the above 
expressions, ${{\rm d}n(M,z)} /{{\rm d}M}$ represents the comoving number density of halos, 
which depends on halo mass and redshift, and 
\begin{align}\label{eq:w}
w(M)= \left\{ \begin{array}{ll} 
I(M)   \hspace{1mm}   \hspace{1mm}      \hspace{3mm}  & \mbox{for}  \hspace{2mm}   \tilde u_{\{I,\,\po\} }  ~;\\[10pt]  
\cfrac{I(M)}{\bar I}    \hspace{5mm}   \hspace{1mm}        \hspace{3mm}  & \mbox{for}    \hspace{2mm} 
\tilde u_{\Pi}   ~,\end{array}  \right.
\end{align} 
where $\bar I = \int{\rm d}M \frac{{\rm d}n} {{\rm d}M} \,   I(M)$. 
The use of $I(M)/\bar I$ for the polarization degree is motivated by the fact 
that, in practice, the polarization fraction from observations  (simulations) in a given pixel (cell) 
$i$ is obtained by adding the contribution of all halos $j$ as $\Pi_i=\sum_j \po_{ij} /  \sum_j I_{ij}$. 
We equate this expression in our formalism by weighting the halos by their intensity as
$\Pi_i=\sum_j \Pi_{ij} I_{ij}/  \sum_j I_{ij}$, where we have used that for an individual halo 
$\po_{ij} = \Pi_{ij} I_{ij}$.
Finally, $P_{\rm lin}(k)$ in Eq.~\ref{eq:h12} denotes the linear 3D matter density power spectrum and 
\begin{align}
b_{\{I, \po, \Pi \}} = \left\{ \begin{array}{ll} 
\cfrac{\int {\rm d}M \frac{{\rm d}n} {{\rm d}M} \, b(M)\,   I(M)}{\bar I} &~ \mbox{for}  \hspace{4mm}   {I, \po }  ~;\\[12pt]  
b(M) &~ \mbox{for}    \hspace{4mm} 
{\Pi}   ~,\end{array}  \right.
\end{align} 
where $b(M)$ denotes the bias for a halo of mass $M$. We adopt an intensity-weighted bias for 
$I$ and $\po$. However, because in our formalism the extent of $\Pi$ is mostly related to the mass 
of the halo through the virial radius (\S~\ref{sec:toy}), we simply use the halo bias in this case. 
Biases weighted according to other parameters and properties, e.g., star-formation rate, may be 
also appropriate depending on the characteristics of the analysis. 

     An additional consideration in the power spectrum calculation is 
the shot noise, or Poisson noise, that arises from the discrete sampling of a continuous field.
In our models we assume that the \lya emission is nearly a continuous field, owing to the fact that 
although \lya photons are sourced by halos, the signal is diffused away from the 
central source due to scattering. This extended diffuse 
emission is described by the \lya profile in our `one-halo term', which strictly speaking, should be 
regarded as arising from the Poisson shot noise of discrete sources convolved with the \lya profile
\citep[see the discussion in][for the case of 21cm studies]{Wolz2019}. We have tested that 
the strictly-defined $I$ and $\po$ shot-noise terms overall match the amplitudes of the respective 
one-halo terms of the power spectra at large scales. 

   Our calculations, therefore, do not include an additional term in the power spectra accounting for the 
shot noise, as this is essentially our one-halo term. We emphasize that the \lya one-halo term does 
not explicitly encapsulate the usual non-linear structure of matter. However, since we use a profile 
directly from observations that do not resolve small-scale structure, some amount of contribution 
from faint galaxies populating the dark matter halo may have already been captured 
\citep[e.g., see the impact of clustered sources on the extended profiles in][]{Masribas2017b}. 
We leave a more detailed study to future work.

\subsection{The $E$ and $B$  Modes of \lya Polarization}\label{sec:eb}

   We derive now the so-called $E$ and $B$ modes for the case of \lya polarization. These 
two quantities are related to the $Q$ and $U$ Stokes parameters, but they allow us to obtain  
polarization information in a coordinate-system-independent manner. The $E$ and $B$ formalism 
was introduced for CMB analysis by \cite{Kamionkowski1997} and \cite{Zaldarriaga1997}, 
and is briefly summarized below. 

    The total and linearly-polarized intensities, as well as the degree of 
polarization, $I, \po$ and $\Pi$, respectively, are scalar (spin $s=0$) quantities, and, therefore, 
their values are invariant under rotations of  the coordinate system that defines them. 
The Stokes $Q$ and $U$ 
parameters, however, depend on the fixed coordinate system, and transform under a rotation of the 
coordinate axes by an angle $\alpha$ in the plane of the sky as 
\begin{align}
Q^\prime =& ~Q \cos 2\alpha + U \sin 2\alpha ~, \\ \nonumber  
U^\prime =& -Q \sin 2\alpha + U \cos 2\alpha ~ .
\end{align} 
Equivalently to the Stokes parameters, polarization can be described using complex numbers, by 
means of the spin $s=\pm2$ fields $_s f = (Q\pm iU)$, that transform under rotations as 
$_s f^\prime = e^{- i s \alpha }  {_s}f$. These fields 
are invariant under a rotation of angle $2 \pi / s = \pm \pi$, owing to the value of their spin, 
but the exact value, as it was the case for $Q$ and $U$, still depends on the orientation of the 
coordinate system. 

     To avoid the dependence on coordinate system, \cite{Zaldarriaga1997} introduced 
two new rotationally-invariant (spin $s=0$) quantities, a.k.a. the $E$ and $B$ modes, that are a 
combination of the $Q$ and $U$ parameters, but independent of the coordinate system. 
In brief, $E$ modes are scalar quantities with similar properties as those of the divergence 
of the electric field, 
while $B$ modes are pseudo-scalars related to the curl of the magnetic field. 
The $E$ and $B$ nomenclature, thus arises from the respective connections to the electromagnetic 
field, and the scalar and pseudo-scalar nature of the modes relies on the sign conservation, or not,   
under parity transformation, respectively. We refer the interested reader to \cite{Zaldarriaga1997} 
and \cite{Zaldarriaga2001} for the quantitative derivation of the CMB $E$ and $B$ modes, and to 
\cite{Kamionkowski1997} for an equivalent approach.   

    To derive the {\it astrophysical} $E$ and $B$ modes of \lya here, it is convenient to make use of the 
small-scale limit (or flat-sky) approximation, which assumes that the sphere denoting the sky can be 
locally treated as a plane. This is valid in our case, since we mostly focus on small distances 
($\ell \gg 1$), 
where curvature effects are small. In the flat-sky approximation, the 
decomposition of a quantity into spherical harmonics can be replaced by a simple expansion in 
plane waves  \citep[e.g.,][]{Seljak1997}, which allows us to write the $E$ and $B$ modes 
 as a simple rotation of the $U$ and $Q$ parameters in Fourier space as  
\citep{Zaldarriaga2001,Kamionkowski2017} 
\begin{align}
\tilde E({\bm \ell}) =& ~ \tilde Q ({\bm \ell}) \cos 2 \psi + \tilde U({\bm \ell}) \sin 2 \psi~, \nonumber  \\
\tilde B({\bm \ell}) =& - \tilde Q ({\bm \ell}) \sin 2 \psi + \tilde U({\bm \ell}) \cos 2 \psi ~ ,
\end{align} 
where $\psi$ represents the angle between the multipole ${\bm \ell}$ and the $\uvec x$ Cartesian axis.
Let us next express the Stokes parameters in real space, by accounting for their (inverse) Fourier 
transform, and considering the tangential and parallel components with respect to the (radial) direction 
toward the center of the source (i.e., $Q_{\rm r}$ and $U_{\rm r}$, respectively)\footnote{We use the 
nomenclature $Q_{\rm r}$ and $U_{\rm r}$ here because of the similarities with the gravitational lensing 
approach, where these quantities describe the {\it tangential} and {\it cross} components of the shear, 
respectively \citep[see, e.g.,][]{Schneider2005}.} for reasons that will become clear below. 
For the case of $\tilde E$, this new expression equates 
\begin{align}
\tilde E(\bm\ell) =  \iint \theta\, {\rm d}\theta\, {\rm d}\phi \,e^{-i \bm\ell \bm{\theta}}  &  \left[ Q_{\rm r} ({\theta})\cos 2 \phi \cos 2 \psi - U_{\rm r}({ \theta}) \sin 2 \phi \cos 2 \psi \right.  \nonumber \\
& \left. +\: Q_{\rm r} ({\theta}) \sin 2 \phi\sin 2 \psi + U_{\rm r}({ \theta}) \cos 2 \phi \sin 2 \psi \right]~,
\end{align}     
where $\theta$ and $\phi$ represent the angular 
colatitude and longitude, respectively, on the sphere (note that ${\bm \theta}$ and ${\bm \ell}$ become 
$\theta$ and $\ell$ because we assume spherical and circular symmetry for the sky and projected halos, 
respectively). Grouping now the $Q_{\rm r}$ and $U_{\rm r}$ terms, and applying trigonometric 
relations, we can write 
\begin{align}
\tilde E(\ell) &=  \int \theta\, {\rm d}\theta\int {\rm d}\phi \,e^{-i \ell {\theta} \cos(\phi - \psi)}   \left[ Q_{\rm r} ({\theta})\cos 2 (\phi- \psi)   + U_{\rm r}({ \theta}) \sin 2 (\phi -  \psi) \right] ~. 
\end{align}  
The second integral above, vanishes for the term containing $U_{\rm r}$ when integrated over $2\pi$. 
For the term containing $Q_{\rm r}$, it can be expressed as a Bessel function of the first kind and 
second order,  $J_2(\ell \theta)$.  A similar derivation, now for the case of $\tilde B$, results in 
reversed surviving and vanishing $U_{\rm r}$ and $Q_{\rm r}$ terms. Thus, the final expressions for the 
two quantities, and for an individual halo of mass $M$ and redshift $z$, are  
\begin{align}\label{eq:ebjqu}
\tilde E(\ell| M,z) =& - \int {\rm d} \theta  \, 2\pi \theta\, J_2(\ell \theta)  \,Q_{\rm r}({ \theta}, M) ~, \nonumber \\
\tilde B(\ell| M,z) =& - \int {\rm d} \theta \, 2\pi \theta\, J_2(\ell \theta)  \,U_{\rm r}({ \theta}, M)  ~.
\end{align} 
The above equations show that each polarization mode is contributed uniquely by 
one of the Stokes parameters, integrated over a circle around the source, analogously to the CMB 
case \citep{Zaldarriaga2001}. In detail, the expression for $\tilde E$ resembles that of the Fourier 
transform of $u_\po$ in Eq.~\ref{eq:fouri}, but with $J_2(\ell \theta)$ instead of $J_0(\ell \theta)$. 
This Bessel function term is the only difference between the final expressions for the power spectra of 
$\po$ and $\tilde E$, and it gives rise to the different power spectra for these quantities displayed 
in \S~\ref{sec:results}.

   Finally, the \lya $\tilde E$ and $\tilde B$ modes just derived above can be used to obtain two additional 
power spectra for polarization, similarly as for $I, \po$ and $\Pi$, via the expressions  
\begin{align}
C_{\ell,\{ \tilde E, \tilde B \}}^{\rm 1h} & = \int {\rm d}z \cfrac{{\rm d}^2V}{{\rm d}z\,{\rm d}\Omega} \int  {\rm d}M \frac{{\rm d}n} 
{{\rm d}M}  \, | \{ \tilde E, \tilde B \}(\ell|M,z)|^2 ~, 
\end{align}
and 
\begin{align}
C_{\ell,\{\tilde E, \tilde B \}}^{\rm 2h} & = \int {\rm d}z \cfrac{{\rm d}^2V}{{\rm d}z\,{\rm d}\Omega}\, P_{\rm lin} \left( k=\frac{\ell + 1/2}{\chi(z)},z \right)  \left[ \int  {\rm d}M \frac{{\rm d}n} {{\rm d}M} \, b_I\,   \{ \tilde E, \tilde B \}(\ell|M,z) \right]^2~ . 
\end{align}
Here we have not included the term $w$, because the amplitude is incorporated into 
$\tilde E$ and $\tilde B$ through $Q_{\rm r}$ and $U_{\rm r}$ (Eq.~\ref{eq:ebjqu}), respectively, and 
we have considered the intensity-weighted bias.

\subsection{The Cross-correlation Power Spectra of \lya Polarization}\label{sec:cross}

      One can further calculate the cross power spectra between the $\tilde E$, $I$,  $\po$ and $\Pi$ 
parameters, because all these quantities have even parity.   The cross power between $B$ and any 
of the other quantities, however, is identically zero because $B$ changes sign under parity 
transformations \citep{Newman1966}. 

The cross power for two distinct $X$ and $Y$ polarization quantities is computed as 
\begin{align}\label{eq:h1x}
C_{\ell,\,XY}^{\rm 1h} = \int {\rm d}z \frac{{\rm d}^2V}{{\rm d}z\,{\rm d}\Omega} \int  {\rm d}M \frac{{\rm d}n} {{\rm d}M} \, 
 w_X(M) \,|\tilde u_X(\ell|M,z)| \,w_Y(M) \,|\tilde u_Y(\ell|M,z)| ~, 
\end{align}
and
\begin{align}\label{eq:h2x}
C_{\ell,\,XY}^{\rm 2h} = \int {\rm d}z \frac{{\rm d}^2V}{{\rm d}z\,{\rm d}\Omega}  P_{\rm lin} \left( k=\frac{\ell + 1/2}{\chi(z)},z \right) &
\left[ \int  {\rm d}M \frac{{\rm d}n} {{\rm d}M} \, b_{X} \,   w_X(M) \,\tilde u_{X}(\ell|M,z) \right] \\ \nonumber
 \times \, & \left[ \int  {\rm d}M \frac{{\rm d}n} {{\rm d}M} \, b_{Y} \,   w_Y(M) \,\tilde u_{Y}(\ell|M,z) \right]   ~ ,   
\end{align}
where $w=1$ for $\tilde E$.

\section{\lya Polarization in Galaxy Halos}\label{sec:polalpha}

   This section details the nature and characterization of the polarized \lya emission around galaxies. 
In \S~\ref{sec:lyaintro}, we summarize the theoretical aspects of \lya polarization in astrophysical 
(galaxy) environments. Then, in \S~\ref{sec:toy}, we describe the modeling of  
the \lya emission profiles used in our calculations.

\subsection{Introduction: \lya Scattering and Polarization}\label{sec:lyaintro}

   The scattering process of \lya radiation is constituted by the absorption and subsequent 
reemission of photons by neutral hydrogen, H{\sc i}, atoms. Considering the typical temperature 
of the neutral hydrogen gas of $T=10^4$ K, the probability for a \lya photon 
to be absorbed (scattered) depends mostly on the hydrogen column density, $N_{\rm HI}$, 
through the optical depth, and on the position of the photon within the line profile, defined 
by a dimensionless variable $x$, as \citep{Rybicki1979}
\begin{align}\label{eq:optdepth}
\frac{\tau_x}{\tau_0} = \frac{a}{\pi} \int_{-\infty}^{\infty} \cfrac{e^{-y^2} {\rm d}y}{(y-x)^2 + a^2} = 
\left\{ \begin{array}{ll} 
\sim e^{x^2} &~ \mbox{core}    ~;\\[2pt]  
\sim \cfrac{a}{\sqrt{\pi}x^2 } &~ \mbox{wing}   ~, \end{array}  \right.
\end{align} 
where $x\sim3$ separates the line core and the wings \citep{Dijkstra2008}.
Here, $x \equiv (\nu - \nu_0) / \Delta {\nu_{\rm D}}$, with $\nu$ denoting 
the photon frequency, and where $\nu_0 = 2.47 \times 10^{15}\,{\rm Hz}$ is the \lya resonance frequency. 
The term $\Delta \nu_{\rm D}\equiv \nu_0 v_{\rm th}/c $ is the thermal or Doppler line width, 
and $v_{\rm th}=\sqrt{2k_{\rm B}T/m_{\rm p}}$
is the thermal velocity of the hydrogen atoms, where $k_{\rm B}$ and $m_{\rm p}$ 
are the Boltzmann constant and the proton mass, respectively. Finally,  
$a=A_{21}/4 \pi \Delta \nu_{\rm D} = 4.7 \times 10^{-4}\,(v_{\rm th}/13 \, 
{\rm km\,s^{-1}})^{-1}$ is the Voigt parameter, where $A_{21} = 6.25 \times 10^{8}\,{\rm s^{-1}}$ 
is the (spontaneous de-excitation) Einstein A coefficient for the \lya transition, and 
\begin{align}
\tau_0 = 5.9\times10^6 \left( \frac{N_{\rm HI}}{10^{20}\,{\rm cm^{-2}}} \right) \left( \frac{T}{10^4\,{\rm K}} \right)^{-0.5}
\end{align}
denotes the optical depth at the line center. In summary, Eq.~\ref{eq:optdepth} implies that when  
\lya photons reach the wing of the line profile, they have a high (low) probability to escape 
(be absorbed by) the neutral medium that they inhabit. 

    In the scattering process, the thermal motion of the atoms, and the possible additional 
velocity component from the bulk motion of the medium that these atoms inhabit (e.g., galactic inflow or 
outflow), introduce a Doppler shift in the photon frequency and, therefore, a change in the photon 
energy measured before and after scattering in the observer frame.  As a result of this effect, 
the \lya photons undergo scattering events in the neutral hydrogen medium until the Doppler 
shift places them far enough in the wings of the line profile that they escape the medium freely.  
The escape of \lya photons from an optically 
thick medium typically occurs after $\sim \tau_0$ scattering events \citep{Osterbrock1962,Auer1968}. 

    In the scattering event, the \lya photon acquires a degree of polarization that depends on the 
angle of scattering, $\beta$, defined by the directions of the incoming and outgoing photons, and 
the position of the photon within the line profile as \citep{Dijkstra2008}
\begin{align}\label{eq:poleq}
\Pi (\beta) = \left\{ \begin{array}{ll}  
\cfrac{\sin^2\beta}{\frac{11}{3} + \cos^2\beta}    &~ \mbox{core}    ~;\\[18pt]  
\cfrac{\sin^2\beta}{{1} + \cos^2\beta}  &~ \mbox{wing}   ~. \end{array}  \right.
\end{align}
The expressions above show that wing scattering \citep[equivalent to Rayleigh 
scattering;][]{Stenflo1980} can introduce a degree of polarization about three times larger 
than that in the core \citep[described by the superposition of Rayleigh and isotropic 
scattering;][]{Brandt1959,Brasken1998}, and as high as $100\,\%$ 
for a scattering angle of $\beta = \pi/2$ \citep{Chandrasekhar1960,Lee1998}. 
Therefore, because wing scattering also 
implies a high escape probability, the observed scattered \lya radiation  can 
carry a large (detectable) degree of polarization. 

    \cite{Dijkstra2008} performed  \lya radiative transfer calculations in a spherically-symmetric outflowing 
H{\sc i} shell of fixed column density around a central source, representing an idealized environment 
surrounding star-forming galaxies at high redshift. They found two results especially relevant to 
our work: (i) the angle of polarization is perpendicular to the impact parameter vector connecting 
the radiation source and the point of last scattering -- before the photons escape the medium 
toward the observer.  (ii) The degree of polarization increases with impact parameter, from a few 
percent at the center of the galaxy, up to tens of percent at larger projected distances. The 
exact polarization fraction values at large distances depend strongly on the column density of 
the scattering gas, and they fluctuate between $\sim 20\,\%$ and $\sim 40\,\%$ for typical galaxy 
columns of $\log(N_{\rm H{\sc I}}/{\rm 
cm^{-2}}) = 19$ and $\log(N_{\rm H{\sc I}}/{\rm cm^{-2}}) = 20$, respectively \citep{Dijkstra2008}. 
The increase of the degree of polarization with impact parameter can be understood by  
considering the {\it effective} angle of scattering of photons observed at a given distance.    
Radiation observed at the center of the source is mostly contributed by photons that have an effective 
(almost) null or $\pi$ (backscattered) scattering angle, resulting in low levels of polarization due 
to the angular dependence of the polarization degree (Eq.~\ref{eq:poleq}). Radiation observed far from 
the center generally has scattering angles closer to $\pi/2$, where the polarization is maximized 
\citep[see figure 1 in][]{Bower2011}.  
The dependence on column density is a consequence of the number of scattering events 
before the escape of the \lya photons.  Large column densities, $\log(N_{\rm H{\sc I}}/{\rm cm^{-2}}) 
\gg 20$, result in many scattering events, which yields to a more isotropic radiation field and, 
in turn, reduces the overall polarization \citep[e.g.,][]{Lee1998}. When the column densities are small, 
$\log(N_{\rm H{\sc I}}/{\rm cm^{-2}} \sim 18-19$), a few scattering events per photon occur and 
thus the average polarization value is closer to the values introduced by the wing scattering prior 
to the escape of the photons \citep[e.g.,][]{Kim2007}. 
The velocity of the gas has a similar effect on the degree of polarization as the column density. High 
velocities produce large Doppler shifts and therefore many photons scatter in the wings of the line 
profile in the first scattering event. This introduces high polarization and high probability 
of escape. Finally, the sign of the velocity vector (inward or outward) makes no difference for this effect, 
so outflows and inflows contribute the same to the polarization of the radiation field \citep{Dijkstra2008}.

   In summary, \lya scattering in a spherically-symmetric outflowing medium around an isotropic 
radiation source will result in an increasing degree of polarization with impact parameter, with values 
that give information about the column density and the motion of the gas, and with the polarization 
angle perpendicular to the radius vector between the center and the last-scattering position.  
We model the polarization around galaxies following this idealized scenario in the following section.

\subsection{Extended \lya Emission Modeling}\label{sec:toy}

   We describe here the modeling of the extended \lya emission in galaxy halos. 
Traditional intensity mapping studies typically model the one-halo term with shot noise, but we 
show in \ref{sec:ps} that using a realistic one-halo term is important 
when considering polarization. Furthermore, \cite{Visbal2018} showed that the resonant nature 
of \lya radiation results in extended emission at high redshifts that can yield inaccurate  
shapes of the power spectra when not taken into account. 

   In \S~\ref{sec:lyaem} below, we detail the calculation of the projected profile for the total \lya 
emission in halos, and we present the calculations for the profile of the polarization fraction in 
\S~\ref{sec:lyapo}. In \S~\ref{sec:masslum}, we detail the halo-mass function and the 
relation between halo mass and luminosity used in the calculations. Spherical symmetry 
around the sources is assumed in all cases. 

\subsubsection{Projected Profile for the Total \lya Emission }\label{sec:lyaem}

   For the projected total \lya emission, we use the analytical surface brightness profile  shape 
derived in \citealt{Masribas2016b} and \cite{Masribas2017}, for a galaxy at the center of the halo. 
This profile shape is based on the numerical simulations of the H{\sc i} distribution around 
Lyman break galaxies at $z\sim 3$ by \cite{Rahmati2015}.  In \cite{Masribas2017}, 
we showed that this profile broadly matched the observed extended \lya surface brightness profiles 
of \cite{Momose2014} at redshift $z=5.7$, and the (compact) \lya profiles of \cite{Jiang2013} at 
$z=5.7$ and $z=6.6$. 
We explore the impact of variations in the general profile shape used here in \S~\ref{sec:dep} in the 
Appendix, but we leave more detailed calculations considering the possible dependence of the profile 
shape on halo mass and redshift to future work using numerical radiative transfer and cosmological 
simulations.

  The \lya surface brightness profile shape is expressed as 
\begin{align}
S_{Ly\alpha}(r_{\perp}) \propto \int_{r_{\perp}}^\infty \cfrac{r\,{\rm d}r}{\sqrt{r^2 - r_{\perp}^2}} \,f_c(r)\,
f_{\rm esc}^{\rm ion}(r)\, \cfrac{1}{4\pi r^2} ~.
\end{align}
Here, the integral is over the \lya emission along the line-of-sight at a given impact parameter 
$r_{\perp}$, the term $1/4\pi r^{2}$ is the geometric dimming effect, and 
$f_c(r)$, and $f_{\rm esc}^{\rm ion}(r) = {\rm exp}\left[ -\int_0^\infty f_c(r)\,{\rm d}r \right]$, are the 
radial H{\sc i} covering factor, and the escape fraction of ionizing photons, respectively. In detail, 
the term $f_c(r)$ denotes the number of H{\sc i} gas clumps  along a differential length at a 
distance $r$ from the center of the source, and it is obtained after applying an inverse Abelian 
transformation to the two-dimensional neutral gas covering factor in \citealt{Rahmati2015} (see 
\citealt{Masribas2016} for details in the calculations, and the dashed curves in Figure 1 of 
\citealt{Masribas2017} for a visualization of these profiles). For the current calculation, we  
disregard the potential impact of the origin of the \lya emission, i.e, fluorescence in this case, on the 
polarization signal \citep[see][for a discussion on these origins]{Masribas2017b}. We simply use this 
profile shape because it is consistent with observations, and assume that the \lya photons 
result in the polarization profile described below. Future radiative transfer simulations will 
explore departures from this idealized case.

\begin{figure*}\center 
\includegraphics[width=0.45\textwidth]{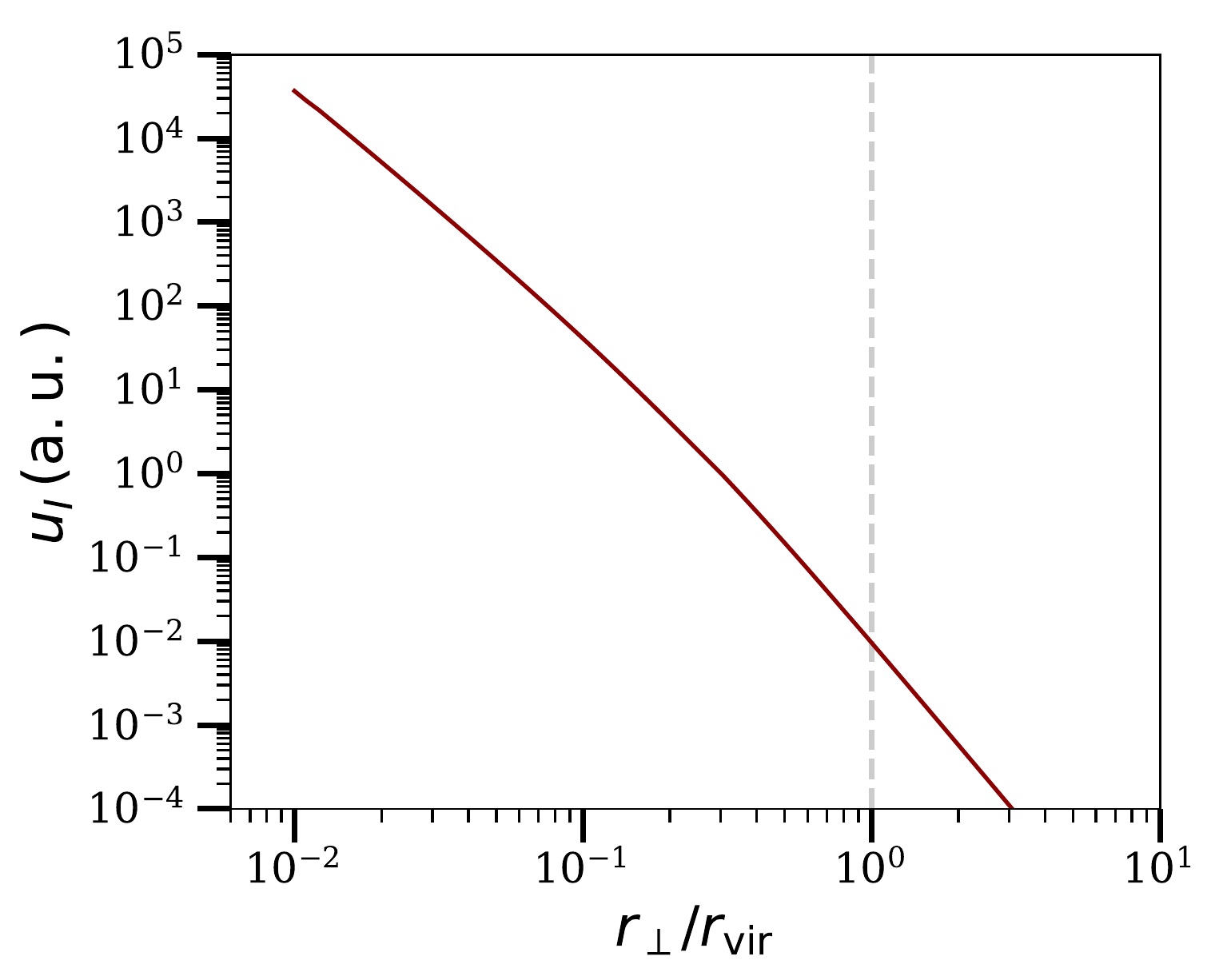}\includegraphics[width=0.45\textwidth]{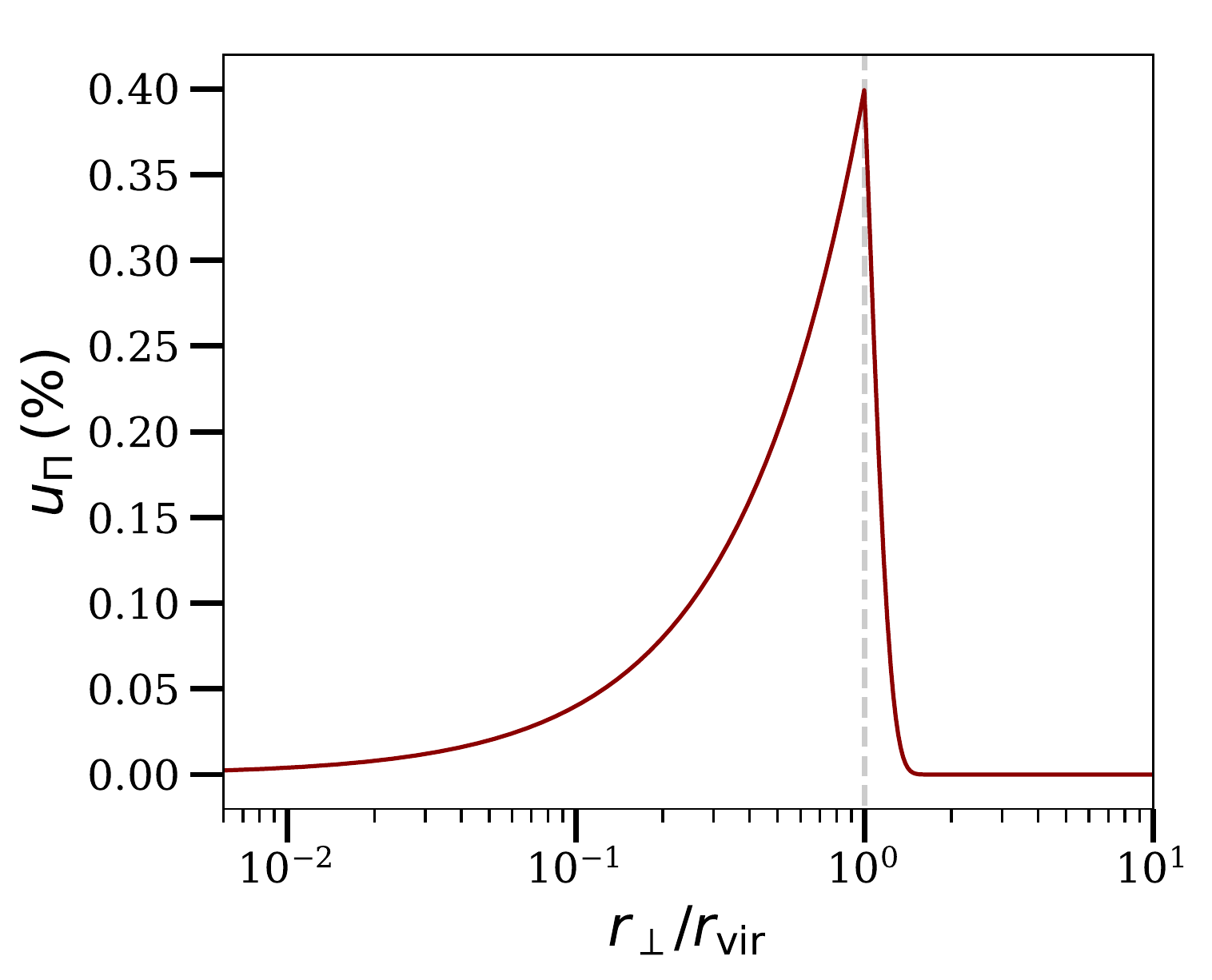}
\caption{Projected profiles for the normalized total \lya intensity ({\it left panel}), and the 
polarization fraction ({\it right panel}), with impact parameter. The {\it vertical dashed lines} denote 
the position of the virial radius. } 
\label{fig:profiles}
\end{figure*}

   Finally, the \lya intensity profile can be written as 
\begin{align}\label{eq:lyaem}
u_{I} (r_{\perp}|M,z) =  \cfrac{S_{Ly\alpha}(r_{\perp})}{\int_0^\infty 2 \pi r_{\perp} {\rm d}r_{\perp} S_{Ly\alpha}(r_{\perp})}  ~, 
\end{align}
where the denominator acts as a normalization constant. Although the intensity profile can depend 
on halo mass and redshift, note that, with this derivation, the profile is independent on these 
quantities. The {\it left panel} in Figure \ref{fig:profiles} shows the resulting normalized profile, 
where the {\it dashed line} denotes the position of the virial radius, for reference.

\subsubsection{Projected Profile for the Polarization Fraction }\label{sec:lyapo}
   
         We model the projected profile of the polarization degree around the halos as a linear increase 
with impact parameter, followed by a steep decrease after peaking at the virial radius, as 
\begin{align}\label{eq:lyapo}
u_{\Pi} (r_{\perp}|M,z) = \Pi_{\rm max} \times \left\{ \begin{array}{ll}  
 \cfrac{r_{\perp}}{r_{\rm vir}}    &~ r_{\perp} \leq r_{\rm rvir}(M,z)   ~;\\[18pt]  
 e^{\left[ 1- \left( r_{\perp}/r_{\rm vir} \right)^5 \right] }  &~ r_{\perp} > r_{\rm rvir}(M,z)   ~,  \end{array}  \right.
\end{align}
where  $\Pi_{\rm max} = 40\,\%$ is the maximum polarization fraction value 
at the virial radius, and $r_{\rm vir}(M,z)$ introduces the dependence on halo mass and redshift. 
The exact dependence on impact parameter is set arbitrarily, and variations are explored in 
\S~\ref{sec:dep} in the Appendix. However, the shape and maximum polarization fraction value for this 
profile agree with those found in the radiative transfer simulations of \cite{Dijkstra2008} and 
\cite{Dijkstra2012}. Furthermore, the profile is also broadly consistent with the simulations and 
observations of the polarization degree around the giant \lya nebula  LAB1 \citep{Steidel2000} 
by \cite{Trebitsch2016} and \cite{Hayes2011}, respectively. The {\it right panel} of Figure 
\ref{fig:profiles} illustrates this polarization fraction profile with impact parameter. 

     A strong assumption in our model is the sharp cut-off of the polarization signal at a given impact 
parameter. In reality, the polarization signal will extend out in the halo as long as the scattering of \lya 
photons exists.  At large impact parameter, however, the number of photons is largely reduced, the 
exact number depending on the slope adopted for the surface brightness, and only a few photons will 
contribute to the polarization signal. We explore the impact of a flat surface brightness profile, 
different values for the position of the cut-off, as well as a smoother cut-off slope in  \S~\ref{sec:dep} in 
the Appendix. Overall, as we will show, these changes have little effect on the one-halo terms of the 
power spectra of $\Pi$. In detail, the peak of the one-halo terms can be broader or narrower, but it is 
always well resolved at large multipole values. This is because the sharp shape of the one-halo 
terms at large $\ell$ arises mostly from the fact that the polarization fraction profile increases with 
impact parameter, contrary to the case of the surface brightness for which the signal decreases with 
distance. The sharp cut-off only impacts the shape of the low-multipole side of the power spectrum 
peaks. 

\begin{figure*}\center 
\includegraphics[width=0.5\textwidth]{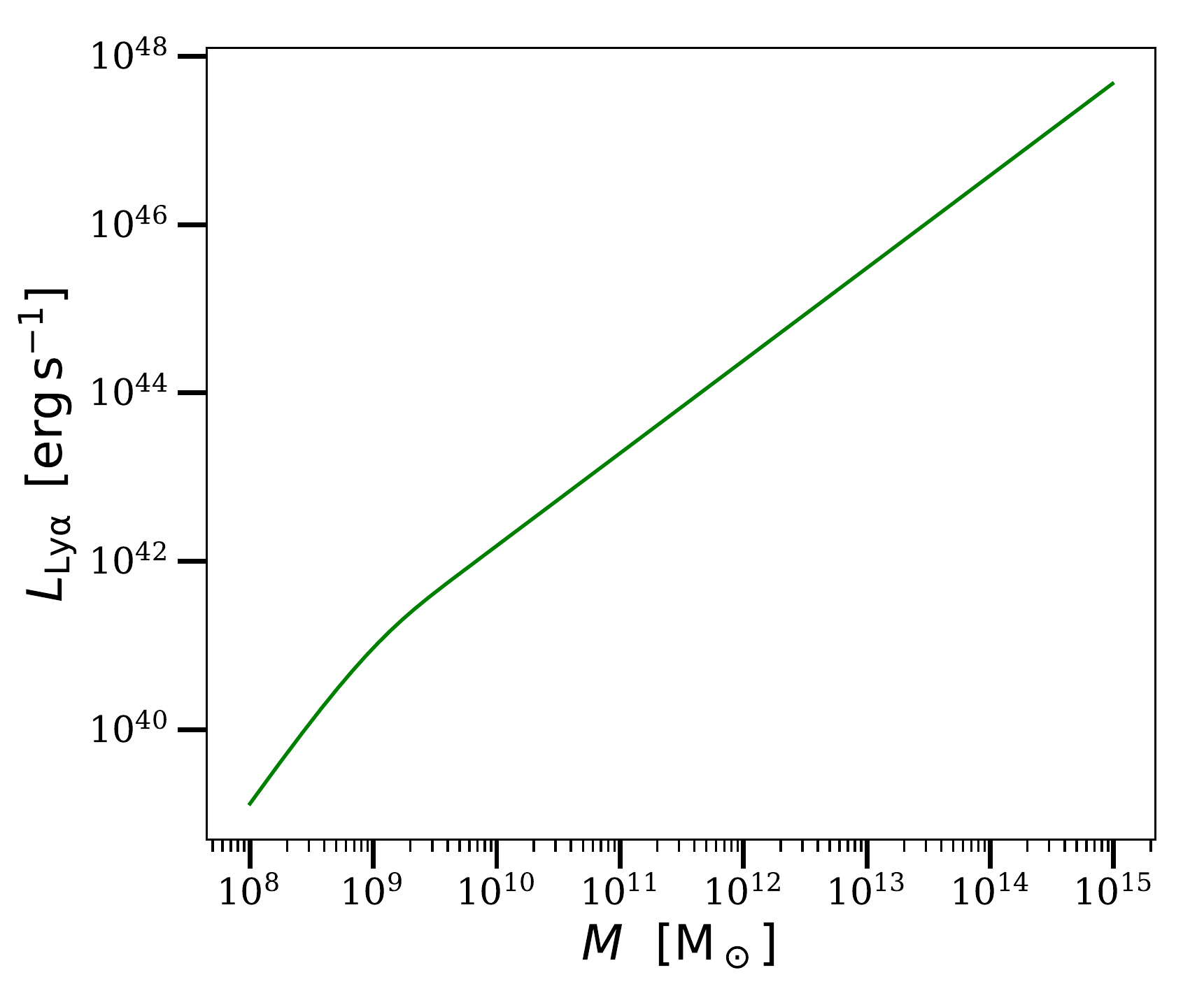}
\caption{Redshift-independent relation between halo mass and \lya luminosity described by 
Eq.~\ref{eq:inoue}. Note that the halo mass in this panel does not include the reduced Hubble 
constant term.} \label{fig:lmplot}
\end{figure*}

\subsubsection{Halo Mass and Luminosity Relation}\label{sec:masslum}

     For our calculations, we use the \cite{Tinker2008} comoving halo-mass 
functions, covering the mass range $8 \leq \log(M/{\rm M_\odot\, h^{-1}}) 
\leq 15$.  

      To relate the halo mass and the \lya luminosity, we use the expression derived by \cite{Inoue2018},   
\begin{align}\label{eq:inoue}
\frac{L_{\rm Ly\alpha} (M)}{10^{42} \,{\rm erg\,s^{-1}}} =  \left(\frac{M}{\rm 10^{10} \,  M_\odot}\right)^{1.1} \left[1 - \exp{-{\frac{10 M}{\rm 10^{10} \,  M_\odot}}}\right] ~, 
\end{align}
illustrated in Figure \ref{fig:lmplot}, and where the halo mass, $M$, now does 
not carry the reduced Hubble constant term. We use this expression for all redshifts in this work, although 
it was derived regarding observations at $5 \lesssim z \lesssim 7$. Finally, the term $I(M)$ in the power 
spectrum equations corresponds to 
\begin{align}
I(M,z) = \frac{L_{\rm Ly\alpha}(M)}{4\pi \, (1+z)}  ~, 
\end{align}   
which gives rise to the units for the $\nu I_\nu$ power spectra. The term 
$(1+z)^{-1}$ disappears when considering the specific intensity power spectra, $I_\nu$.

    Our formalism assumes that all the \lya photons produced via Eq.~\ref{eq:inoue} will be observed, 
i.e., it ignores the effect of the escape fraction of \lya photons. However,  in our model, only 
dust  contributes to the escape fraction value, not neutral gas. The neutral hydrogen gas can diffuse 
the \lya emission far from the source via scattering, but the number of \lya photons is conserved, 
contrary to the case of dust where the photons are mostly destroyed. Therefore, the use of \lya 
escape fraction values that arise from measuring the removal of photons along the line of sight 
covering the central regions of galaxies should be avoided. For simplicity, we have also ignored 
the potential effect from a galaxy duty cycle. Because \lya arises mostly from young stars, this 
effect may be considerable for massive halos with old stellar populations \citep[e.g.,][]{Ouchi2018}. 
Similarly, the effect of varying star-formation rates and efficiencies with redshift, as well as the 
dispersion around the mean values, could be important 
\citep[see, e.g.,][]{Inoue2018,Sadoun2019,Laursen2019} 
but it is not accounted for in Eq.~\ref{eq:inoue}. We defer more detailed calculations to future work.

\section{Results}\label{sec:results}

   This section presents the power spectra obtained with the formalism described above. 
In \S~\ref{sec:ps}, we show the power spectra obtained with our fiducial profile models.  
The distribution of halo masses contributing across redshifts, and for various multipoles, is 
presented in \S~\ref{sec:mass}. In \S~\ref{sec:average}, we extract information about the 
polarization fluctuation in halos, and we show the cross power spectra for the fiducial models 
in \S~\ref{sec:crosspow}

   For the calculations below, we consider a redshift depth of $\Delta z=0.5$, and assume that the 
redshift-dependent quantities are constant over this range.  We note that this assumption is 
less valid at high redshifts, where the quantities evolve more rapidly with time, but we adopt it 
here for simplicity. 

\subsection{Power Spectra of \lya Polarization}\label{sec:ps}

\begin{figure}\center 
\includegraphics[width=0.51\textwidth]{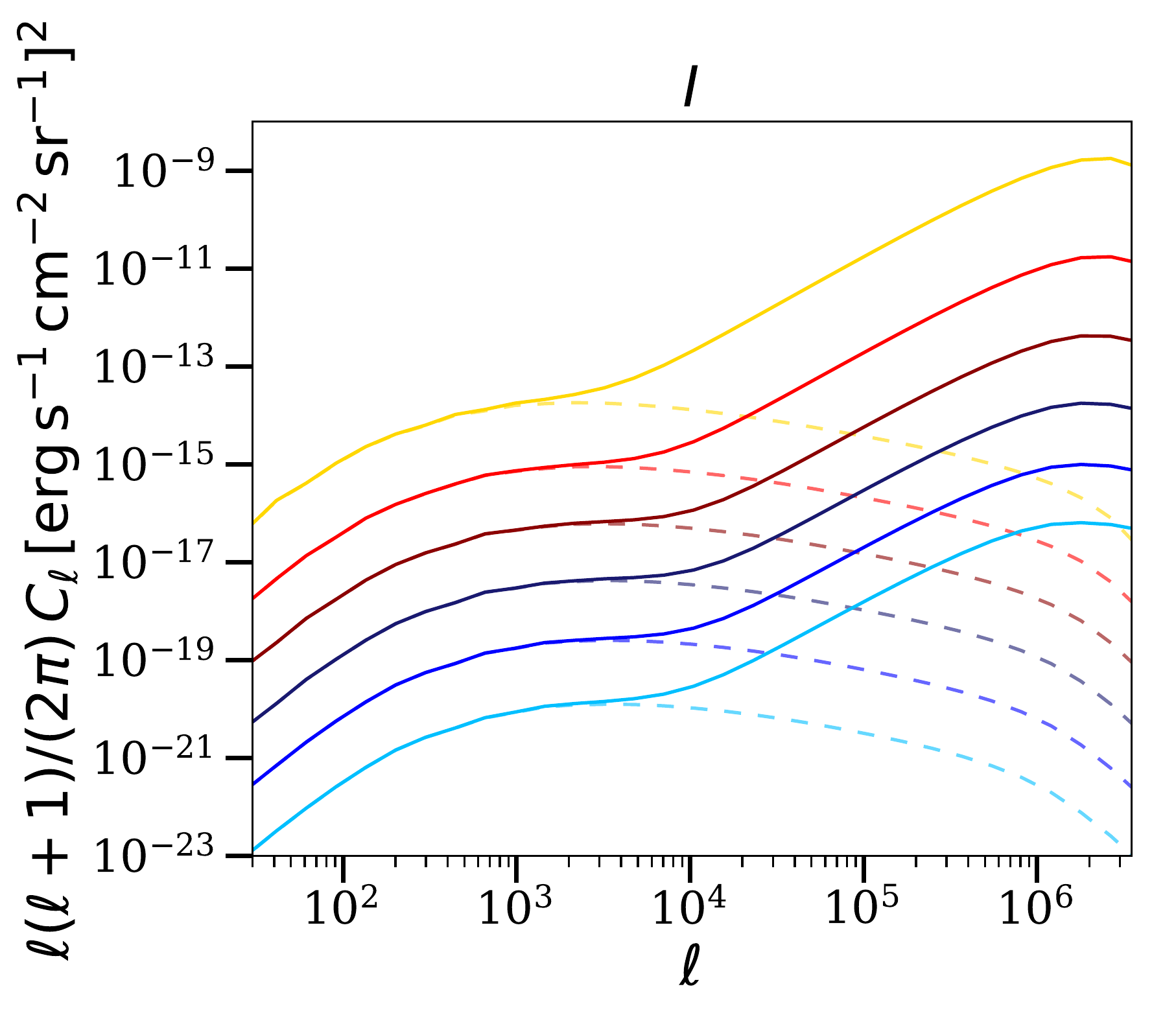}\includegraphics[width=0.51\textwidth]{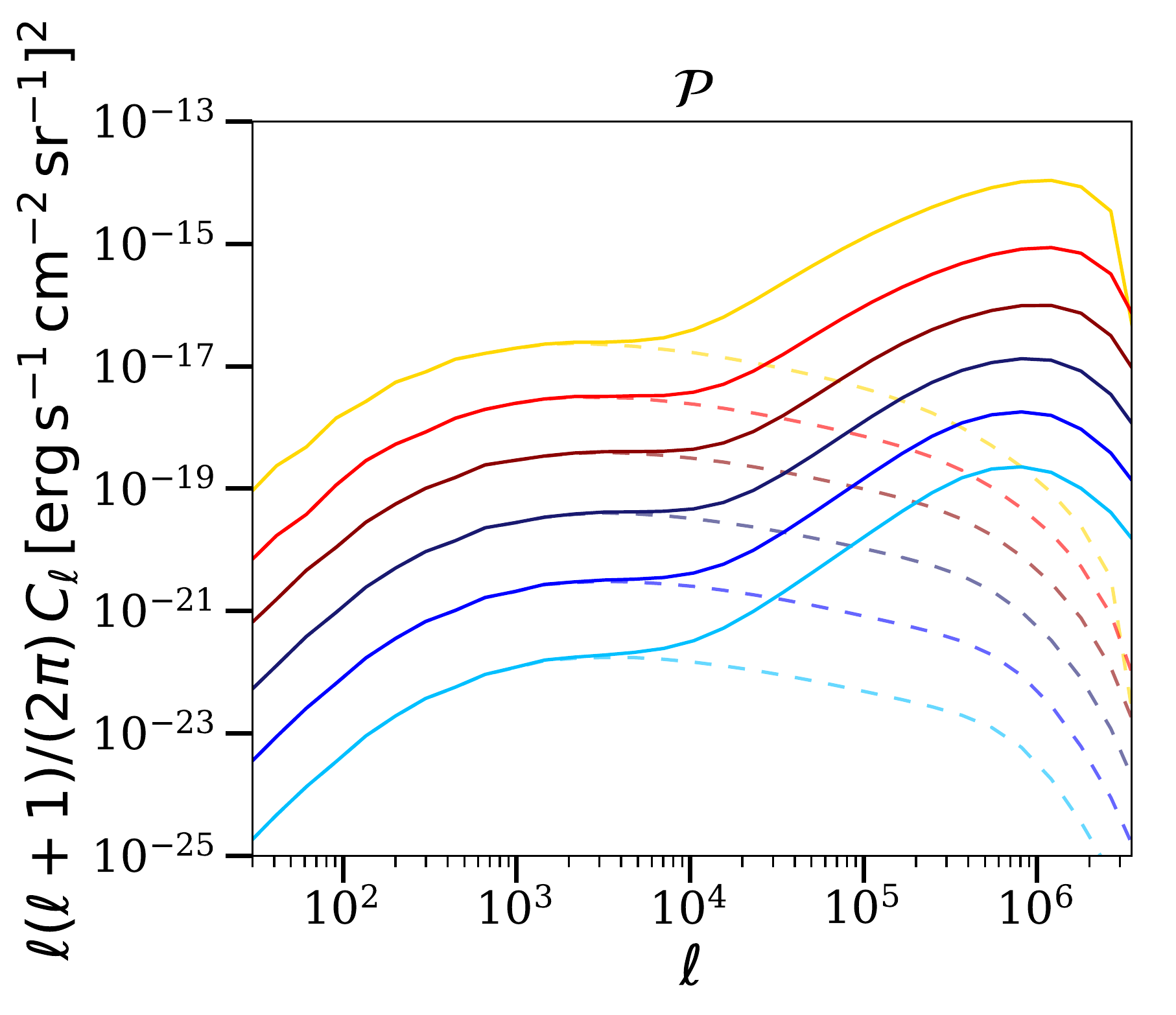}
\includegraphics[width=0.51\textwidth]{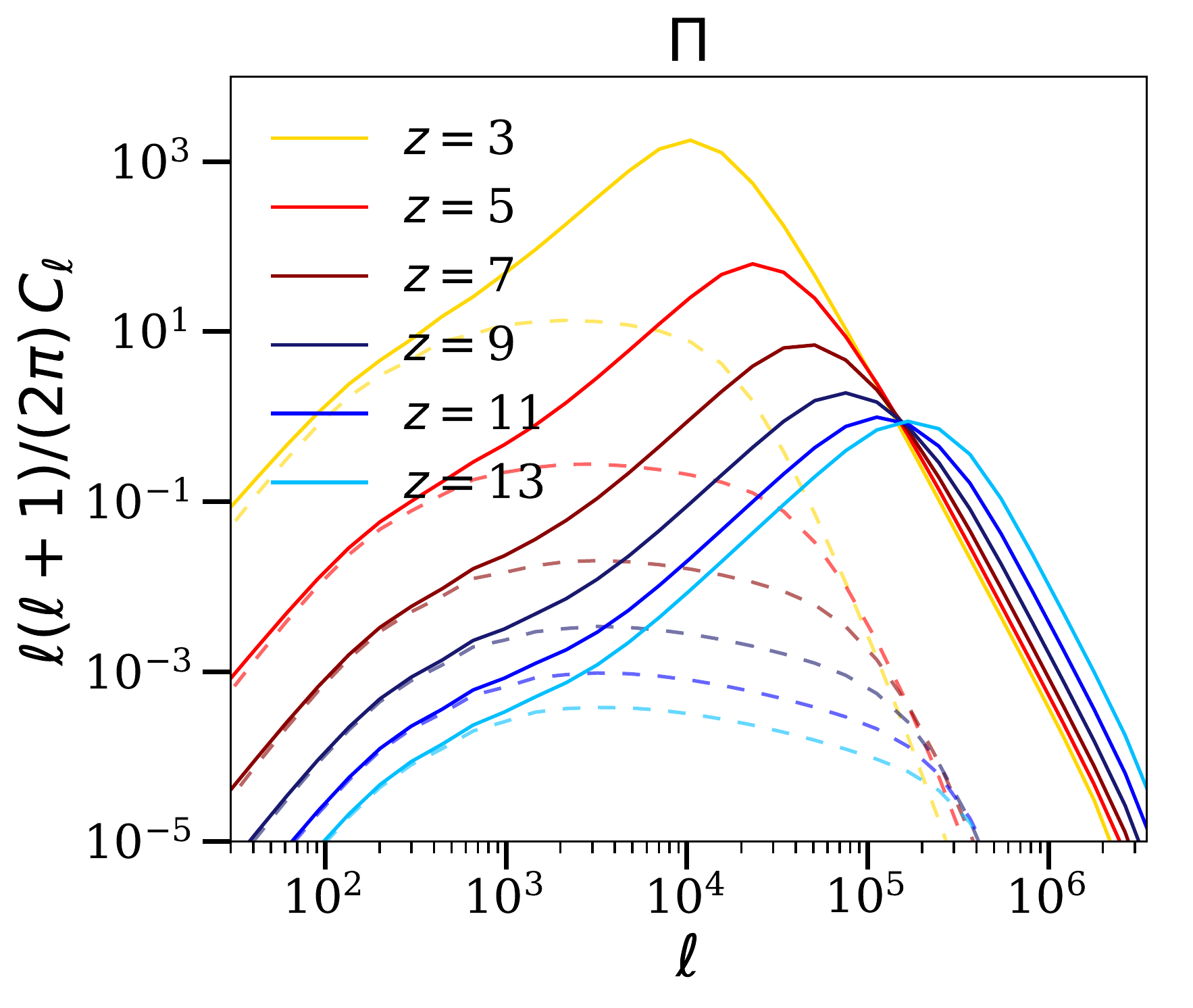}\includegraphics[width=0.51\textwidth]{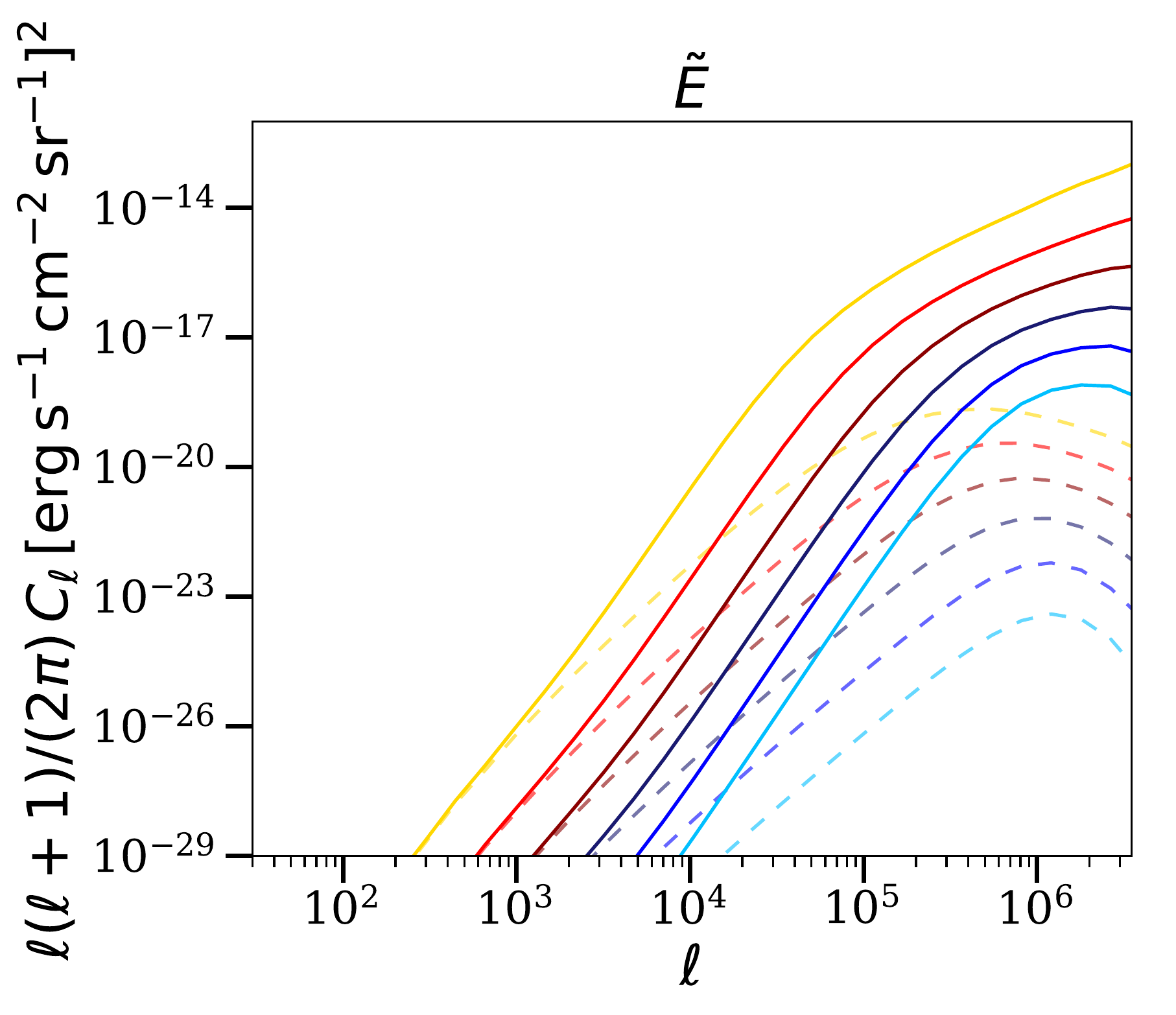}
\caption{From top to bottom and left to right, the panels show the power spectra 
for the \lya (total) intensity, $I$,  polarized intensity, $\po$, polarization degree, $\Pi$, and 
$\tilde E$ modes at redshifts $z=3,\,5,\,7,\,9,\,11$ and $13$. The signal from $B$ modes is null 
by construction in our formalism. The sums of the one- and two-halo terms are denoted by 
the {\it solid lines}, and the two-halo terms are represented by the {\it dashed ones}. 
The power spectra of $\Pi$ and $\tilde E$ present sharp features (peaks or 
knees) whose positions depend on redshift. These peaks are related to the average size of the halos 
dominating the signal at a given redshift, while $I$ and $\po$ mostly only change the amplitude with 
redshift. } 
\label{fig:psplot}
\end{figure}
    
    Figure \ref{fig:psplot} displays the power spectra for the \lya polarization quantities $I$, $\po$, 
$\Pi$, and $\tilde E$,  at redshifts $z=3,\,5,\,7,\,9,\,11$ and $13$, and for the fiducial 
parameters described in \S~\ref{sec:toy}. By construction, the  $B$ mode signal is null in our formalism 
(see \S~\ref{sec:bmodes}). The figure shows that the spectra of the quantities $\Pi$ and $\tilde E$ yield 
more information on the halo population than that  accessible by $I$ alone, due to the peaks (or knees) 
present in the power of these quantities ({\it solid lines}). The position of the 
one-halo peaks for  $\Pi$ varies with redshift, from $\ell \sim10^4$ at $z=3$ to 
$\ell \sim10^5$ at $z=13$. The peak position indicates the 
average size of the halos dominating the power. Because there are few massive halos at high 
redshift, the power peaks at high $\ell$ values (small distances), while the 
increasing number of massive halos when decreasing redshift shifts the peak toward lower 
multipoles. Thus, the measurement of the peak position at a given redshift reveals the mean 
halo size (and in turn mass) dominating the polarization signal, provided that other potential effects 
are known. The position of the knees in the power spectra of $\tilde E$ is also related to the average 
size of the halos, although this relation is complex due to the Bessel function term in Eq.~\ref{eq:ebjqu}. 
None of these measurements is possible with intensity alone. 

     In \S~\ref{sec:dep} in the Appendix, we address the impact of  variations in the fiducial model 
parameters on the power spectra of the polarization quantities. We test changes in the spatial 
extent of the 
polarization signal, and in the intensity and polarization profile shapes. In general, these variations 
result in changes in the amplitudes of the power spectra, as well as in the positions of the peaks. 
Different behaviors are observed for different quantities and redshifts, which indicates that the 
analysis of various quantities and redshifts could be used to constrain the shape and extent of the 
real-space profiles more reliably than with one quantity (e.g., intensity) alone \citep[see][for a 
methodology to extract physical -- small-scale -- information from the intensity mapping power 
spectra]{Sun2019}. 

   The comparisons in \S~\ref{sec:dep} in the Appendix show that the slope of the surface brightness 
profile in the halos is a crucial parameter for extracting polarization information from the power spectra. 
When the surface brightness profile is very steep, variations in the polarization profile (especially at 
large impact parameters) have little effect on the overall power spectra, because they are contributed 
by a small number of photons. This implies smooth one-halo peaks in general, for quantities 
other than $\Pi$. Variations in the polarization profile are most visible as effects in the power spectra 
of the polarization quantities when many photons contribute to the scales of interest. This 
occurs for surface brightness profiles that remain significantly flat out to the impact parameters 
corresponding to those scales. Large neutral gas regions illuminated by (various) bright sources, 
such as  \lya blobs or nebulae \citep[e.g.,][]{Geach2016}, as well as galaxy overdensities 
\citep[e.g.,][]{Steidel2011,Matsuda2012}, can 
keep extended and slowly decreasing surface brightness profiles, while isolated galaxies are expected 
to have steeper slopes, similar to our fiducial calculations, \citep[e.g.,][]{Leclercq2017}.

\subsection{Halo Mass Distribution across Redshift}\label{sec:mass}

    We assess here the halo masses that dominate the power at given multipoles and 
redshifts. For this calculation, we consider only the 
one-halo term, since it dominates the power at high $\ell$ values, where the peak of the power occurs 
in most cases. The halo-mass dependence is obtained via the partial derivative 
\begin{align}
\frac{{\rm d}\ln C_\ell (z)}{{\rm d}\ln M} \equiv \frac{M}{C_\ell (z)}  \frac{{\rm d} C_\ell (z)}{{\rm d} M}  = 
\frac{M\,  \frac{{\rm d}n} {{\rm d}M}  \, w^2(M)\, | \{ \tilde u, \tilde E, \tilde B \}(\ell|M,z)|^2  }{\int  {\rm d}M \frac{{\rm d}n} 
{{\rm d}M}  \, w^2(M)\, | \{ \tilde u, \tilde E, \tilde B \}(\ell|M,z)|^2}  ~,
\end{align}
where $\tilde u \equiv \tilde u_{\{I, \po, \Pi \}}$, and $w$ is the same as in Eq.~\ref{eq:w},  with 
$w=1$ for $\tilde E$ and $\tilde B$.

\begin{figure}\center 
\includegraphics[width=0.51\textwidth]{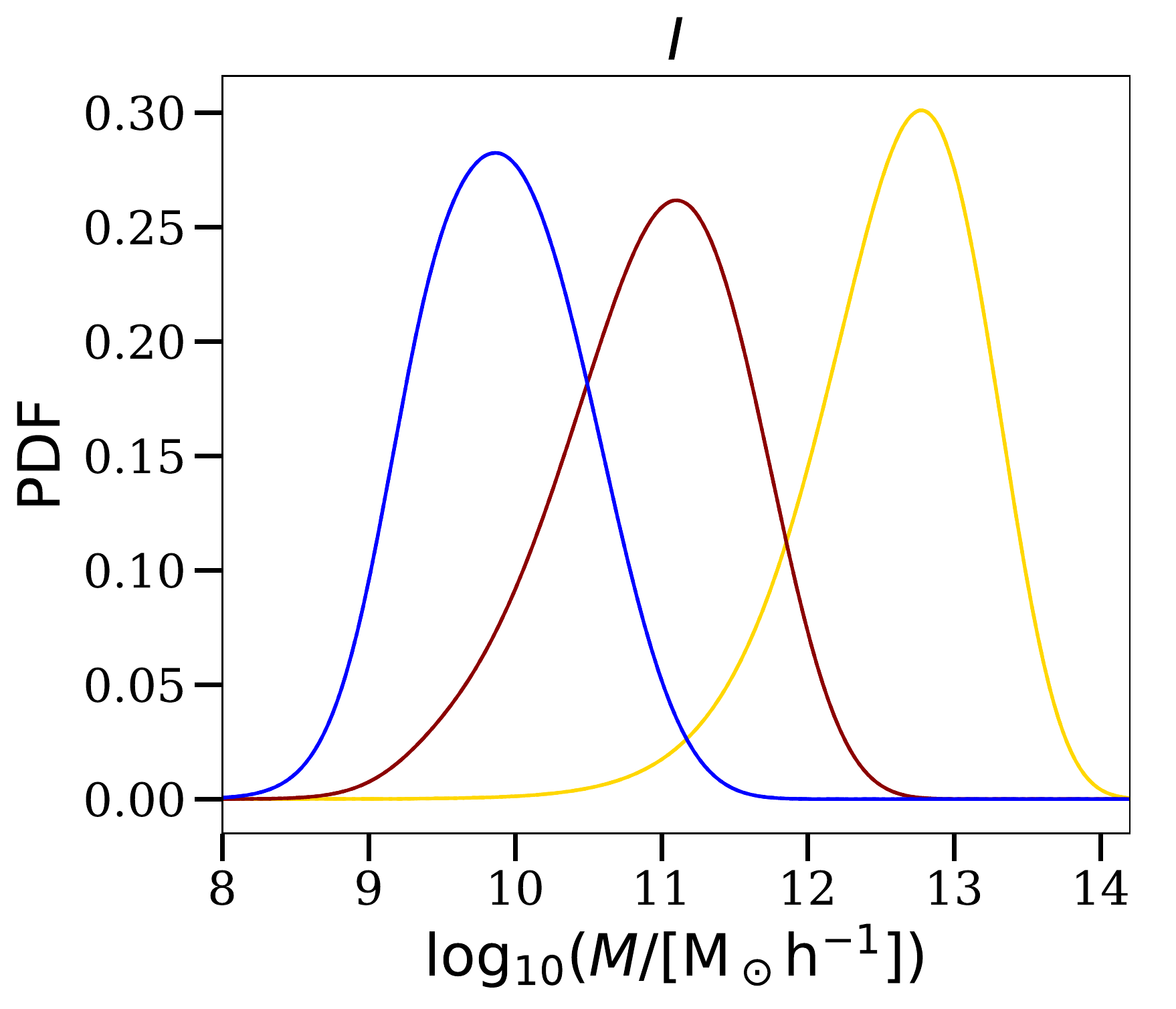}\includegraphics[width=0.51\textwidth]{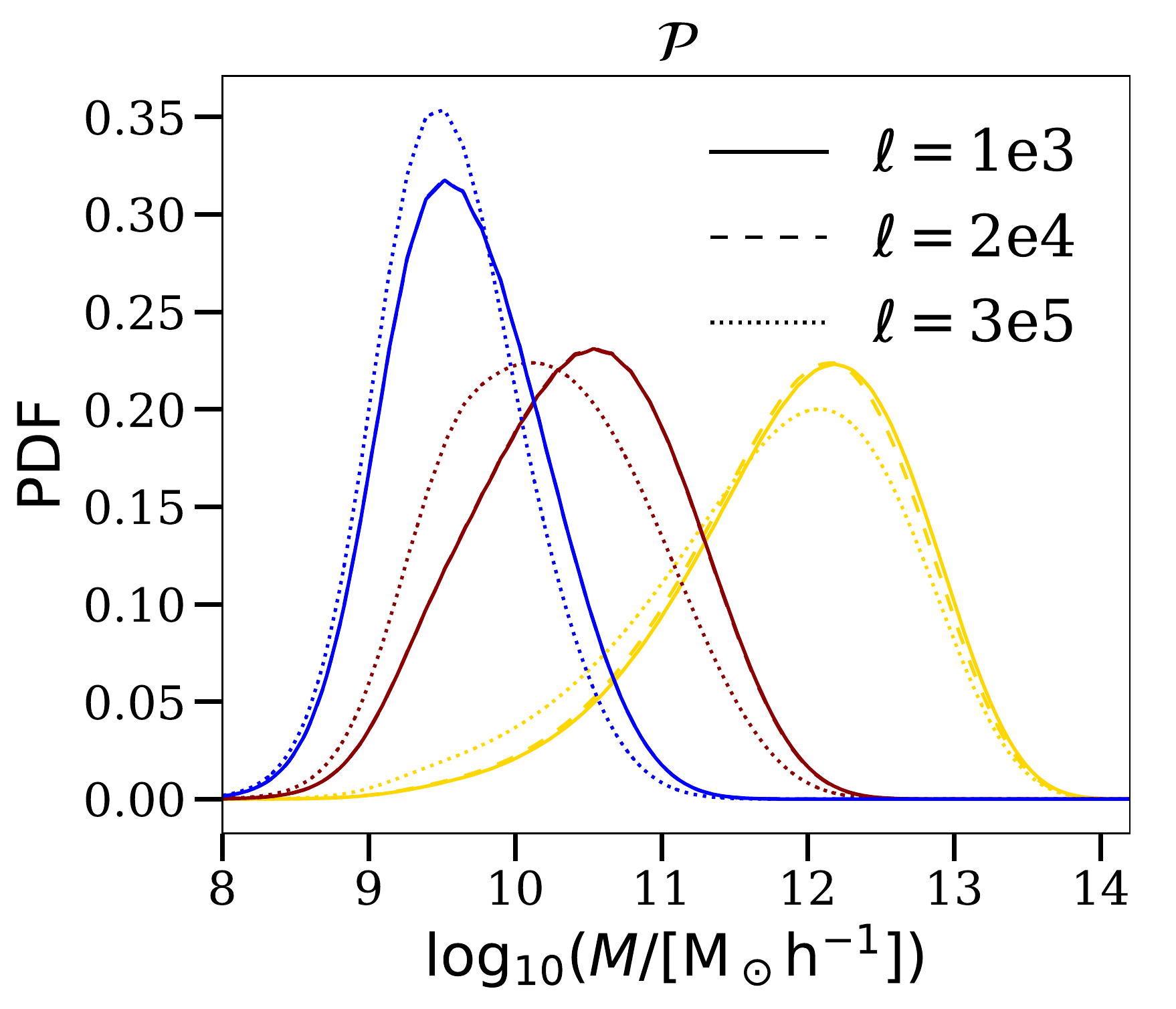}
\includegraphics[width=0.51\textwidth]{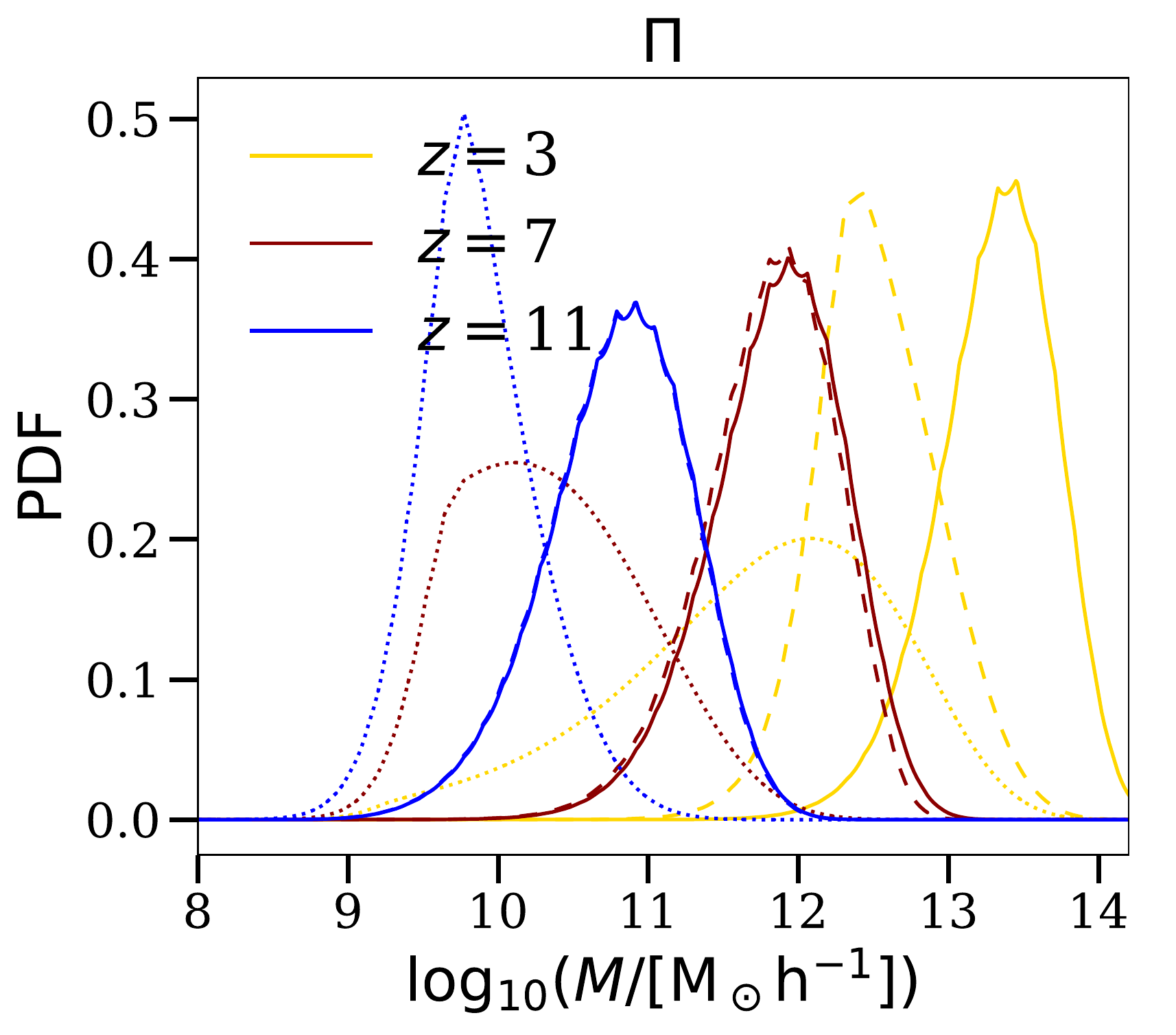}\includegraphics[width=0.51\textwidth]{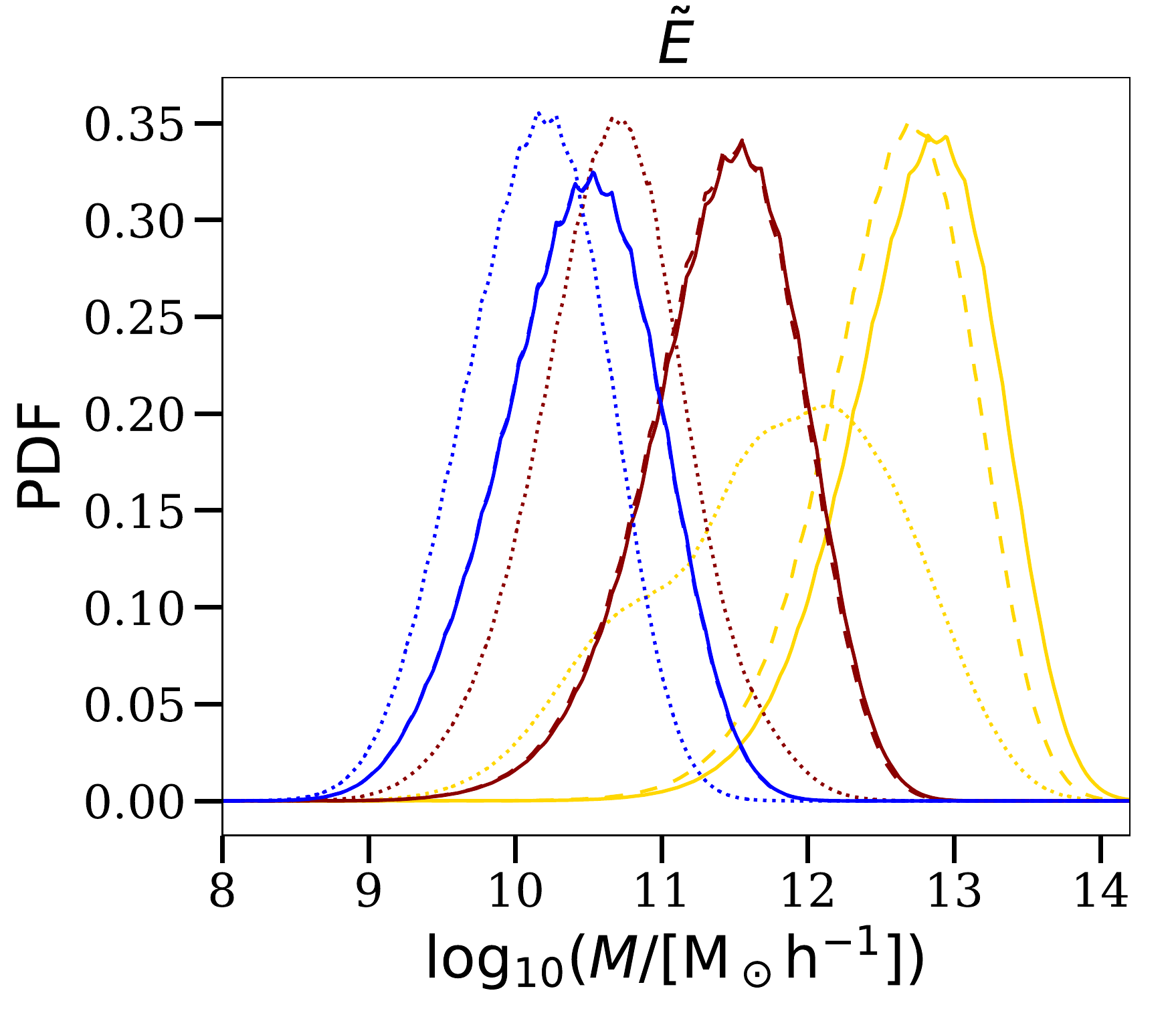}
\caption{Halo mass contribution to the \lya polarization quantities at redshifts $z=3$ ({\it yellow}), 
$z=7$ ({\it dark red}), and $z=11$ ({\it blue}), and at $\ell=10^3$ ({\it solid lines}), $\ell=2\times 10^4$ 
({\it dashed lines}), and $\ell=3\times10^5$ ({\it dot-dashed lines}). In general, the distributions shift 
toward lower halo masses as redshift increases due to the lower number of massive halos at early 
cosmic times. For the case of intensity, halo masses contribute equally to all $\ell$ in our models by 
construction and, therefore, differences between masses are not observed. The contribution of different 
halo masses to different multipoles is most significant for the quantities $\Pi$ and $\tilde E$.} 
\label{fig:massplot}
\end{figure}

   Figure \ref{fig:massplot} shows the halo-mass distributions for the polarization quantities at 
three multipole and redshift values, $\ell=10^3$, $\ell=2 \times 10^4$, and $\ell=3 \times 10^5$, 
and $z=3,\,7$ and $11$, respectively. Overall, the distributions peak at higher halo masses when 
decreasing redshift, reflecting the increase in the number of massive halos at low redshift dictated 
by structure formation. The signal at high multipoles is dominated by less massive halos, due to 
the relation between halo mass and extent of the polarization signal in our formalism. This is most 
visible for the quantities $\Pi$ and $\tilde E$, whose power spectra in Figure \ref{fig:psplot} was 
already related to the average halo mass through the position of the peaks (knees) with redshift.  
For the case of intensity there is not dependence on multipole, because the normalized 
profile shape of intensity is independent of halo mass by construction in our formalism.

\subsection{Polarization Fluctuations in Halos}\label{sec:average}

    We assess now the polarization information that can be retrieved from the ratio between the power 
spectra of $\po$ and $I$. 

    Let us  consider here the halos as polarized point sources in the limit $\ell \to 0$, 
where the power spectra of the one-halo terms are approximated by those of the shot (Poisson) 
noise. In this case, the shot-noise power spectra for $I$ and $\po$ are proportional to 
\citep{Tegmark1996} 
\begin{align}\label{eq:shot1}
C_{\ell,I}^{\rm \,shot} & \propto \int  {\rm d}M \frac{{\rm d}n} 
{{\rm d}M}  \, |I(M)|^2 ~,
\end{align}
and 
\begin{align}\label{eq:shot2}
C_{\ell,\po}^{\rm \,shot} & \propto \int  {\rm d}M \frac{{\rm d}n} 
{{\rm d}M}  \, |\po(M)|^2 ~, 
\end{align}
respectively. Similarly, the $\po$ power spectrum for the halos with polarization fraction $\Pi$ can 
be expressed as \citep{Lagache2019}
\begin{align}
C_{\ell,\po}^{\rm \,shot}(\Pi) & \propto \Pi^2 \int  {\rm d}M \frac{{\rm d}n} 
{{\rm d}M}  \, |I(M)|^2 ~,
\end{align}
where we have assumed that $\po(M) \equiv \Pi \, I(M)$, with $\Pi$ independent on halo 
mass. Then, the power spectrum of the entire distribution of polarization fraction values, $P(\Pi)$, 
equates  
\begin{align}\label{eq:ratio}
C_{\ell,\po}^{\rm \,shot}& =  \int_0^1 P(\Pi)\, C_{\ell,\po}^{\rm \,shot}(\Pi)  \,{\rm d}\Pi   =   \langle \Pi^2 \rangle   \,  C_{\ell, I}^{\rm \,shot}   ~,
\end{align}
where $\langle \Pi^2\rangle$ denotes the mean squared value of the polarization fraction in halos. 
Thus, the ratio of the one-halo term power spectra of $\po$ and $I$ in the limit $\ell \to 0$ gives 
information about the polarization fluctuation in the entire halo population. 

\begin{figure}\center 
\includegraphics[width=0.63\textwidth]{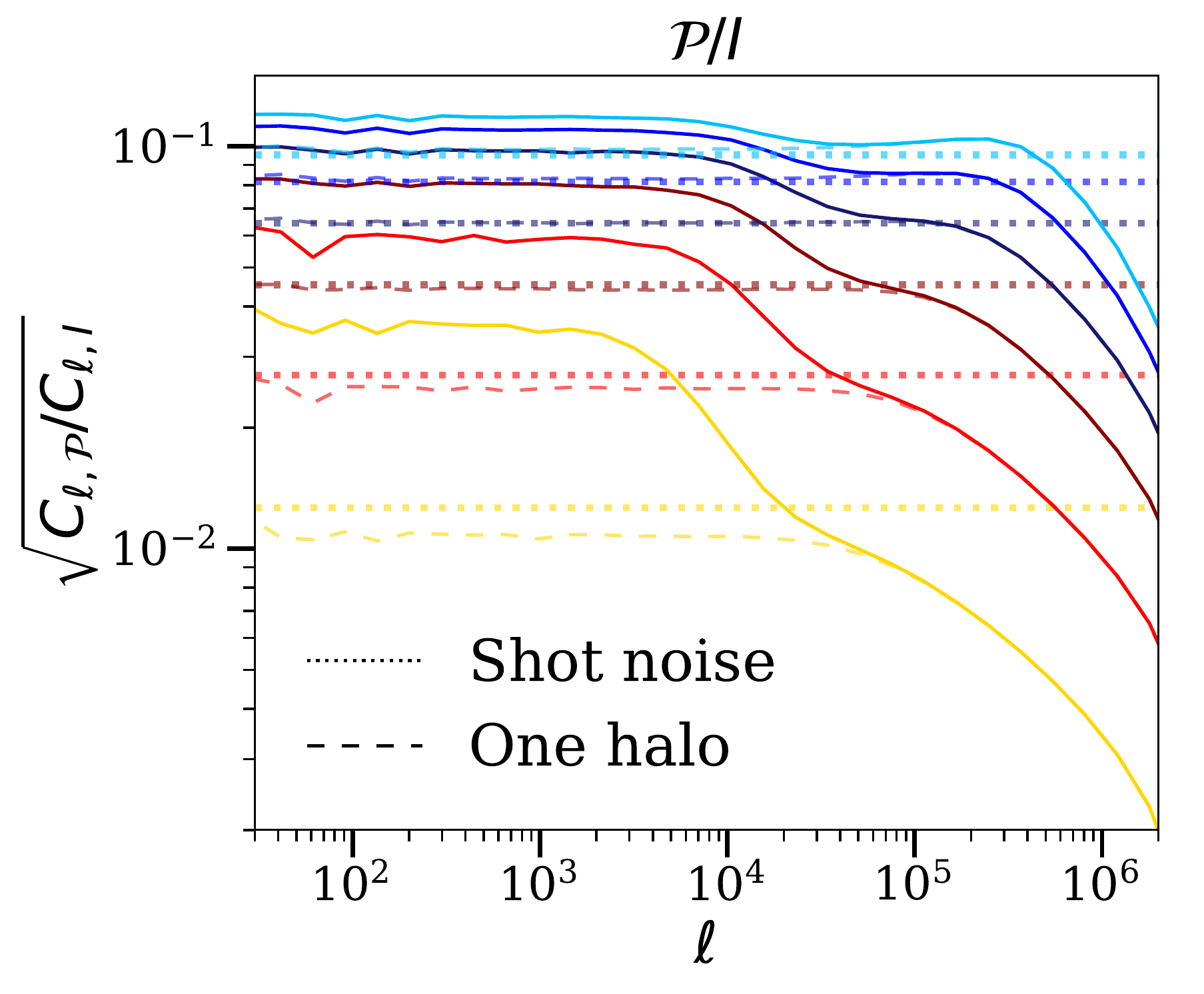}
\caption{Ratio between the power spectra of $\po$ and $I$ at the redshifts of our calculations. 
From top to bottom, the lines denote the redshifts $z=13,$ 11, 9, 7, 5, 3.  
The {\it solid lines} denote the the square root of the ratio, and the {\it dashed lines} account for the 
case of the one-halo terms alone. The {\it dotted horizontal lines} show the square root of the ratio 
between the shot-noise power of $\po$ and $I$, i.e., ${\langle \Pi^2\rangle}^{1/2}$, which denotes  
the r.m.s of the polarization fluctuation in halos. This ratio broadly compares to the one-halo term 
values in the limit $\ell \to 0$, the differences arising from assuming a polarization fraction per halo 
independent on halo mass.  The increase of ${\langle \Pi^2\rangle}^{1/2}$ with redshift is due to the 
more compact signal (less number density of massive halos) at earlier times. The steepness of the 
decay of the one-halo terms with redshift is an indicator of the distribution of $\Pi$ values. At high 
redshifts, most halos are small, and thus the distribution of $\Pi$ is narrow, which yields a steep slope. 
Instead, a broader distribution of halo sizes and, therefore, of $\Pi$ values at lower redshifts, results in 
a smoother decay of the ratio. The same dependence on halo sizes can be inferred from the position 
from where the one-halo terms begin their rapid decrease toward high $\ell$ values.} \label{fig:pooveri}
\end{figure}

   Figure \ref{fig:pooveri} shows the ratio between the power spectra of $\po$ and $I$ at the redshifts of 
our calculations, from top to bottom, $z=13,$ 11, 9, 7, 5, 3. The {\it solid lines} denote the square root 
of the ratio, and the {\it dashed lines} represent the one-halo terms alone. The {\it dotted horizontal lines} 
show the square root of the ratio between the shot-noise power of $\po$ and $I$ (Eq.~\ref{eq:ratio}), i.e., 
${\langle \Pi^2\rangle}^{1/2}$, which is the r.ms. fluctuation of polarization fraction values 
between halos. For the case of measured power spectra, the value of ${\langle \Pi^2\rangle}^{1/2}$ can 
be estimated from the flattening of the one-halo terms, where the dashed lines approximate the dotted 
lines. The differences between the flattening of the one-halo term and the ratio of power spectra arises 
from our assumption that the halo polarization fraction is independent on the halo mass. 
The increase of fluctuations with redshift, from ${\langle \Pi^2\rangle}^{1/2} \sim 0.01$ at 
$z=3$ to ${\langle \Pi^2\rangle}^{1/2} \sim 0.1$ at $z=13$, arises from the smaller average 
halo size at early times than at low redshift. In our formalism, the polarization fraction increases quickly 
with impact parameter in small sources because of the small virial radius (Eq.~\ref{eq:lyapo}), and it is 
more sensitive to variations of this slope, and in turn of the halo mass. 
The reduced variation of ${\langle \Pi^2\rangle}^{1/2}$ at high redshifts indicates that the distribution of 
halo sizes is similar at these epochs, while the distribution of halo sizes evolves more rapidly at low 
redshift. 

     Additional information can be inferred from the slope of the decay of the one-halo terms toward high 
multipoles in Figure \ref{fig:pooveri}. A steep decay, or a pronounced knee, signals that the polarization 
degree profiles are similar for the entire halo population. The steep decay of the power at high 
redshift indicates that the halos have a narrower size distribution compared to low redshift, where the 
decay shape is smoother. Note that this redshift evolution arises due to the dependence of the 
polarization fraction signal with halo size, through the virial radius, in our formalism. Finally, the position 
of the knee in the one-halo term of the power spectra can be used as an estimator of the average size 
for the polarization signal, similarly to the case in the auto power spectra of $\Pi$ and $\tilde E$ 
previously discussed in \S~\ref{sec:ps}.

\begin{figure}\center 
\includegraphics[width=0.51\textwidth]{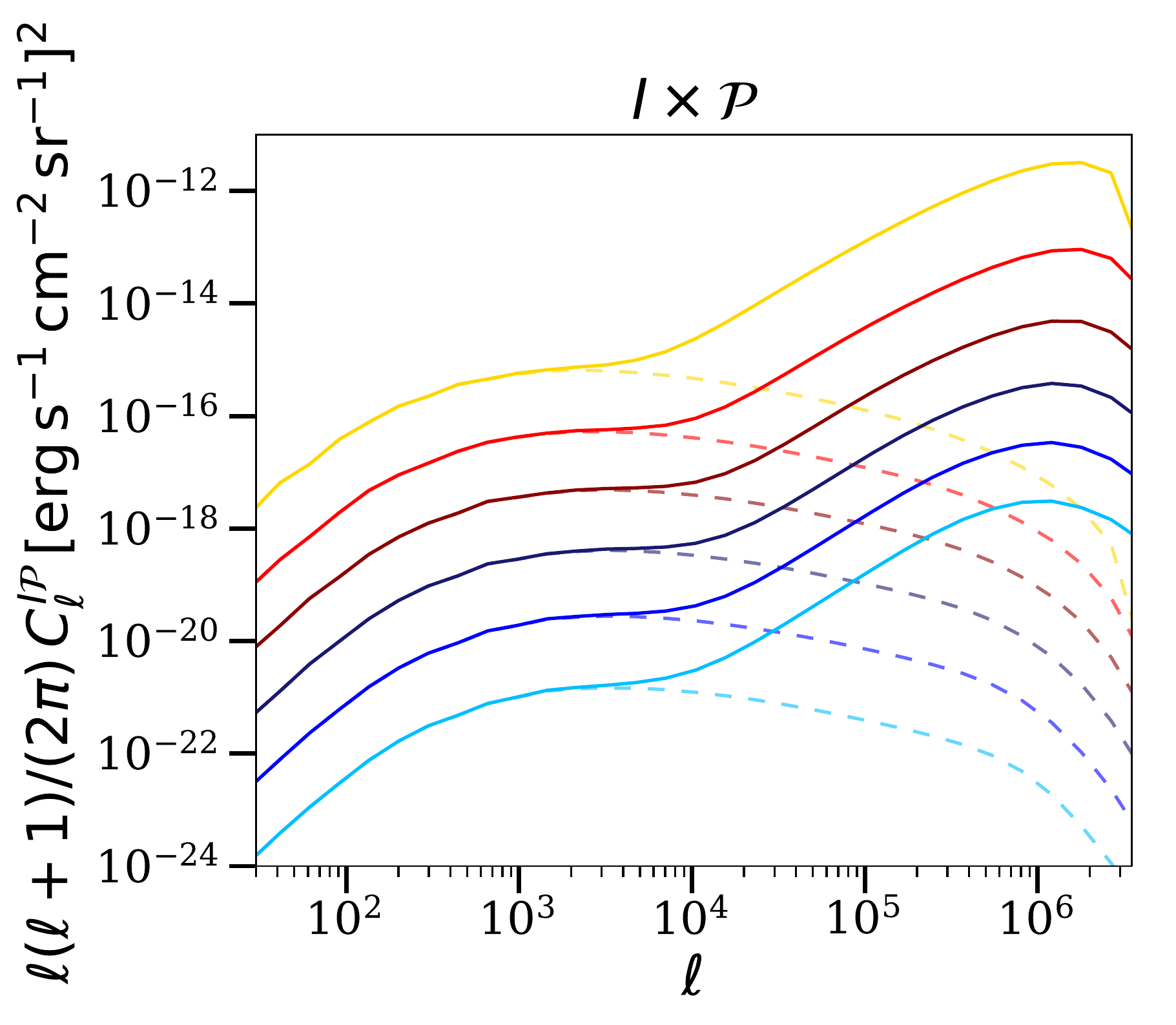}\includegraphics[width=0.51\textwidth]{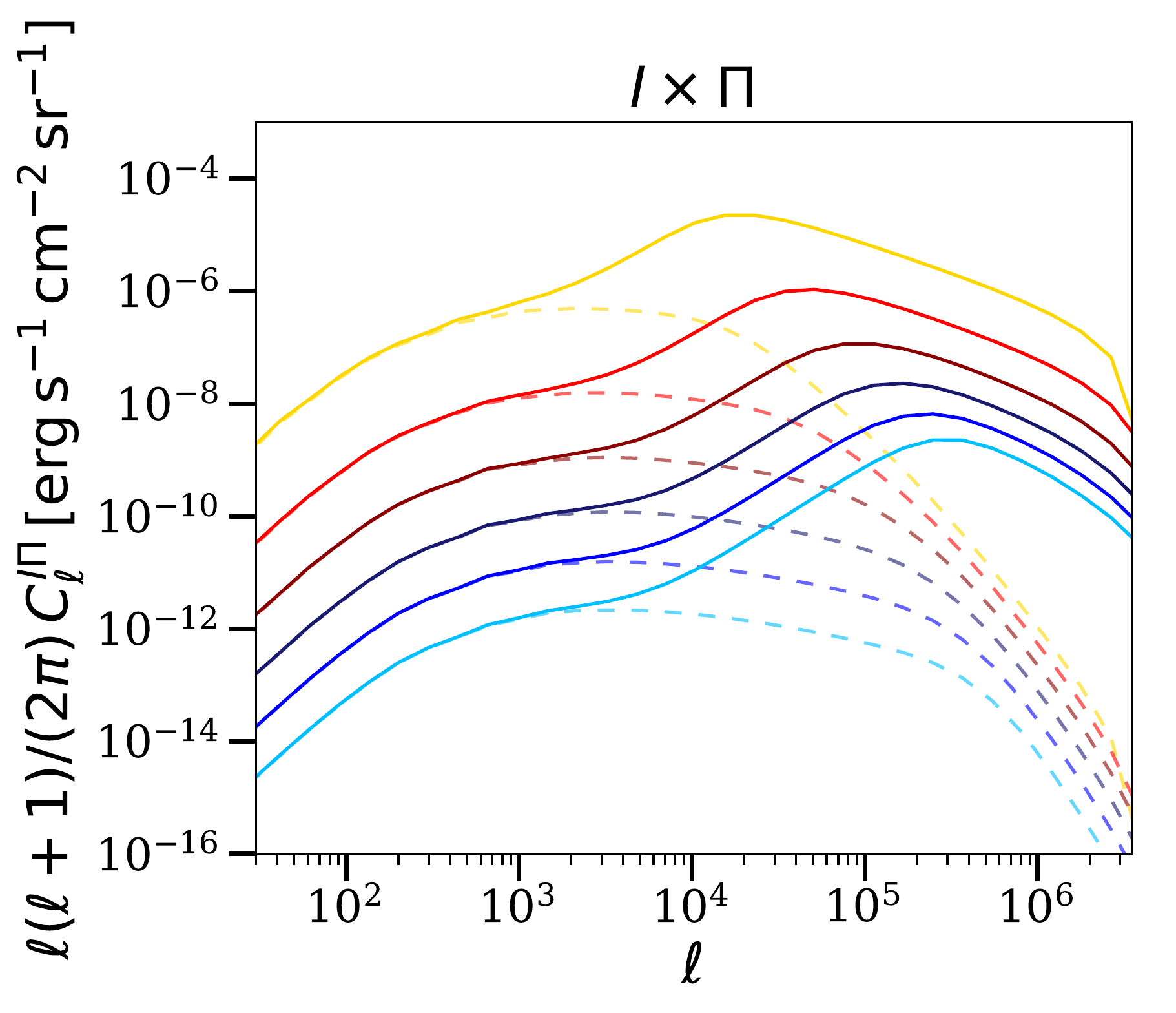}
\includegraphics[width=0.51\textwidth]{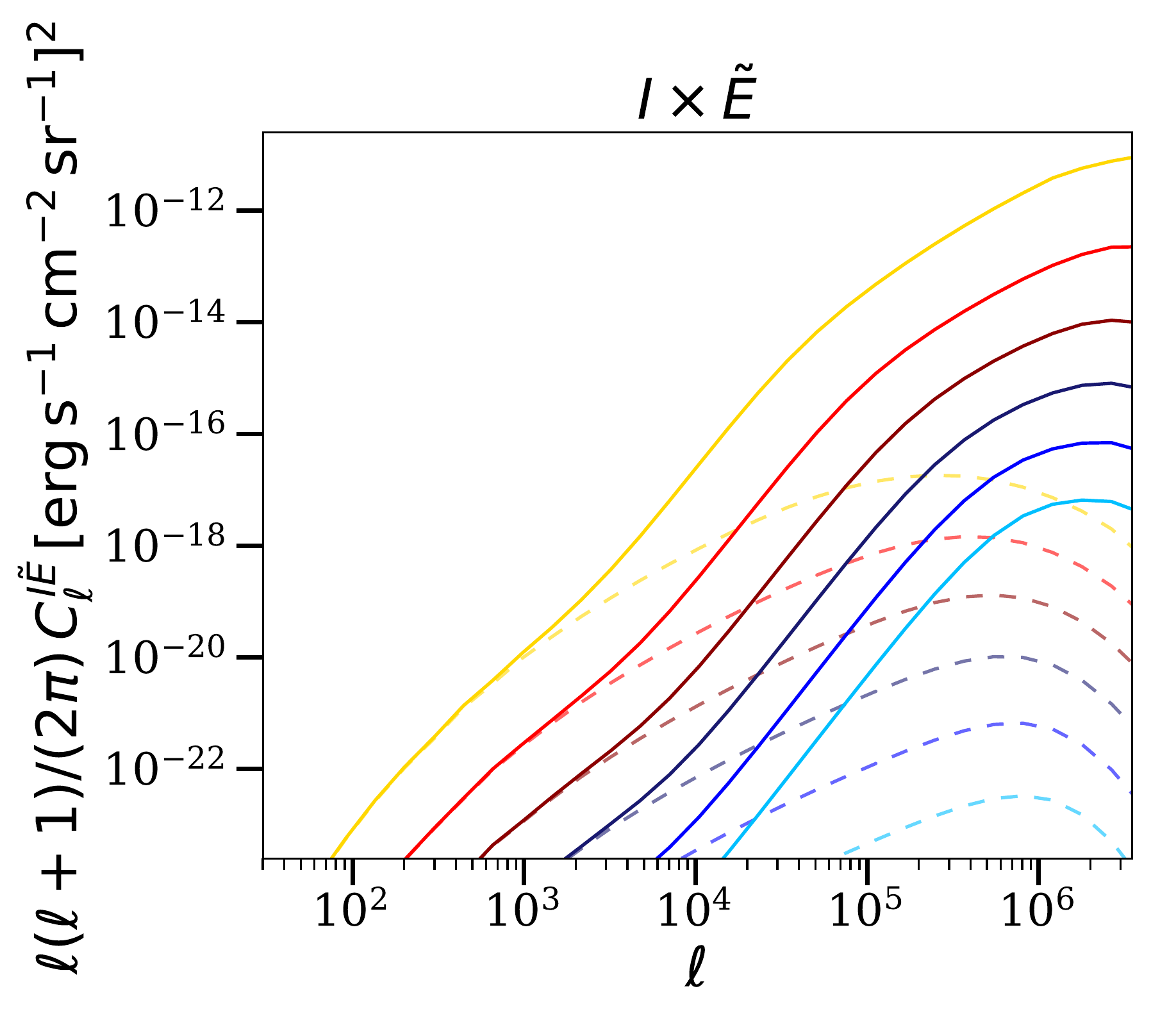}\includegraphics[width=0.51\textwidth]{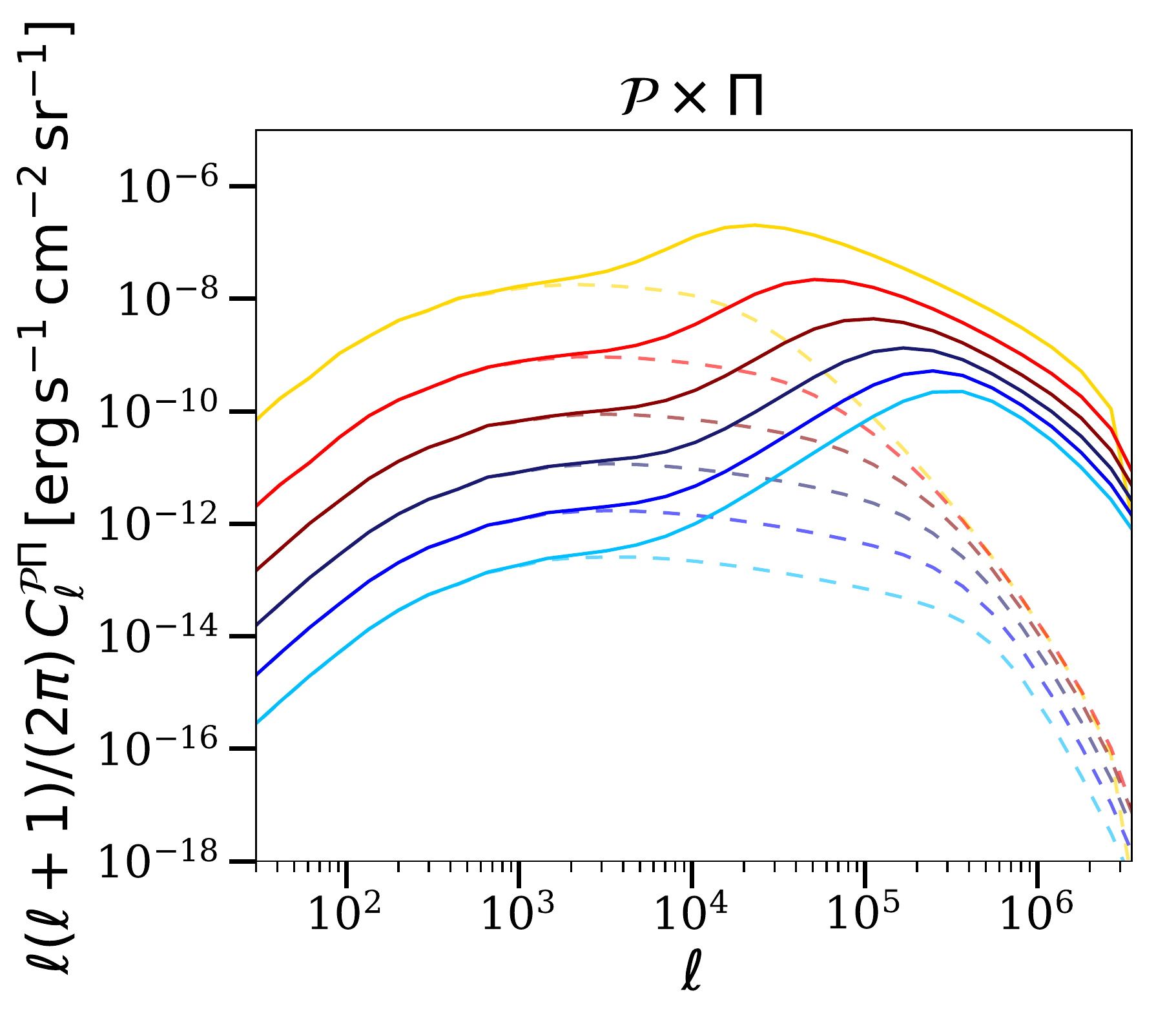}
\includegraphics[width=0.51\textwidth]{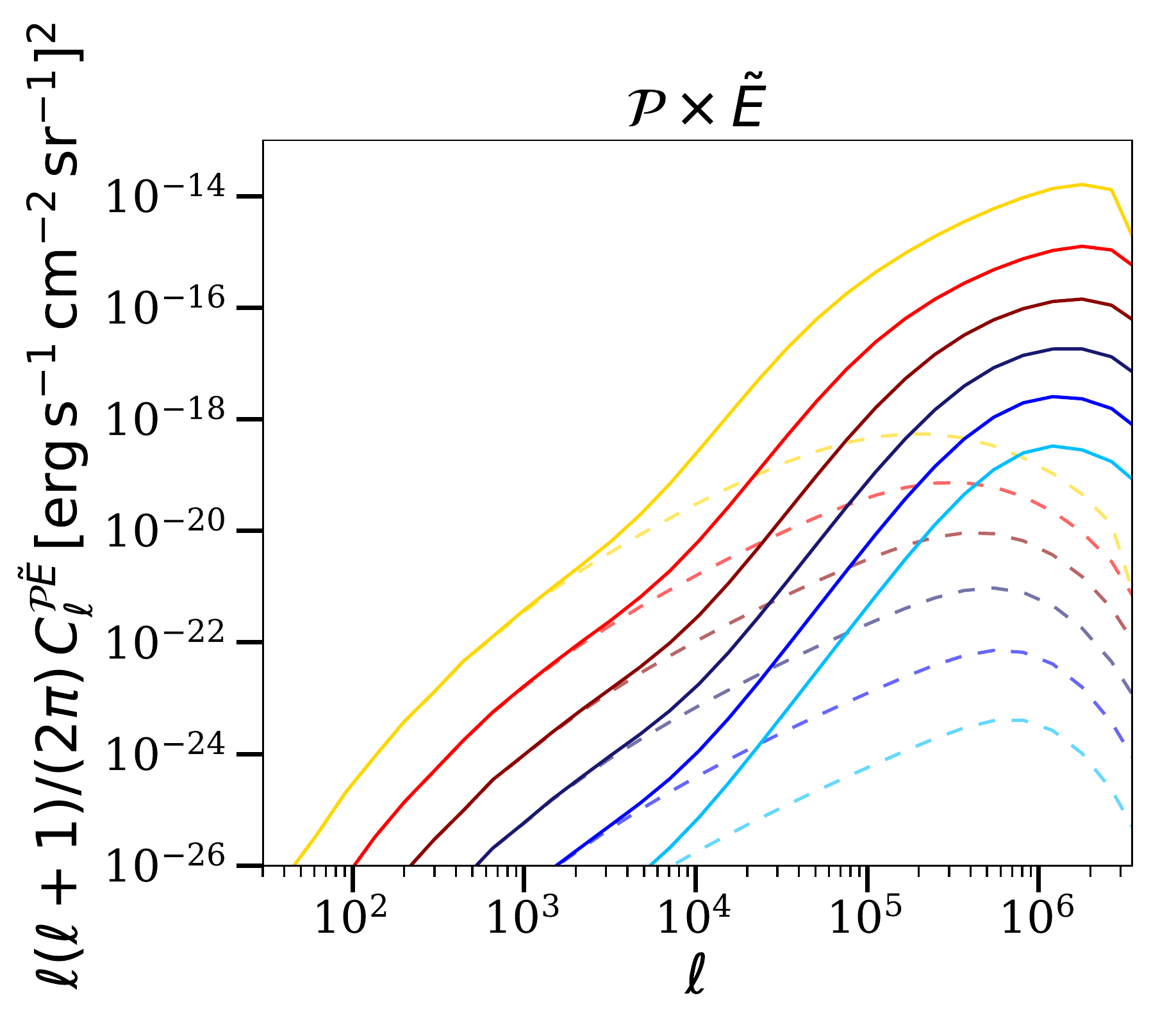}\includegraphics[width=0.51\textwidth]{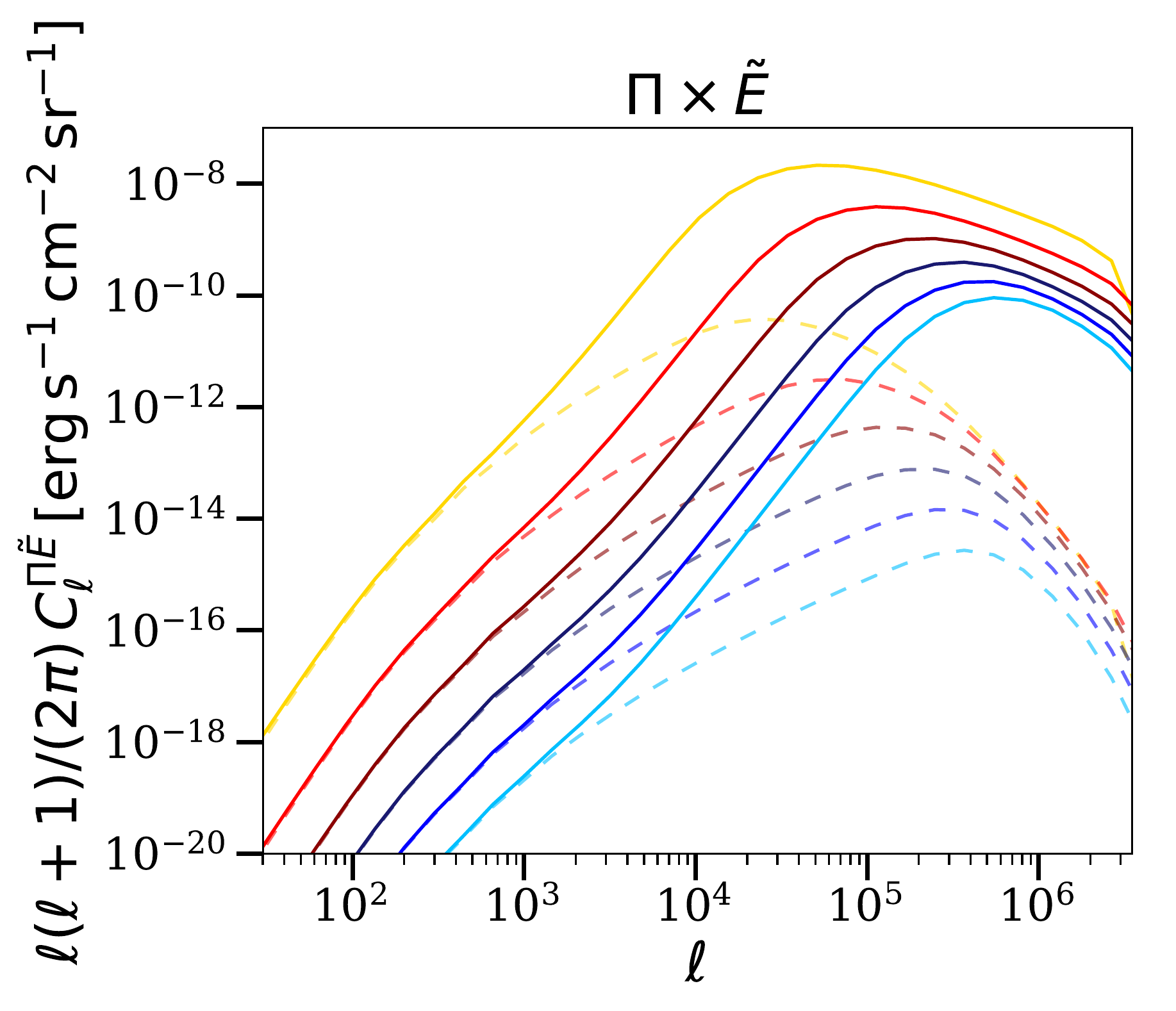}
\caption{Cross power spectra for the polarization quantities $I$, $\po$, $\Pi$, and $\tilde E$, and for 
the fiducial model parameters.} 
\label{fig:cross}
\end{figure}

\subsection{Cross Power Spectra of \lya Polarization}\label{sec:crosspow}

   Figure \ref{fig:cross} shows the cross power spectra between the quantities $I$, $\po$, $\Pi$, and 
$\tilde E$, taking into account the fiducial model parameters. Overall, the cross power spectra that 
consider the polarization fraction, $\Pi$, present the sharpest peaks in the one-halo terms, which makes 
their identification easier than in other cases. Furthermore, the position of the peaks changes with 
redshift, which indicates that the position is connected to the average size of the polarization signal, 
i.e., the dominant size of the halos in our formalism.

\section{Detectability Estimates}\label{sec:detection}
   
   We perform calculations for the detectability of the \lya polarization signal below, and 
we discuss the impact of foregrounds in \S~\ref{sec:foregrounds}.
  
   This section presents estimates on the detectability of the power spectra, assuming 
that the signal is Gaussian, for simplicity. In practice, the signal may be highly non-Gaussian 
due to the small (non-linear) galaxy scales where the polarization is maximized. In this later 
case, a full covariance matrix calculation would be required \citep[see, e.g.,][]{Komatsu2002}.  

    For a Gaussian statistic, the S/N can be computed following \cite{Knox1995} as
\begin{align}
S/N^2 \,(\ell) = \frac{{C_\ell}^2}{(\Delta C_\ell)^2}~,
\end{align} 
where 
\begin{align}\label{eq:varicl}
(\Delta C_\ell)^2 \equiv {\rm Var}(C_\ell)  =  \frac{2}{2 \ell + 1} \frac{1}{f_{\rm sky}}  \left( C_\ell  + w^{-1} 
e^{\ell^2\, \sigma_b^2}  \right)^2   ~.
\end{align} 
The first summand in the above expression describes the sample (cosmic) variance, and the second one 
represents the instrumental (thermal) noise, where $W_\ell = e^{\ell^2\, \sigma_b^2}$ is the window 
function for a Gaussian beam of size $\sigma_b$, and $f_{\rm sky}$ denotes the fraction of the sky 
covered by the observations\footnote{The inclusion of the term denoting the fraction of sky 
covered by the survey, $f_{\rm sky}^{-1}$, is valid as long as the sampling of the spectra accomplishes 
a binning of size $\Delta \ell \gtrsim 2\pi / \Theta$, where $\Theta$ is a linear dimension of the observed 
field \citep{Knox1997}.}. 
The term 
\begin{align}
w\equiv \left(\sigma_{\rm pix}^2 \, \Omega_{\rm pix}   \right)^{-1}  ~ 
\end{align} 
represents a weight per solid angle, and $\sigma_{\rm pix}$ and $\Omega_{\rm pix}$ are the  
pixel uncertainty and solid angle, respectively. 
The pixel uncertainty can be calculated as   
\begin{align}\label{eq:sigpix}
\sigma_{\rm pix} = \frac{s}{\sqrt{t_{\rm pix}}}~, 
\end{align} 
where $t_{\rm pix} = (N_{\rm feeds} \, \Omega_b  / 4\pi \,f_{\rm sky})\, t_{\rm survey}$ is the observing time per pixel, 
with $N_{\rm feeds}$ denoting the number of spectro-polarimeters (spatial channels) simultaneously 
observing the sky, and $t_{\rm survey}$ is the total observing time of the experiment\footnote{For 
simplicity, we ignore here that measurements of polarized light can require the observation of 
the sky at different directions, which therefore divide the total time typically in two.}. The 
numerator in Eq.~\ref{eq:sigpix} describes the sensitivity, and can be 
accounted for via the noise equivalent flux density (NEFD) as 
\citep{Sun2019}
\begin{align}
s = \frac{\rm NEFD}{\Omega_b} ~, 
\end{align}  
where 
\begin{align}
\Omega_b = \Theta_{\rm FWHM}^2 = (2\sqrt{2\ln 2}\, \sigma_b)^2 ~  
\end{align} 
denotes the beam solid angle, and 
$\Theta_{\rm FWHM}$ describes the 
full-width at half maximum for the beam. 

  For the case of the polarization degree, $\Pi$, we calculate the uncertainty by accounting for 
the propagation of the uncertainties in $I$ and $\po$ as 
\begin{align}\label{eq:errorprop}
\cfrac{\Delta C_{\ell,{\Pi}}}{C_{\ell,\Pi}} = \sqrt{\left(   \cfrac{\Delta C_{\ell,{\po}}}{C_{\ell,{\po}}}  \right)^2 + \left( \cfrac{\Delta C_{\ell,{I}}}{C_{\ell,I}} \right)^2} ~.  
\end{align}
This is motivated by the fact that, in practice, $\Pi$ will be derived from the separate 
measurements of these two quantities. This approach yields an uncertainty 
higher by a factor $\sim 1.5$ compared to that from simply using Eq.~\ref{eq:varicl}. 

    We estimate the sensitivities required to detect the polarization signal, and compare them to 
the sensitivity levels of real ground- and  space-based instruments. The ground-based case is 
compared to the HETDEX experiment \citep{Hill2008}, and the space-based estimate considers CDIM 
\citep{Cooray2019}.  None of these instruments, however, are (presently) designed to perform 
polarization observations.  

    Briefly, HETDEX is a ground-based experiment, equipped with a spectrograph and $150$ 
integral field units \citep[IFUs; ][]{HillT2014}, that will perform a blind wide-field spectroscopic 
survey. HETDEX is expected to detect $\sim 0.8$ million LAEs in the redshift range 
$1.9 < z < 3.5$, and over an area of $\sim 400$ ${\rm deg^2}$ on the sky for three years. 
However, \cite{Fonseca2017} already noted that HETDEX can also be used for \lya intensity studies, 
because the IFUs will take data from several patches of the sky blindly, i.e., regardless of the number 
or position of known \lya sources in them. This data, 
therefore, will contain a number of bright sources, but will also include the faint diffuse emission 
from undetected and/or extended objects that are the target of intensity mapping. 
Furthermore, the sensitivity of HETDEX is designed to detect the \lya emission line flux at high 
spectral resolution, i.e., over a redshift depth of $\Delta z \approx 0.006$, in order to resolve the 
\lya line profile. Because this high spectral resolution is not required for intensity studies (e.g., we 
consider here $\Delta z = 0.5$), in practice, we can add the flux from many spectral HETDEX 
bins and thus reduce the pixel uncertainty, $\sigma_{\rm pix}$, by a factor of $\sqrt{N_z} \approx 9$, 
where $N_z$ is the number of spectral bins.

      CDIM is a proposed intensity mapping space observatory designed to study the epoch of cosmic 
reionization via the \lya emission in the wavelength range $0.75 \lesssim \lambda / \mu {\rm m} 
\lesssim 7.5$, covering a sky area of $\sim 300$ ($\sim 15$) ${\rm deg^2}$ for a wide (deep) survey, 
and with a spectral resolution of $R=300$. 

      Table \ref{ta:detect} quotes the parameters adopted for our  calculations with a 
hypothetical ground-based \lya polarization experiment, Lyapol-G, and a space-based experiment, 
Lyapol-S. The {\it first column} refers to the experiment, and the {\it second column} is the spectral 
resolution assumed in the calculations. The {\it third column} quotes the pixel uncertainty resulting  
from the observing times and individual characteristics of the experiments at the redshift 
of \lya stated in the {\it fourth} column. The {\it fifth} and {\it sixth columns} are the pixel and beam solid 
angles, respectively.  The fraction of the sky covered by the surveys is quoted in the {\it seventh column}. 
Overall, the sensitivities quoted in the third column of Table \ref{ta:detect} are a factor of $\sim 10$ (for 
Lyapol-G) and of $\sim 100$ (for Lyapol-S) higher than the nominal values of HETDEX and CDIM, 
respectively. Although the total intensity can be detected at the nominal values for these instruments, 
we show that the higher sensitivities are required to reach the polarization signal in a broad redshift 
range. We have also reduced the pixel and beam sizes for Lyapol-S compared to the case of CDIM 
in order to achieve the small physical scales where the polarization power is significant at high redshifts.

\begin{table*}\movetableright=-1in    
\ra{1.3}
	\begin{center}
	\caption{Instrumental Parameters}\label{ta:detect}
	\begin{threeparttable}
		\begin{tabular}{lcccccc} 
		\toprule 
		 Instrument  &$R$  &$\sigma_{\rm pix}\,{[\rm erg\,s^{-1}\, cm^{-2}\,sr^{-1}]}$  &$z_{\rm Ly\alpha}$	&$\Omega_{\rm pix}\,{\rm [{arcsec}^2]}$ &$\Omega_{b}\,{\rm [{arcsec}^2]}$	   &$f_{\rm sky}$   \\ \hline
        Lyapol-G       &$700$      &$2.6\times10^{-6}$	  &$3$   &$5.40$	 &$9$    &$0.010$     \\ 
        Lyapol-S       &$300$      &$1.6\times10^{-7}$	&$9$     &$0.50$	 &$2$    &$0.008$      \\ \hline        
		\end{tabular}	
	\end{threeparttable}
	\end{center}
\end{table*}

\begin{figure}\center 
\includegraphics[width=0.51\textwidth]{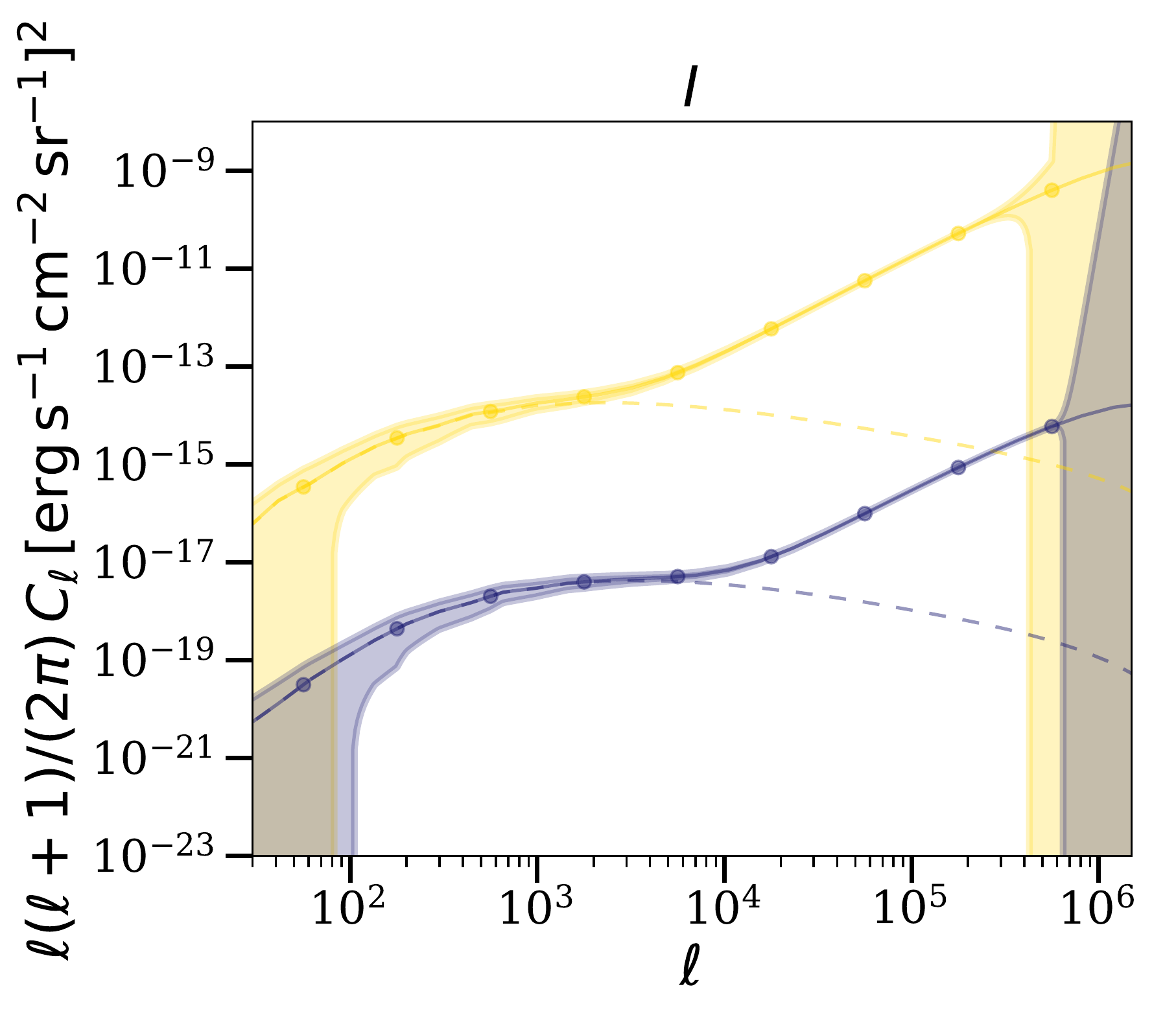}\includegraphics[width=0.51\textwidth]{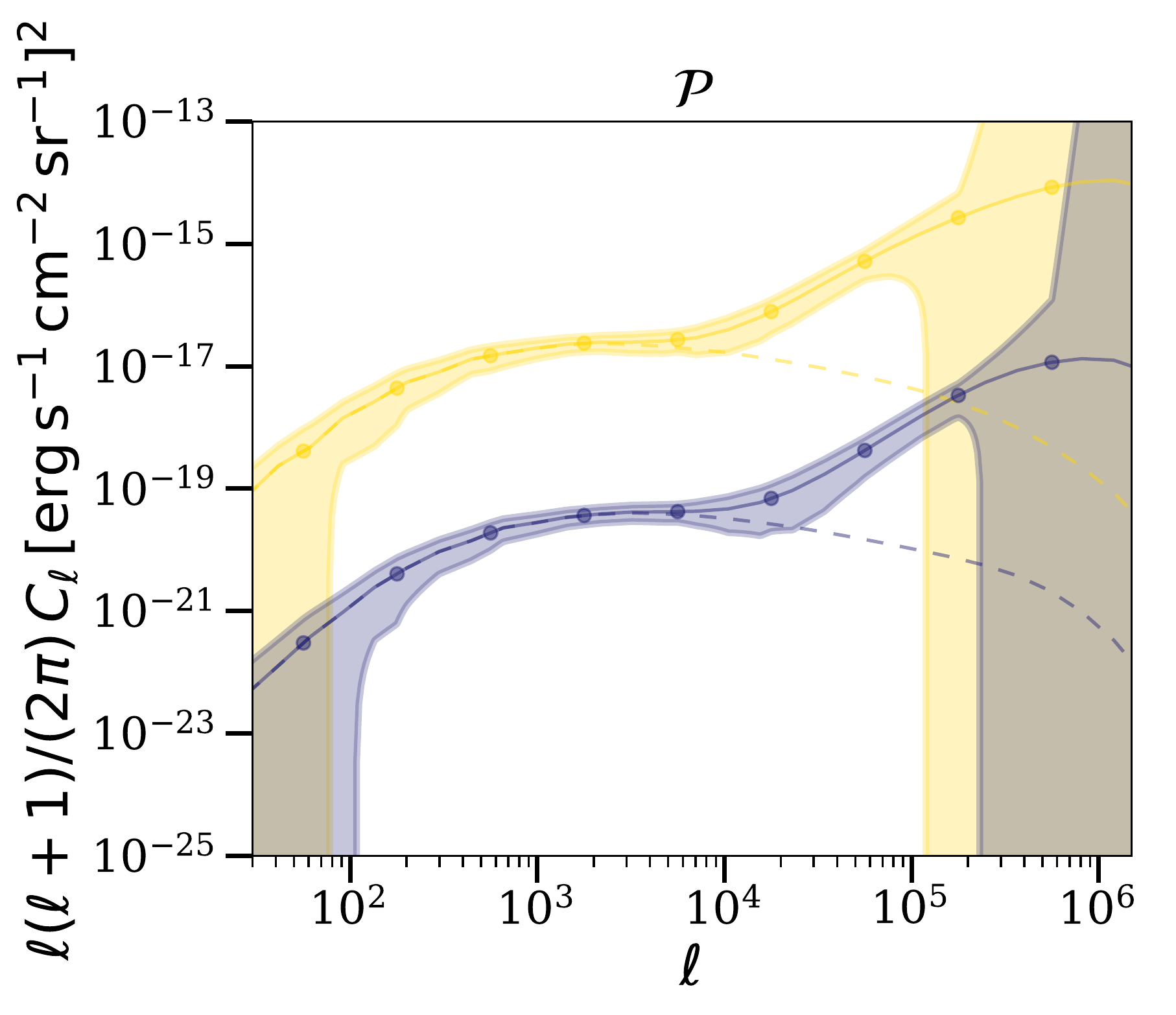}
\includegraphics[width=0.51\textwidth]{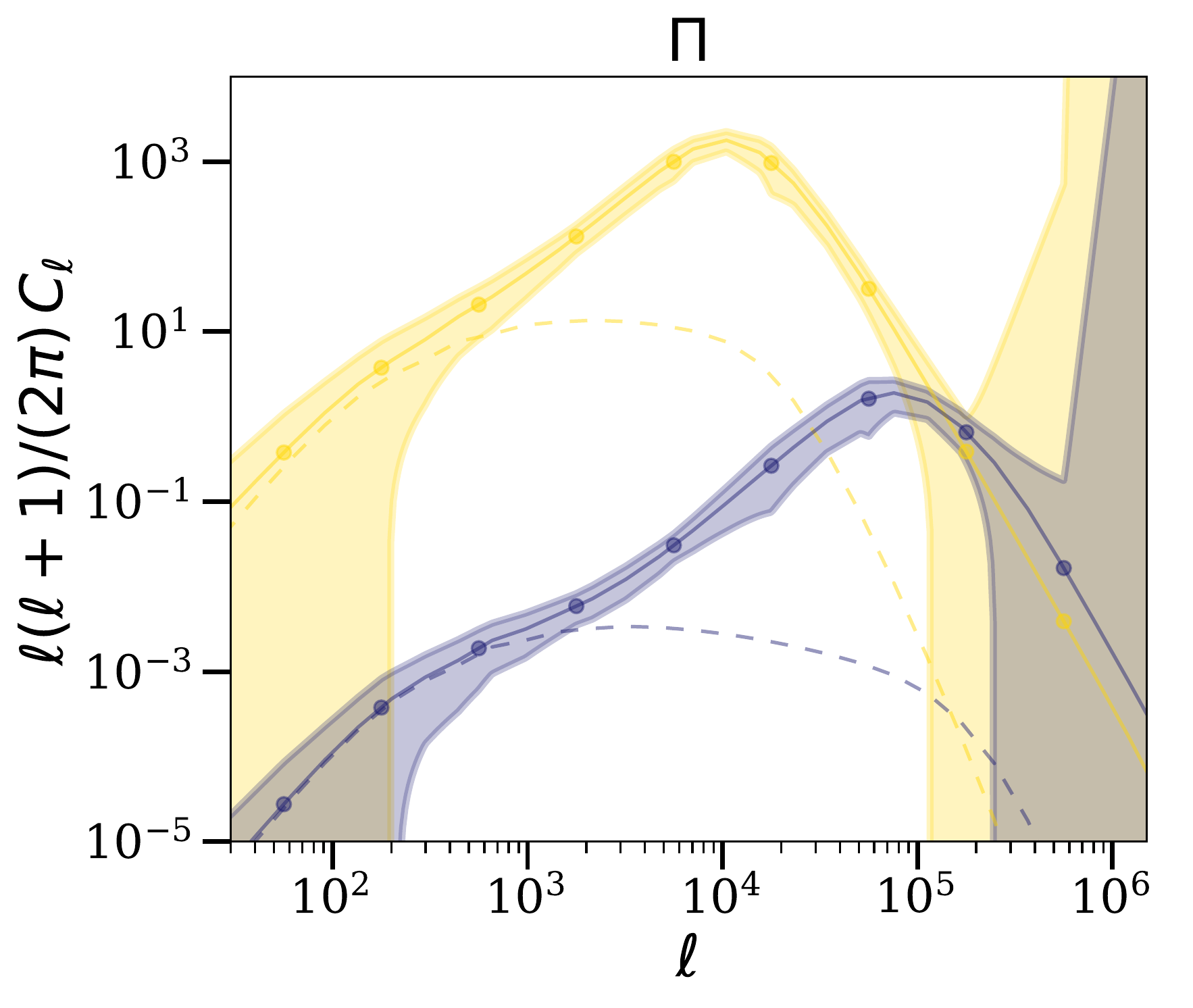}\includegraphics[width=0.51\textwidth]{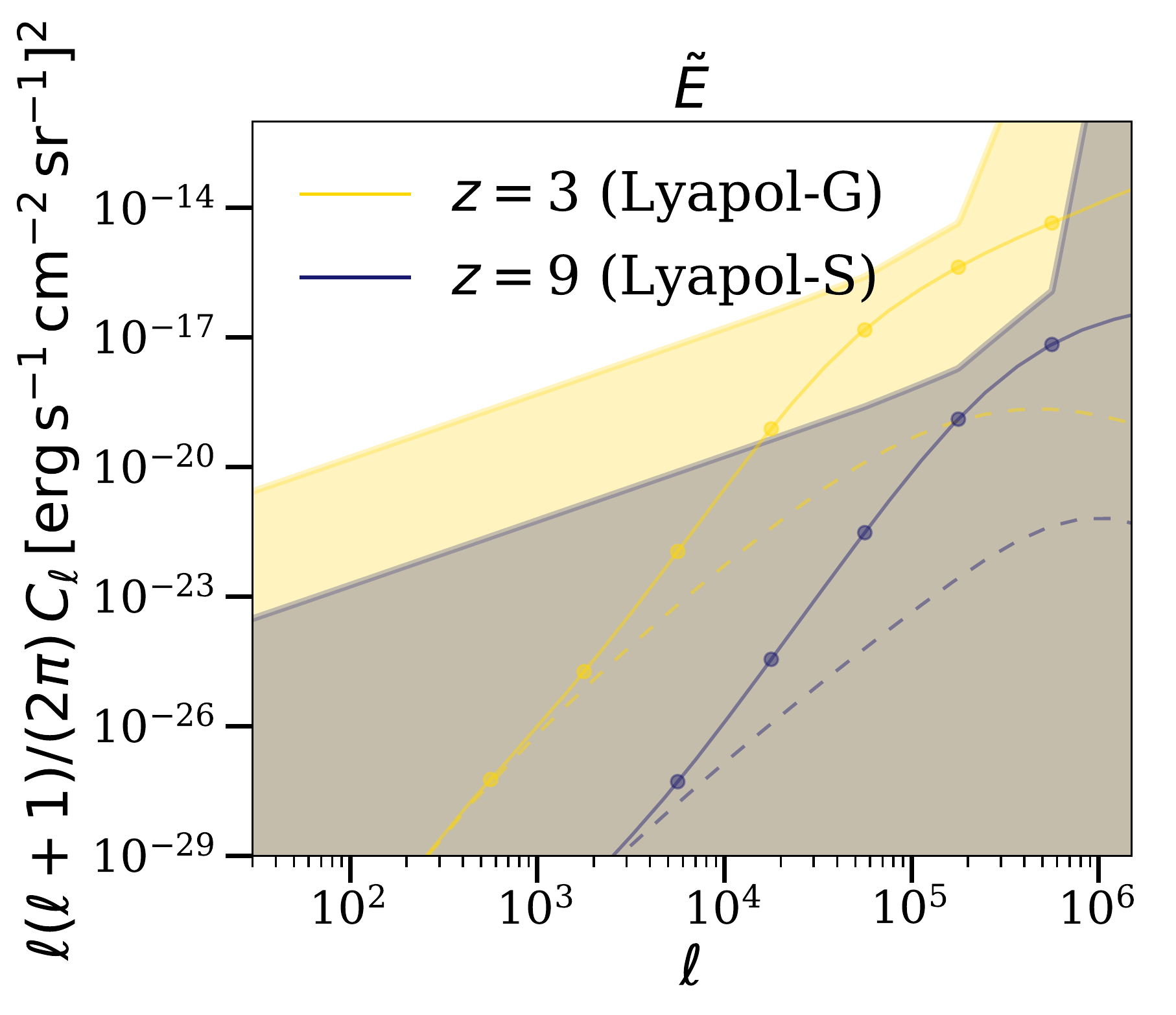}
\caption{Detectability estimates for the fiducial power spectra of Figure \ref{fig:psplot} at 
redshifts $z=3$ ({\it yellow lines}) and $z=9$ ({\it blue lines}) with the parameters of Lyapol-G and 
Lyapol-S described in Table \ref{ta:detect}, respectively, and  considering a redshift depth 
$\Delta z = 0.5$ in all cases. The {\it dots} 
represent the positions where the variance is calculated, and the {\it shaded areas} represent the 
uncertainty in the power by interpolating between points. The amount of signal collected by the large 
redshift depth enables precise measurements of the power spectra between $\ell \gtrsim 
10^2$  and $\ell \sim 10^5$ for all but the $\tilde E$ quantities. The low values of the $\tilde E$ 
power at small multipoles and the beam and pixel sizes are the reasons behind the non detection.}
\label{fig:error}
\end{figure}

   Figure \ref{fig:error} displays the uncertainties for the fiducial power spectra of Figure \ref{fig:psplot} at 
redshifts $z=3$ ({\it yellow lines}) and $z=9$ ({\it blue lines}) with the parameters of Lyapol-G and 
Lyapol-S described in Table \ref{ta:detect}, respectively, and  considering a redshift depth 
$\Delta z = 0.5$ in all cases. The {\it dots} represent the positions where the variance is calculated, 
and the {\it shaded areas} represent the uncertainty, obtained by simply interpolating between the 
values in the points. Overall, this figure shows that the amount of signal collected by the large redshift 
depth ($\Delta z = 0.5$) enables measurements of the power spectra between $\ell \sim 10^2$  and 
$\ell \sim 10^5$ for all the quantities but $\tilde E$. The steep decay and low values of the $\tilde E$ 
signal toward low multipoles does not allow detecting this power even at the lowest redshift. The 
peak of the power at multipole values $\sim 10^5$ is high enough to be detected, but this would require 
a smaller beam and pixel sizes than the ones quoted in Table \ref{ta:detect}.

\begin{figure}\center 
\includegraphics[width=0.51\textwidth]{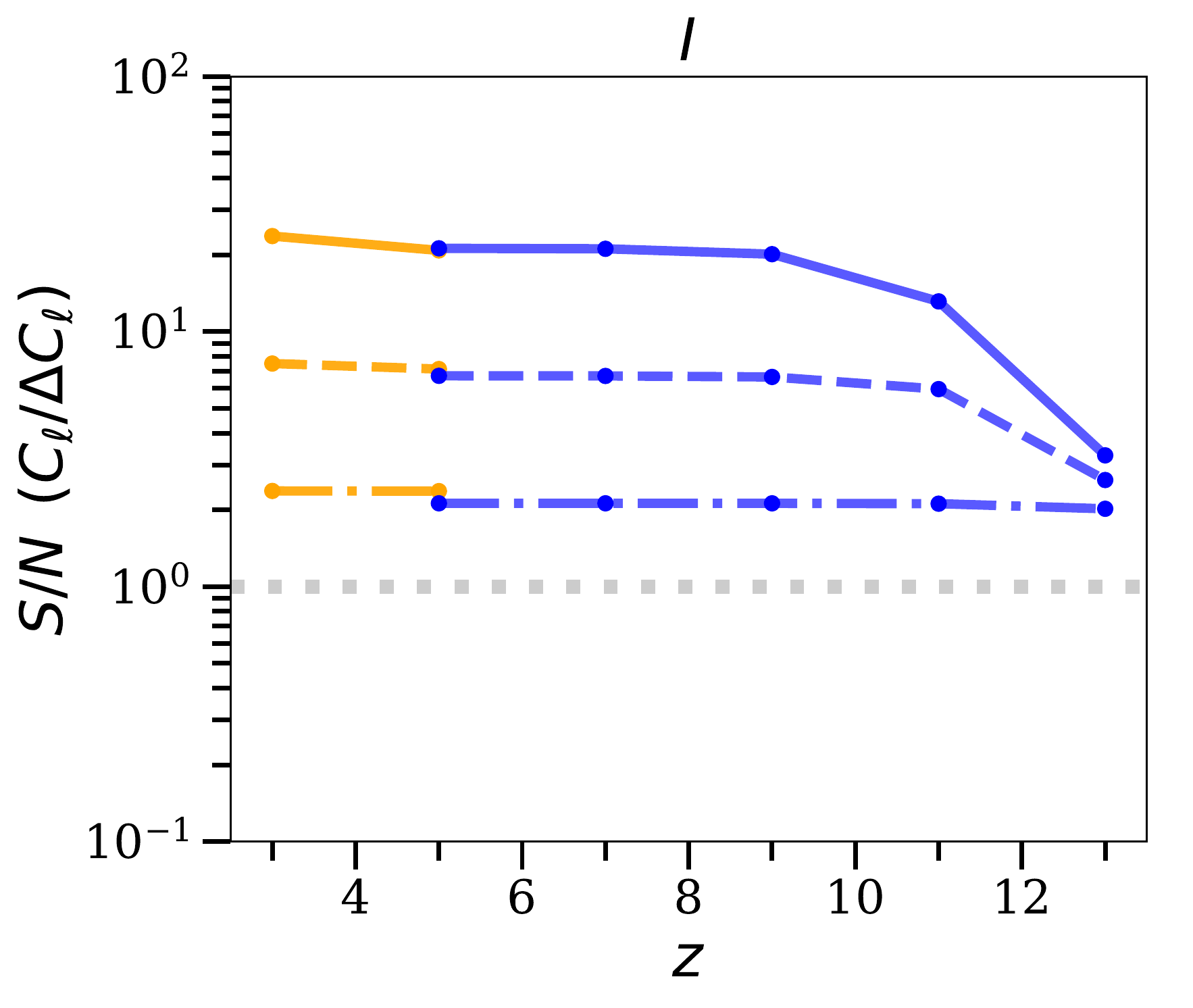}\includegraphics[width=0.51\textwidth]{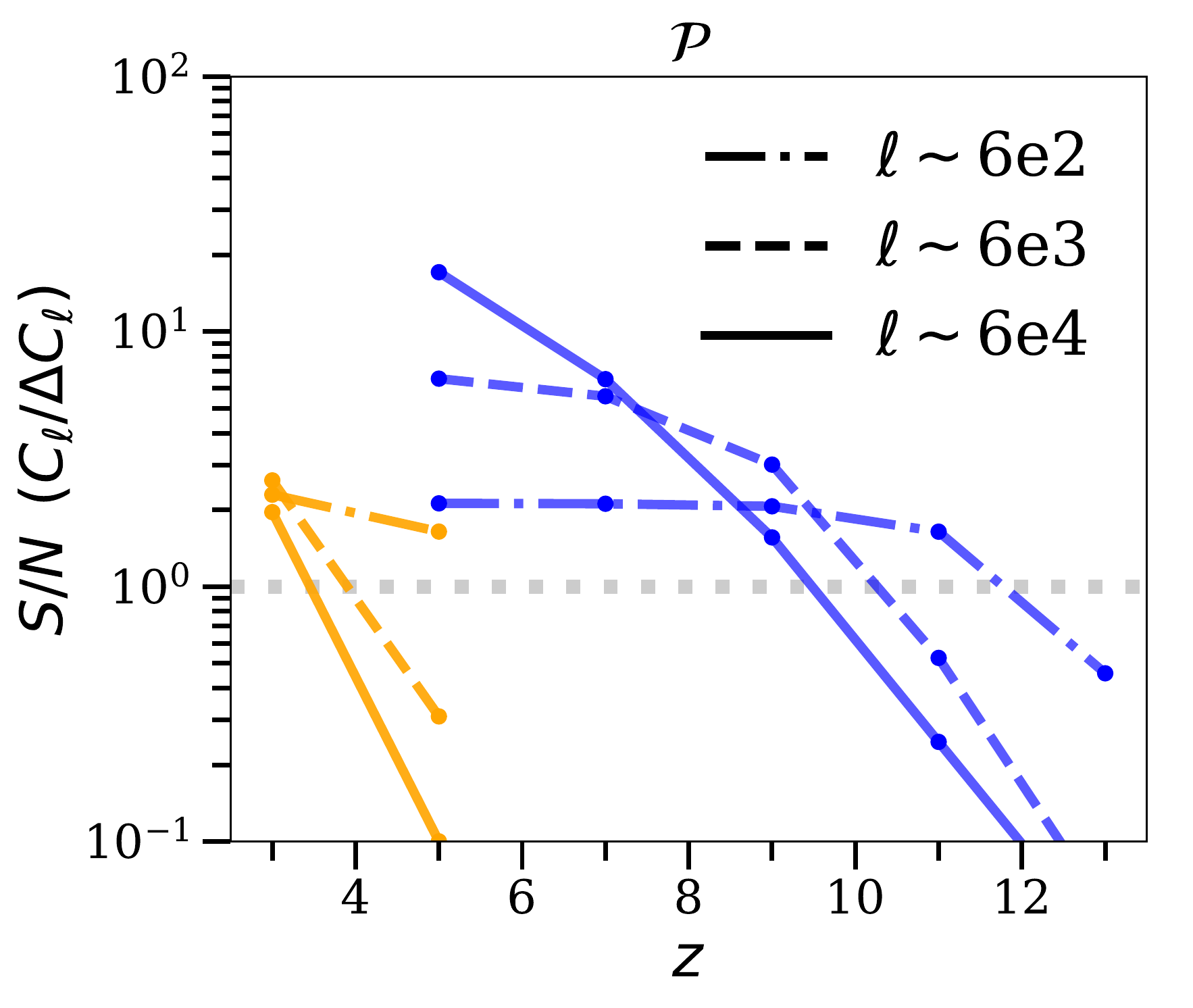}
\includegraphics[width=0.51\textwidth]{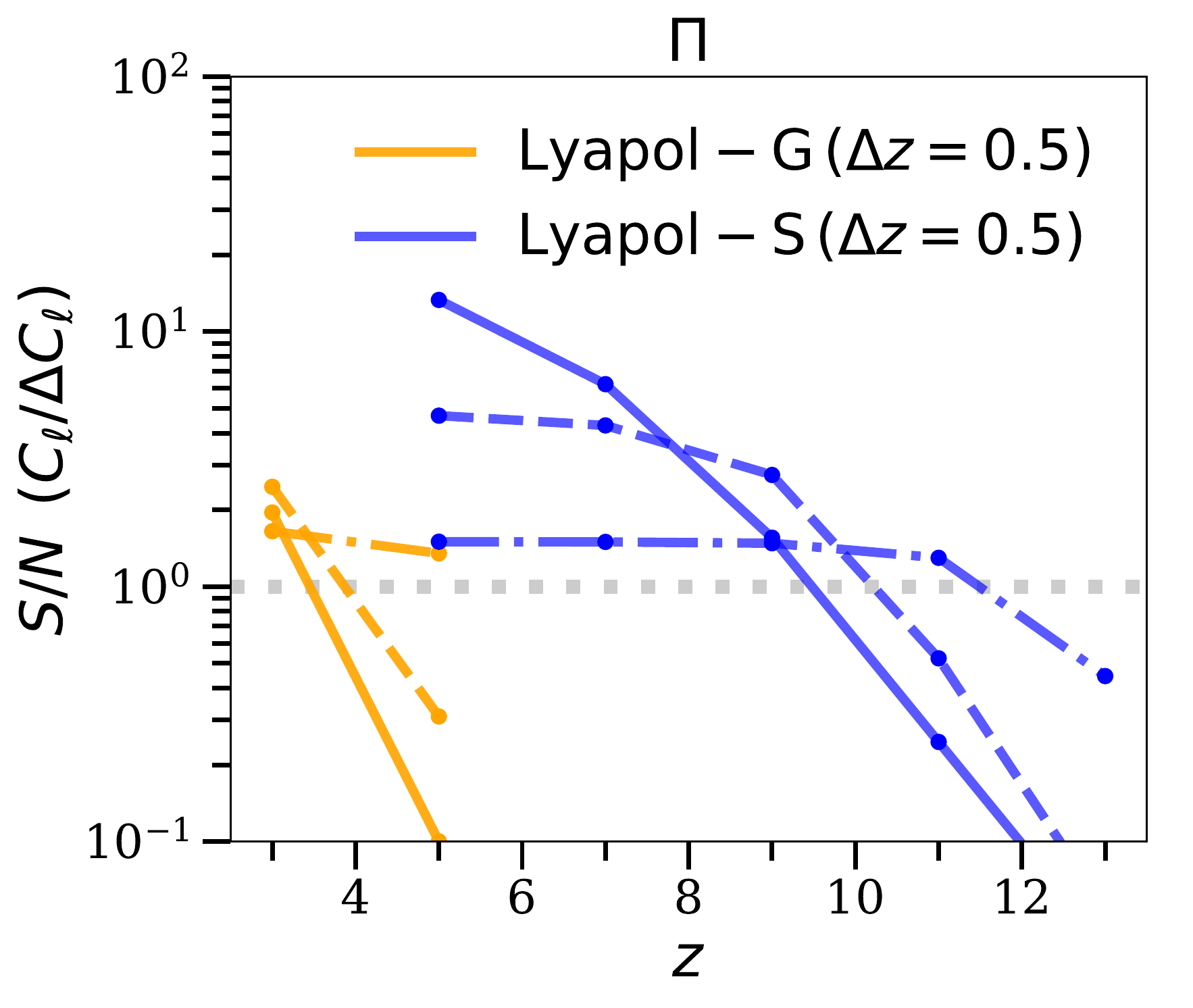}
\caption{S/N estimates for three multipole 
values, and for all but the $\tilde E$ polarization quantities. The {\it orange} and {\it blue lines} denote 
the ground-based Lyapol-G  and space-based Lyapol-S setups, respectively, with a redshift depth of 
$\Delta z = 0.5$. We have used the same sensitivity for $z=5$ and $z=3$ for Lyapol-G. The 
sensitivity evolution of Lyapol-S with redshift is taken from the CDIM deep case reported 
in \cite{Cooray2019}, rescaled to our value at $z=9$ in Table \ref{ta:detect}. } 
\label{fig:snr}
\end{figure}

   Figure \ref{fig:snr} illustrates the S/N for the two instrumental setups at three multipole values, 
and for all but the $\tilde E$ polarization quantities. The {\it orange} and {\it blue lines} denote 
the ground-based Lyapol-G  and space-based Lyapol-S setups, respectively, with a redshift depth 
of $\Delta z = 0.5$. We have assumed the same sensitivity for Lyapol-G at $z=5$ and $z=3$. The 
evolution of the sensitivity with redshift for the case of Lyapol-S is taken from the CDIM `deep' case 
shown in the middle panel of figure 2 in \cite{Cooray2019}, rescaled to our value at $z=9$ in Table 
\ref{ta:detect}. The instrument sensitivity increases by a factor of $\sim 3$ from redshift $z \sim 
5$ to $z \sim 13$, but this is counterbalanced by the fact that the number of spectral channels that 
cover $\Delta z = 0.5$ decreases with redshift at a fixed spectral resolution. Overall, the S/N for the 
total intensity remains fairly constant at all three multipoles up to $z \sim 9$, beyond which the 
instrumental sensitivity suppresses the signal rapidly, from the highest to the lowest 
multipoles. Because the power of the polarized intensity is a few orders of magnitude fainter than that 
for the total intensity, the steep decrease in S/N appears already at redshift $z \sim 5-7$.
The turnover of the $\Pi$ power spectra is detectable (S/N$\sim2-3$) at $z=3$ with Lyapol-G, and 
up to $z\sim 9$ with Lyapol-S.

   In summary, these simple estimates suggest that the detection of \lya polarization up to the 
early times of reionization requires sensitivities higher than those of current and near-future experiments. 
We discuss in the next section  the  sources of foreground contamination that need to be 
taken into account for these observations. 

\subsection{Foregrounds}\label{sec:foregrounds}

   The major source of foreground polarization detectable from the ground at wavelengths 
of $\lambda \sim 1\mu$m is the atmospheric Rayleigh-scattered radiation from the Moon and the 
stars \citep{Glass1999}. From space, the atmospheric component is significantly reduced, and the 
primary contamination arises from starlight scattered by the Milky Way dust \citep{Sparrow1972,
Arai2015}. These contributions, however, present a smooth spectrum with a known frequency 
dependence, which one could try to model and subtract from the observations \citep{Brandt2012}. 
Furthermore, the use of cross-correlations would allow the identification of the foreground signal, 
because the latter would not be correlated with the sources (galaxies) of \lya polarization of interest.

   A significant benefit of using polarized emission for the detectability, compared to using 
total emission, is the reduced impact from interlopers. When considering polarized radiation, 
emission at a given frequency that redshifts into the detection window of Ly$\alpha$, will not be 
misidentified as \lya unless it is polarized. This improvement enables one to use large redshift  
depths for the integration of signal, and thus increase the S/N, without suffering from extreme 
interloper contamination. 

        A source of foreground contamination for the polarization signal of interest, 
may be the polarized H$\alpha$ radiation originating from Raman-scattered \lyb radiation 
\citep[][see also the case of scattered O{\sc vi} by neutral hydrogen in 
\citealt{Nussbaumer1989}]{Lee1998}. This process, however, occurs at column densities typical 
of that in the interstellar medium and, therefore, the signal would be, in most cases, compact 
and in the core of the sources. If this process happens below the spatial resolution of the detector, 
the average polarization signal would be null. However, if it is extended, it may be included in the 
observations. The polarization signal from the core of objects such as high-redshift radio galaxies 
\citep[e.g.,][]{Cimatti1998,Vernet2001} or Seyfert II galaxies, is also expected to be 
compact and, therefore, to not introduce significant contamination into the measurements. Finally, 
in \cite{Masribas2018}, we demonstrated that the radiation from a hyperluminous quasar that is 
Thomson scattered by the free electrons (or scattered by dust) in the circumgalactic medium of the 
host galaxy can be detectable. Even though this signal can extend well out into the halo, these bright 
sources are rare and identifiable (maskable) to avoid contamination from electron scattering on the 
\lya polarization signal of interest here.

\section{\lya $B$ Modes as Probes of Halo Anisotropy, Gravitational Lensing and Faraday Rotation}\label{sec:bmodes}

    The \lya $B$ mode power in our formalism is null, because we have considered a radially symmetric 
(isotropic) polarization signal around the halos. In reality, however, the H{\sc i} distribution in galaxy 
halos can present a complex and inhomogeneous geometry, and the emission of \lya radiation from 
the source can be highly anisotropic, which will result in patterns departing significantly from the 
idealized isotropic case. Therefore, a \lya $B$ mode signal is expected to arise from actual galaxies, 
where the $B$ mode amplitude will be an indicator of the amount of `polarization anisotropy' in the 
halos. Measurements of the global \lya $B$ and $E$ mode signals at various redshifts could  
be used as indicators of the evolution of the average inhomogeneity and anisotropy  of halos over 
time. This quantification, in turn, might be a tracer of the major physical processes driving galaxy 
evolution, such as merging rates, or feedback effects impacting the properties of the gas in the halos 
at different redshifts. 

     Other sources of \lya $B$ modes are the effects of gravitational lensing \citep{Zaldarriaga1998},  
and Faraday rotation \citep{Kosowsky1996,Kosowsky2005}, which convert the propagating \lya $E$ 
modes into  $B$ modes. For a high number density of \lya polarization sources, covering a large 
fraction of the sky, weak gravitational lensing $B$ modes may be considerable and of interest 
\citep[e.g.,][]{Foreman2018}. Furthermore, because the \lya sources exist at all redshifts, one could  
perform a tomographic analysis of the lensing signal, separating the contribution of different redshift 
bins. However, we expect the lensing signal to be small when the fraction of the sky covered by 
sources is small, owing to the small size of the polarization signal in the halos. In this case, however, 
one might be able to investigate the galaxy lensing effects by measuring the shear introduced to the 
shape of the $E$ modes around individual objects. The impact from Faraday rotation is uncertain, 
because it depends strongly on the magnetic fields, as well as on the distribution of matter in the Milky 
Way and the intergalactic medium, all quantities difficult to constrain with precision. However, 
\cite{Soma2014} found that the impact of Faraday rotation on pre-reionization polarized 21cm radiation 
is very important due to the large wavelength of this radiation. Because Faraday rotation depends 
on the square of the wavelength, this effect would be about $(10^{5})^2$ times smaller for \lya than 
for 21cm, albeit the other parameters remain the same for both frequencies. Finally, $B$ modes might 
also arise from the clustering or merging, as well as overlap, of halos, which is beyond the capabilities 
of the halo model approach. 

    In addition to these `physical' sources of \lya $B$ modes, it is also possible that there is a 
contaminant signal arising from `ambiguous' modes \citep[e.g.,][]{Lewis2002,Bunn2003}. Ambiguous 
modes appear when only a fraction of the sky is observed. In this case, the decomposition of the 
polarization signal is {\it non-local} and {\it non-unique}, and therefore modes that are simultaneously   
divergence free (like $B$ modes) and curl free (like $E$ modes) appear. In other words, it is not clear 
whether the power of these modes is contributed by $E$ or $B$. This effect can be especially significant 
for the case of a \lya $B$ mode measurement, because we expect the $B$ modes to be subdominant 
compared to $E$ modes. The level of leakage between $E$ and $B$ modes may be significant 
compared  to the signal expected for the $B$ modes, and it can therefore misguide the interpretation of 
the observations.

\section{Future Work}\label{sec:future}

    In our calculations, we have not included the potential effect of Population III galaxies, 
which would result in a significant increase of the \lya emissivity compared to our fiducial 
calculations that consider normal (Population II) galaxies  \citep[e.g.,][]{Raiter2010,Masribas2016}. 
This effect, however, would be significant for the (global) power spectra calculations at redshifts 
above $z\gtrsim 10-15$, where the average star-formation rate may be dominated by Population III 
galaxies, as suggested by recent numerical  \citep{Jaacks2018}, as well as (semi-)analytical 
\citep{Mebane2018,Mirocha2018} star-formation calculations. 

    \cite{Rybicki1999} suggested that an important source of \lya polarization other than galaxies, even 
before cosmic reionization, could be the scattering of photons with intergalactic (IGM) neutral hydrogen 
gas moving with the Hubble flow. This polarization can reach degrees of polarization as high as 
$\sim 70\,\%$, although \citep{Dijkstra2008} noted that this would be the case for gas beyond $\sim 10$ 
virial radii from galaxies. The `static' intergalactic gas closer to galaxies   
would reach lower polarization degrees, on the order of $\lesssim 7 \,\%$. However, the gas at a few 
virial radii may be inflowing toward the halo center due to gravitational collapse, which may introduce 
polarization levels of a few tens of percent. This IGM \lya polarization component can be important at 
redshifts above $z\sim 6$, where the IGM may still present large neutral gas regions. 
We will investigate the impact of this intergalactic polarization via analytical and numerical calculations  
in future work.

     An important aspect that needs to be revisited in future work is the effect of performing cross 
correlations between the polarization signals and other tracers of cosmic 
structure and/or line emission at other frequencies (e.g., galaxies, quasars, or 21 cm, CO, C{\sc ii}, 
and H$\alpha$ emission). For example, because the \lya polarization signal is high at small 
(galaxy) scales, the cross correlation of \lya polarization with galaxies could be used to enhance 
the detectability at those scales.

\section{Conclusions}\label{sec:conclusions}

    We have presented an analytical formalism of \lya polarization, arising from the scattering 
of photons with neutral hydrogen gas around galaxies, for intensity mapping studies. We have 
used the halo-model formalism, as well as \lya profiles based on simulations and observations, for 
modeling the signal. We have estimated the auto and cross power spectra of the \lya quantities total 
intensity, $I$, 
polarized intensity, $\po$, polarization fraction, $\Pi = \po / I$, and the astrophysical \lya $E$ and 
$B$ modes, introduced here for the first time in galaxy studies, and derived from the CMB formalism. 
The dependence on model parameters and the impact of variations in their values has been 
investigated, as well as the detectability of the power spectra for the aforementioned quantities, 
considering the redshift range $3\lesssim z \lesssim 13$. The main findings of this work are 
as follows: 

\begin{enumerate}

\item The power spectra of the polarization quantities $\Pi$ and $E$ present sharper 
         features than the power spectra of $I$ and $\po$ in general, especially for the one-halo 
         terms (Figures \ref{fig:psplot} and \ref{fig:cross}). The position 
         of the one-halo peaks of $\Pi$ and $E$ depends on redshift, and it is related to the average 
         halo size (and mass) dominating the signal at a given time. 
                  
\item The ratio between the power spectra of the polarized intensity and the total intensity gives 
	information of the polarization fluctuations between halos. Furthermore, the 
	distribution of sizes for the polarization signal can be obtained from the ratio of the one-halo 
	terms at high multipoles. Finally, the evolution of the polarization fluctuations with redshift 
	indicates the dependence of the polarization signal with halo size (Figure \ref{fig:pooveri}).

\item The signal from \lya $B$ modes is null by construction in our formalism, because we consider 
	symmetry around the halos. In real data, however, a $B$ mode signal is expected to arise from the 
	anisotropy in the halo gas distribution and the radiation field. The combined measurements of \lya 
	$E$ and $B$ modes for various redshifts will yield information about the physical properties and the 
	evolution of cold gas in halos (\S~\ref{sec:bmodes}).

\item Variations in the amplitudes and shapes of the \lya profiles, especially in the slope of the 
	surface brightness profile, produce different changes for the power spectra of different polarization 
	quantitites, and for different redshifts. Comparisons between various quantities, and at various 
	redshifts, enables one to extract the physical characteristics (slope and extent) of the real-space 
	\lya profiles (\S~\ref{sec:dep} in the Appendix). 

\item The detectability of the polarization signal requires improvements in the sensitivity of 
	current ground- and space-based experiments by factors between $\sim 10 - 100$, depending on 
	redshifts and experiments (Figures \ref{fig:error} and \ref{fig:snr}, and 
	\S~\ref{sec:detection}). Foreground contamination from the atmosphere, and Milky Way 
	dust-scattered radiation, is expected to be important and needs to be modeled and removed 
	 (\S~\ref{sec:foregrounds}). 
	
\item The contamination from interlopers is expected to be smaller when considering polarized radiation 
	than total radiation, because the contaminant radiation needs to also be polarized 
	to impact the measurements. 
         
\end{enumerate}

    We have shown that the use of polarization in intensity mapping studies enables 
extracting more physical information about the galaxies and their environments than total 
emission alone. This first work has presented the general formalism, which will  
be extended, as well as applied to specific cases, via analytical and numerical 
calculations in coming studies.

\section*{Acknowledgements}

    We are grateful to Agn\`es Fert\'e for an inspiring discussion that motivated the idea of considering 
polarization in intensity mapping experiments. We are indebted to Chris Hirata and Siavash Yasini, 
who greatly contributed to the derivation of the Lyman-alpha E and B mode formalism, and to Bryan 
Steinbach and Emmanuel Schaan for noting the nature of the shot-noise terms. We thank our colleagues 
Peter Laursen, Phil Korngut, Jason Sun, Phil Berger,  Marta Silva, Matt Johnson, Chen Heinrich,  
Isabel Swafford, Marlee Smith, Adam Lidz, Fred Davies, Jae Hwan Kang, Jordi Miralda Escud\'e, 
and others, for comments and discussions during this project. We are also thankful to Bin Yue and 
Maxime Trebitsch for noting the effect of weak lensing on the polarization signal. This research was 
carried out at the Jet Propulsion Laboratory, California Institute of Technology, under a contract 
with the National Aeronautics and Space Administration (80NM0018D0004).

\bibliographystyle{aasjournal}
\bibliography{impol}\label{References}


\appendix

Below, 
\S~\ref{sec:dep} shows  the impact on the power spectra of variations in the fiducial model parameters. 
\section{Dependences on Model Parameters}\label{sec:dep}
 
    We explore below the impact of variations in the model parameters on the fiducial power spectra of 
Figure \ref{fig:psplot}.  We first test changes in the extent of the polarization signal, out to three and 
five virial radii, as well as for the hypothetical case for which all halos show the same polarization 
extent. We then assess changes in the shape of the surface brightness and polarization profiles.

\begin{figure}\center 
\includegraphics[width=0.51\textwidth]{psI.pdf}\includegraphics[width=0.51\textwidth]{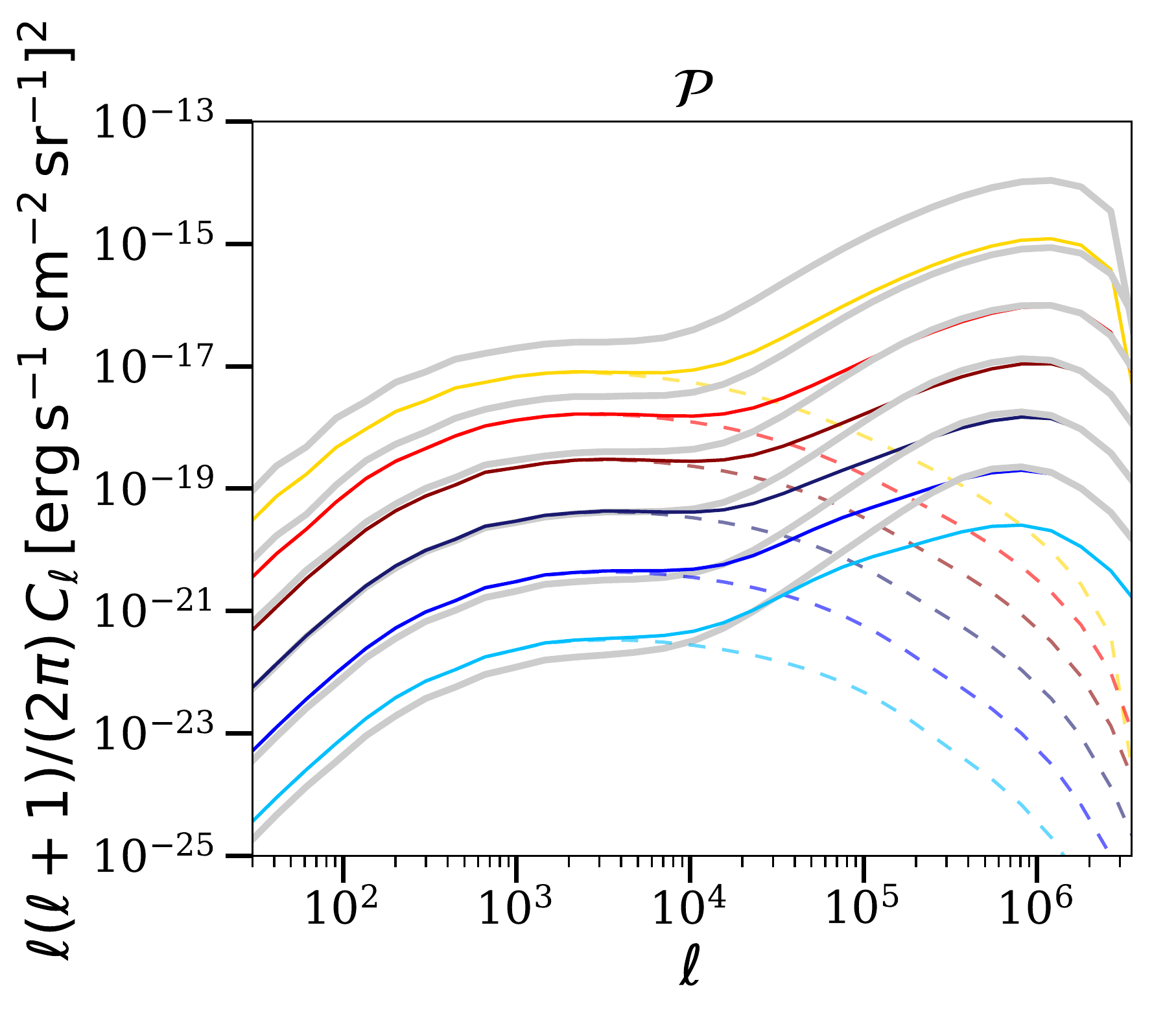}
\includegraphics[width=0.51\textwidth]{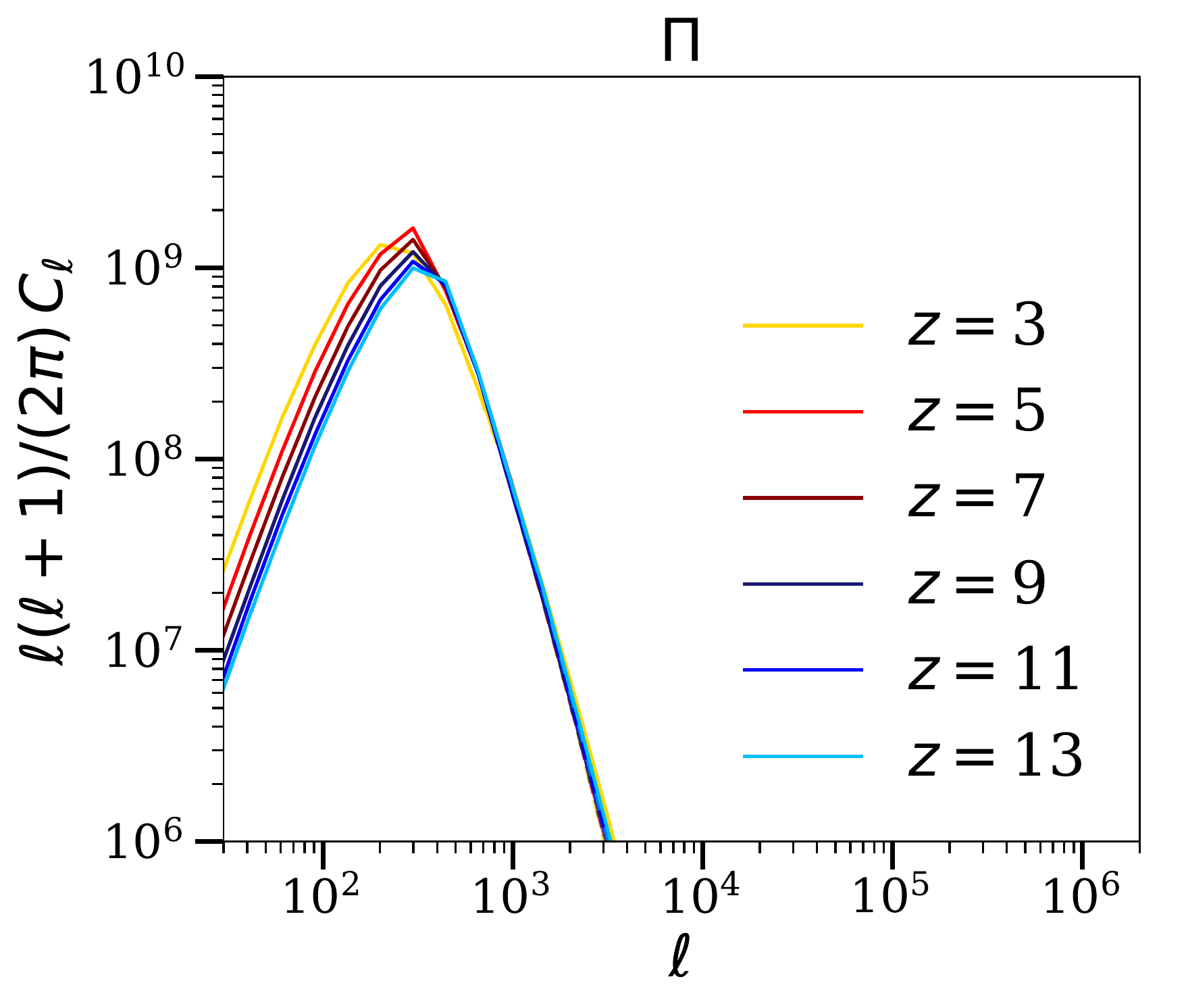}\includegraphics[width=0.51\textwidth]{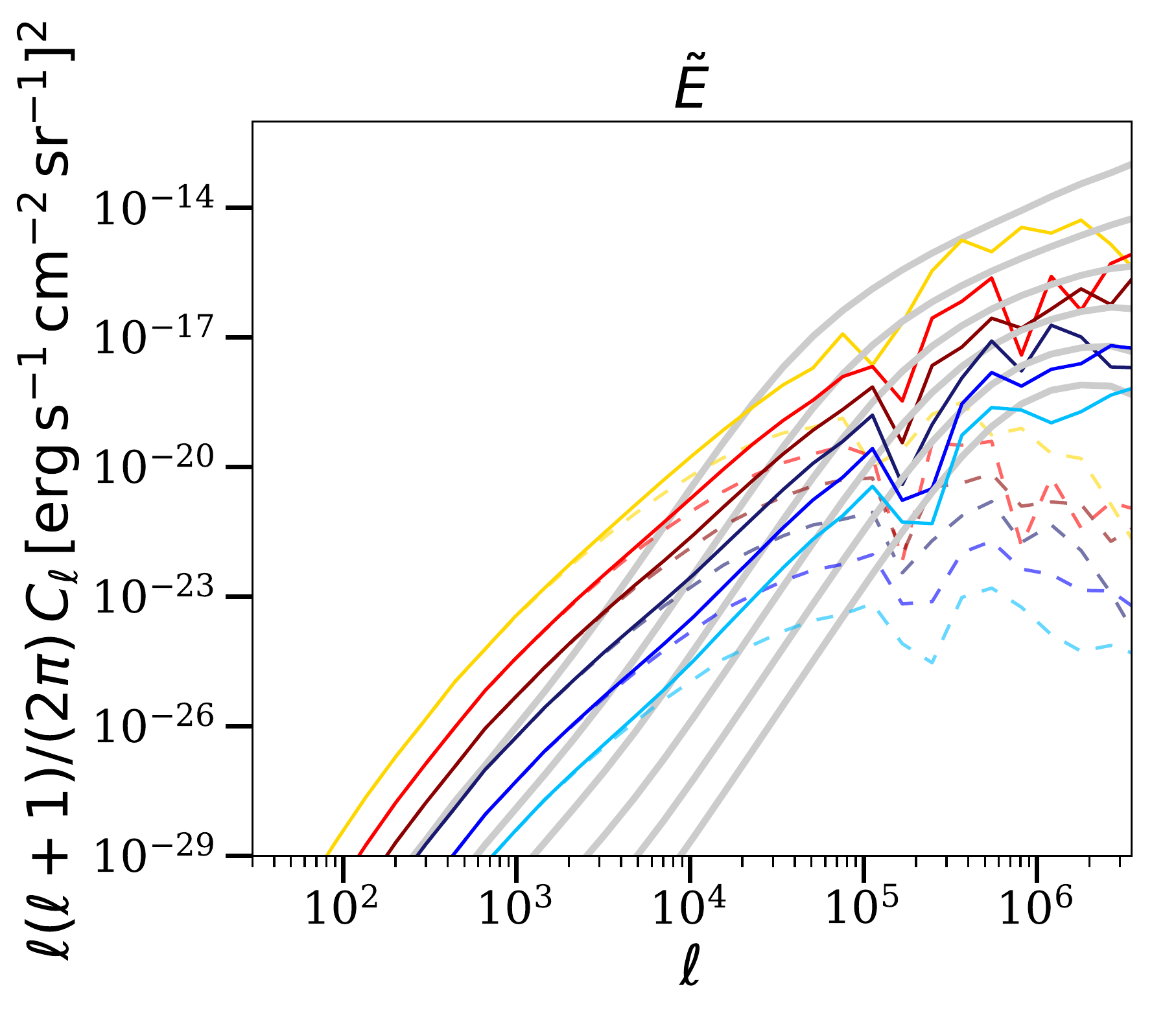}
\caption{Power spectra estimates adopting a position for the maximum polarization fraction at three 
virial radii ({\it color lines}), compared to the fiducial calculations that consider one virial radius 
({\it gray lines}). In general, the spectra for $\po$ and $\tilde E$ show a higher power at low 
multipoles and a lower one at large $\ell$ values compared to the fiducial case, and the one-halo term 
peak in the fiducial calculation is now smoothed out. The power for $\Pi$ shows a sharp peak 
corresponding to the two-halo term at $\ell \sim 10^2 - 10^3$, whose amplitude is now more than three 
orders of magnitude above the noise (not visible). The power does not change for the total intensity, 
$I$, because the signal does not depend on the polarization fraction.} 
\label{fig:rvir3}
\end{figure}

\begin{figure}\center 
\includegraphics[width=0.51\textwidth]{psI.pdf}\includegraphics[width=0.51\textwidth]{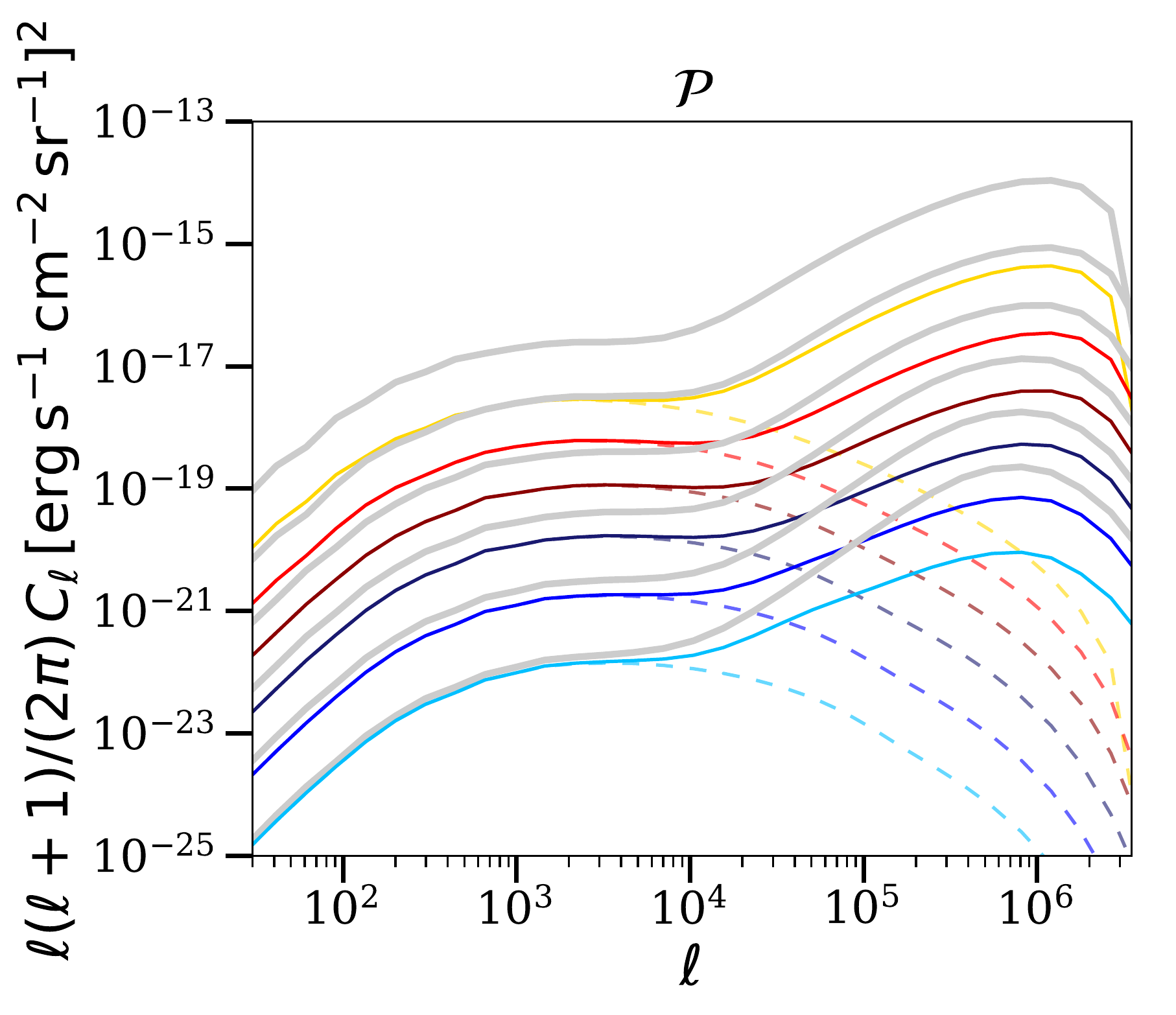}
\includegraphics[width=0.51\textwidth]{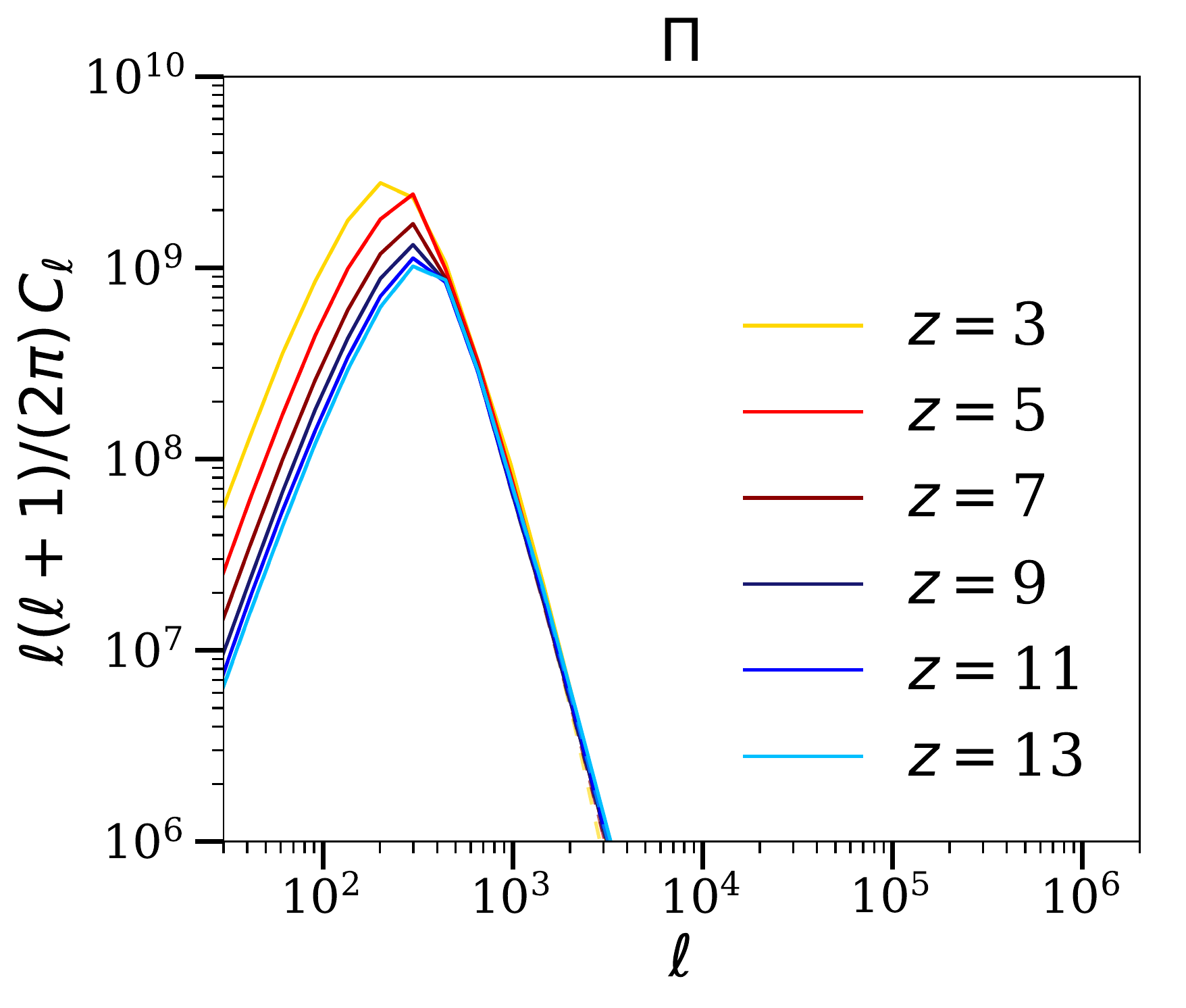}\includegraphics[width=0.51\textwidth]{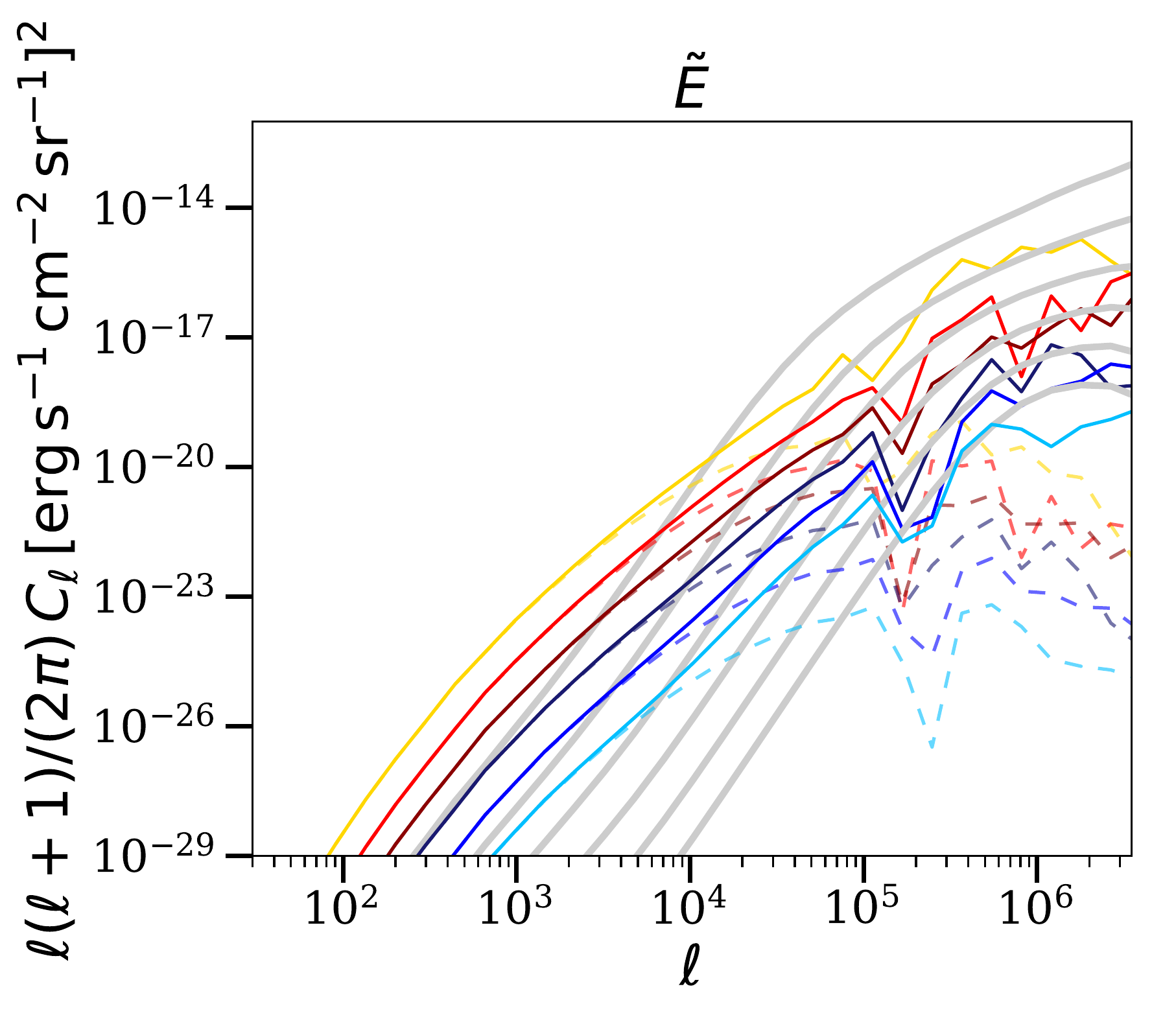}
\caption{Same as Figure \ref{fig:rvir3}, but with the maximum polarization fraction value now at five 
virial radii.} 
\label{fig:rvir5}
\end{figure}

\begin{enumerate}

\item{Extent of the polarization signal}. Our fiducial model assumes that the maximum  
polarization fraction value occurs at the virial radius of the halos, and it becomes zero rapidly after 
that position. Figure \ref{fig:rvir3} illustrates the impact of shifting the peak value out to three 
virial radii ({\it colored lines}) on the fiducial  power spectra of 
Figure \ref{fig:psplot} ({\it gray lines}). 
For the case of the total intensity, this produces no differences  because $I$ does not depend 
on the polarization fraction. For $\po$ and $\tilde E$, the shape of the power spectra is 
smoother. The major impact of changing the extent of polarization 
is visible in the power spectra of $\Pi$ ({\it bottom left panel}), which is now characterized by a 
sharp feature at $\ell \sim 10^2 - 10^3$. 

Figure \ref{fig:rvir5}  shows the case of shifting the maximum polarization fraction value out to 
five virial radii, which results in a similar behavior as for the case of three virial radii just discussed.

We also test the impact of a fix size for the polarization fraction profile, for comparison.  
Figure \ref{fig:fixsize} shows the power spectra obtained by considering the maximum polarization 
degree for all halos occurring at an (arbitrary) impact parameter of $r_{\perp}=50$ comoving kpc 
${\rm h^{-1}}$. At the lowest redshifts, the one-halo terms of $\po$ and $\tilde E$ are enhanced 
compared to the fiducial case, because now all halos are small instead of distributed in a broad 
range of sizes. Because all halos are typically small at high redshifts, the fix (small) impact parameter 
power spectra do not differ significantly from the fiducial calculations. The largest impact is visible 
as a reduction of the power spectra of $\Pi$ at the lowest redshifts. 
All the spectra now peak at the same position, because the extent of the polarization signal 
is constant.

\begin{figure}\center 
\includegraphics[width=0.51\textwidth]{psI.pdf}\includegraphics[width=0.51\textwidth]{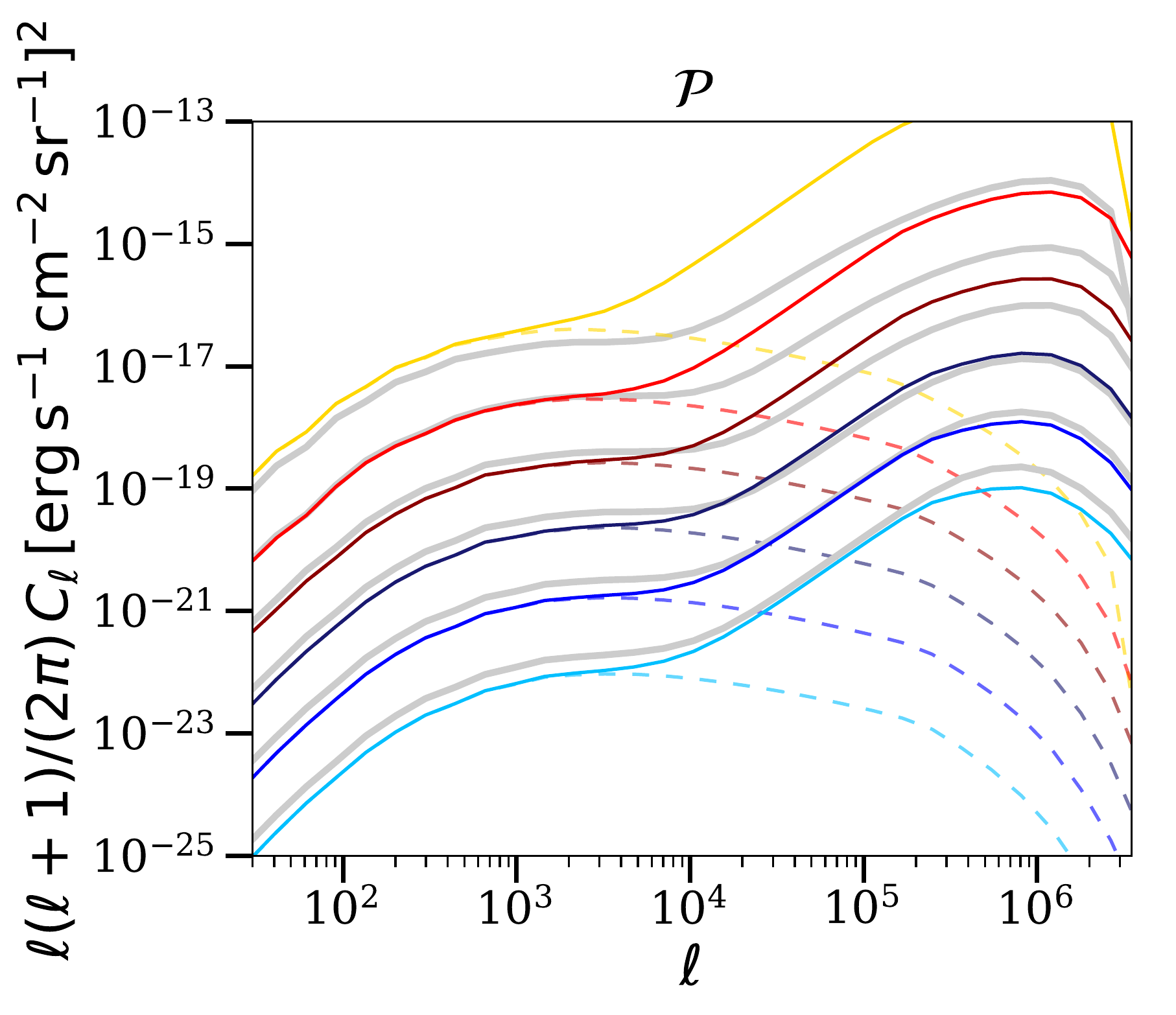}
\includegraphics[width=0.51\textwidth]{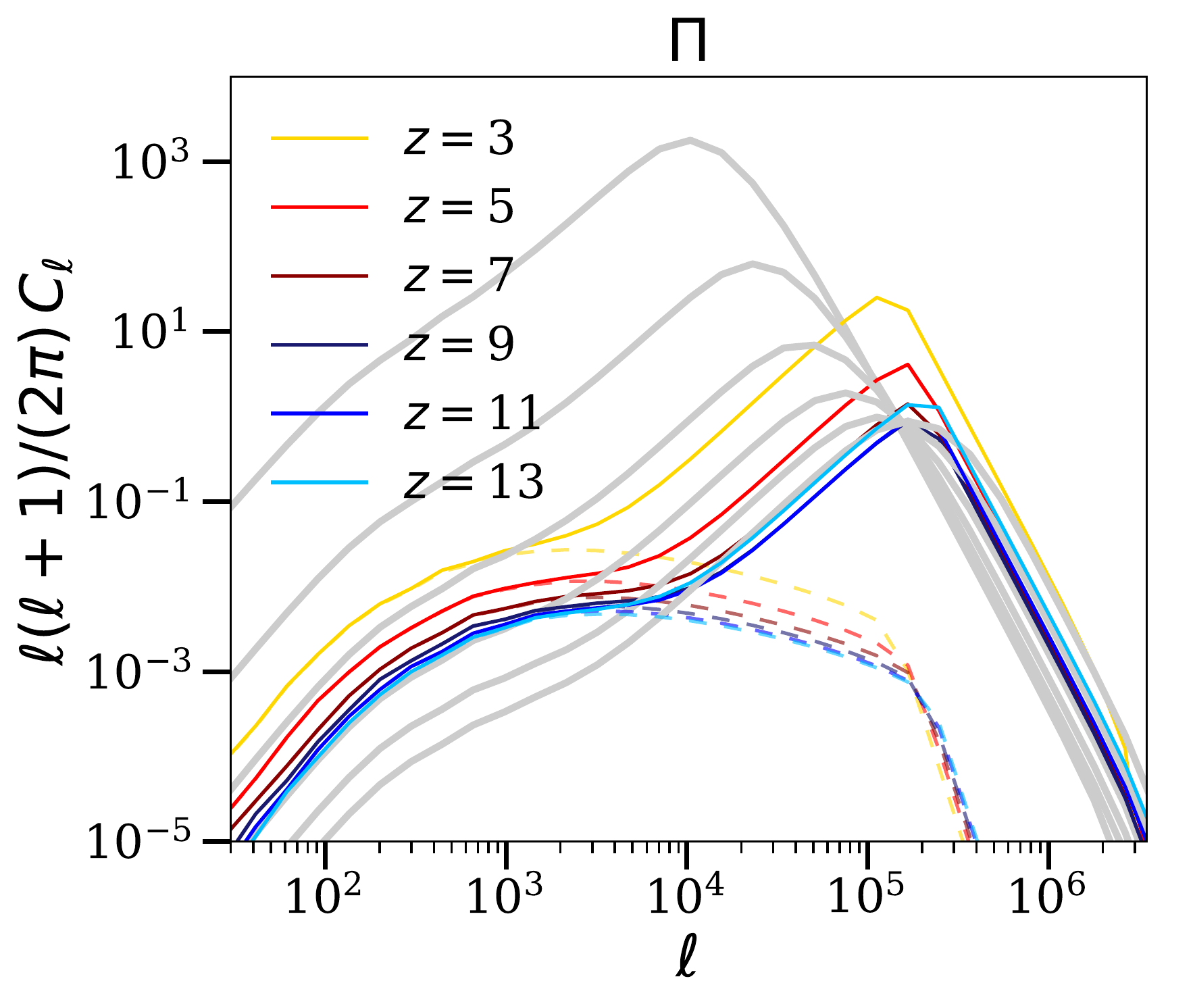}\includegraphics[width=0.51\textwidth]{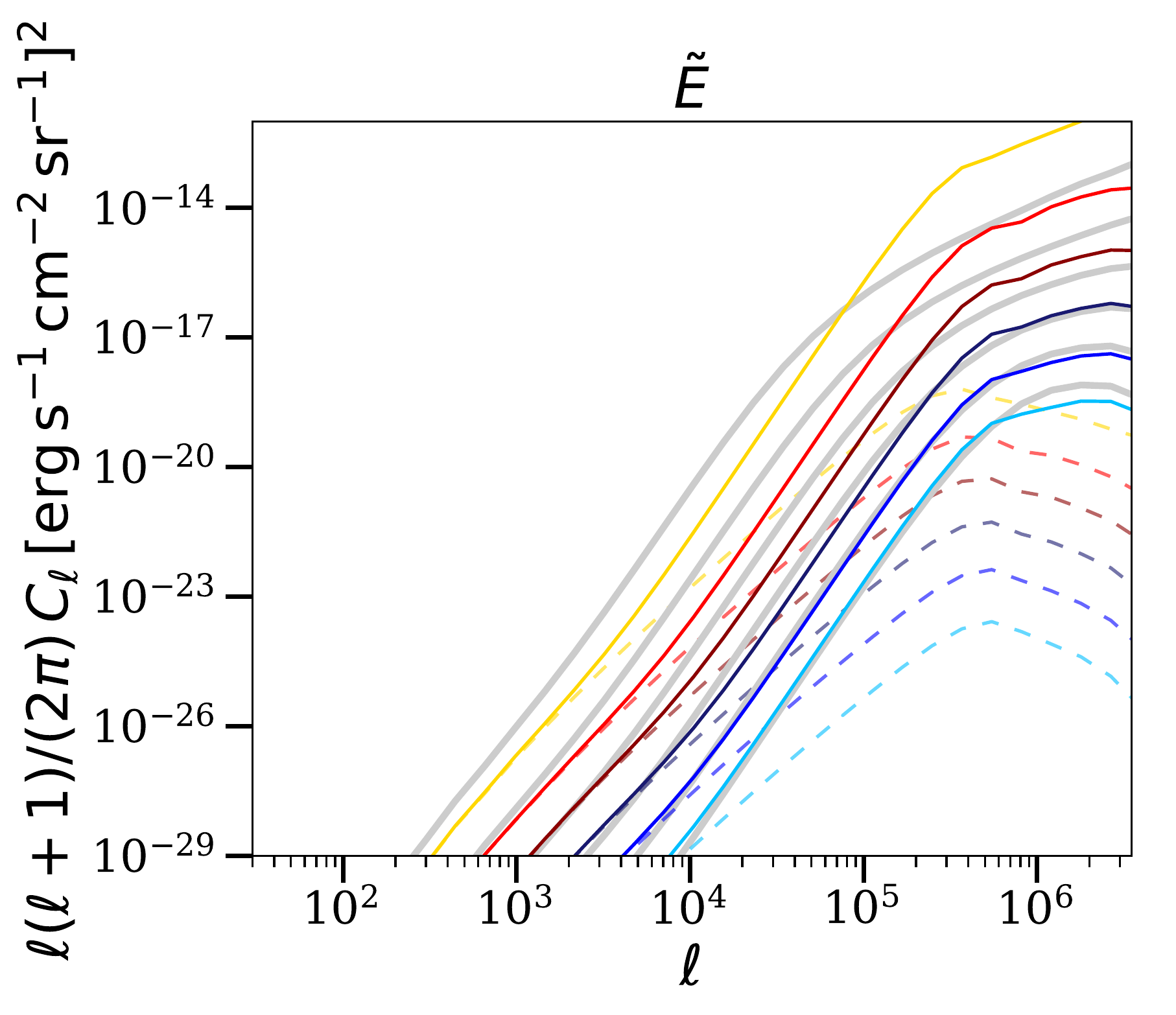}
\caption{Same as Figure \ref{fig:rvir3}, but fixing now the position of the maximum polarization fraction 
to a projected distance of $r_{\perp}=50$ comoving kpc ${\rm h^{-1}}$ from the center for all halos.} 
\label{fig:fixsize}
\end{figure}

\begin{figure}\center 
\includegraphics[width=0.51\textwidth]{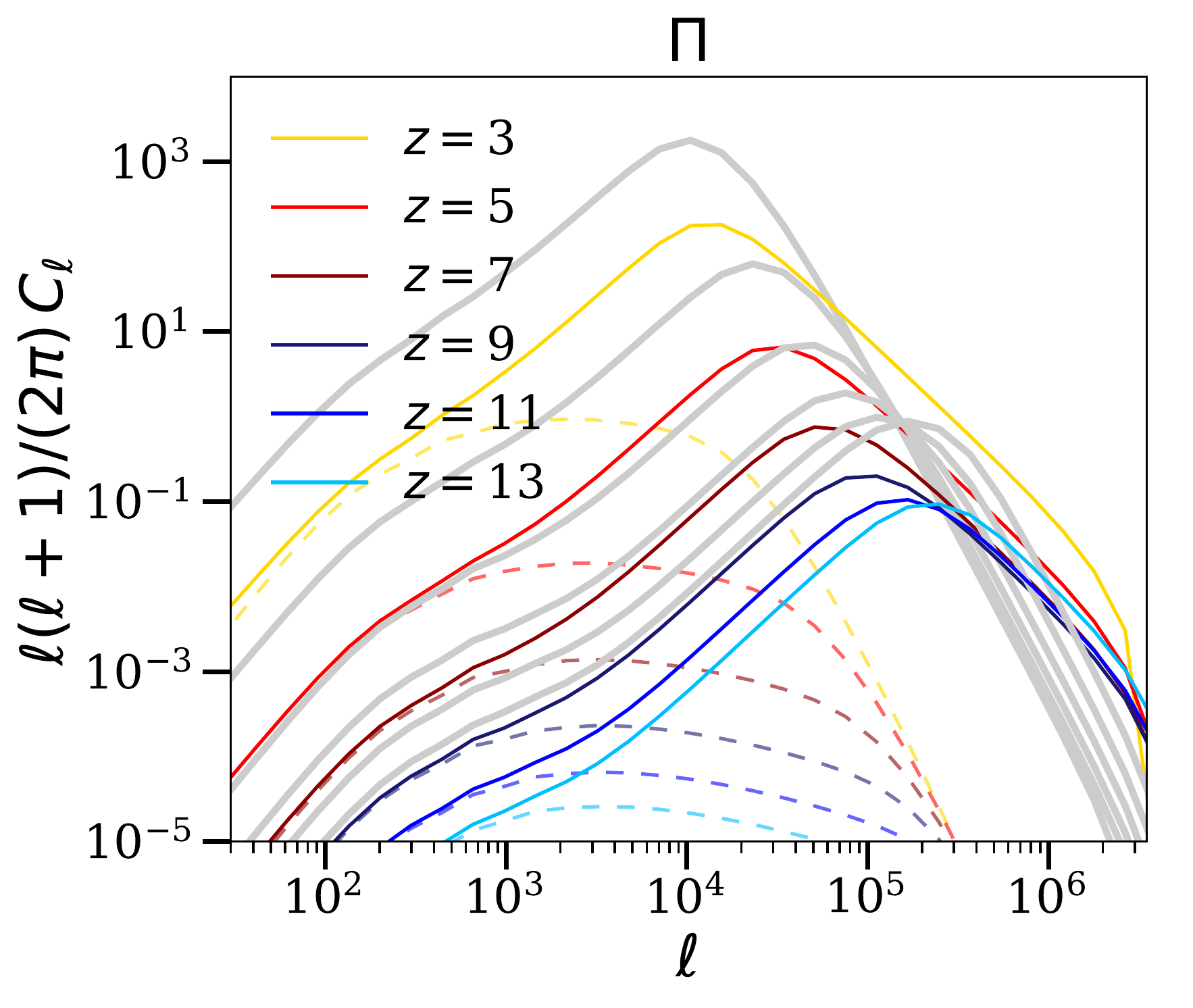}\includegraphics[width=0.51\textwidth]{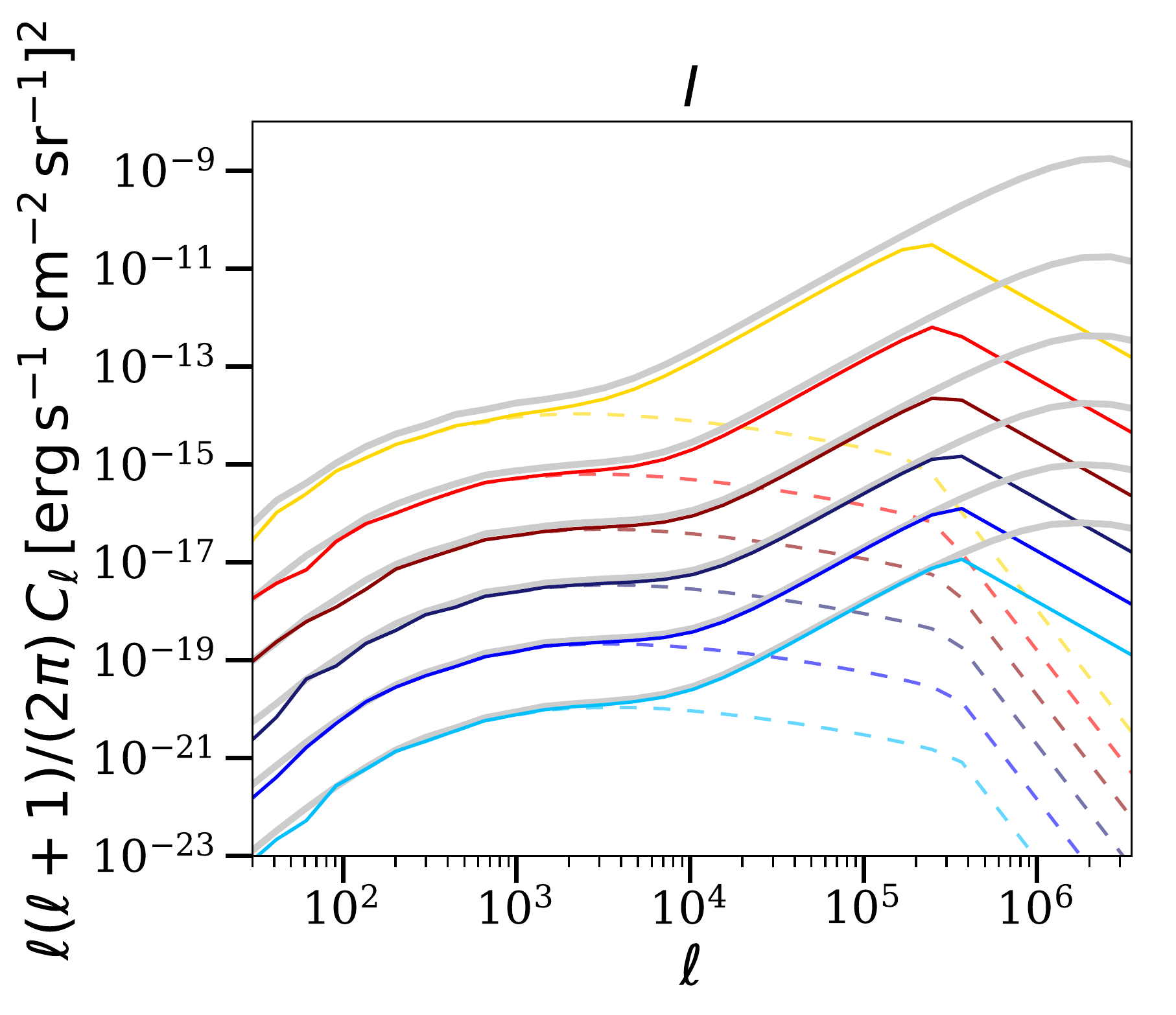}
\includegraphics[width=0.51\textwidth]{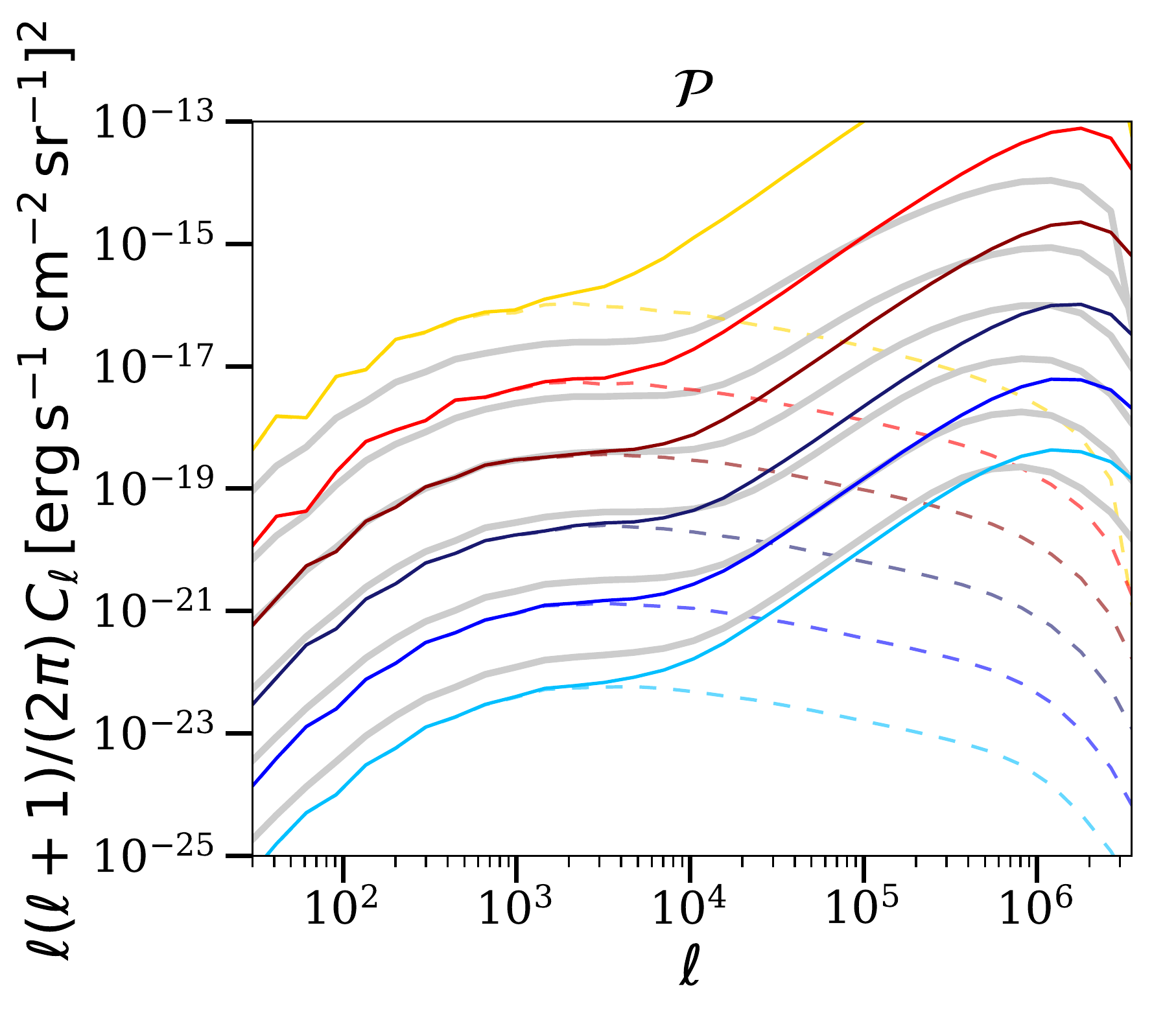}\includegraphics[width=0.51\textwidth]{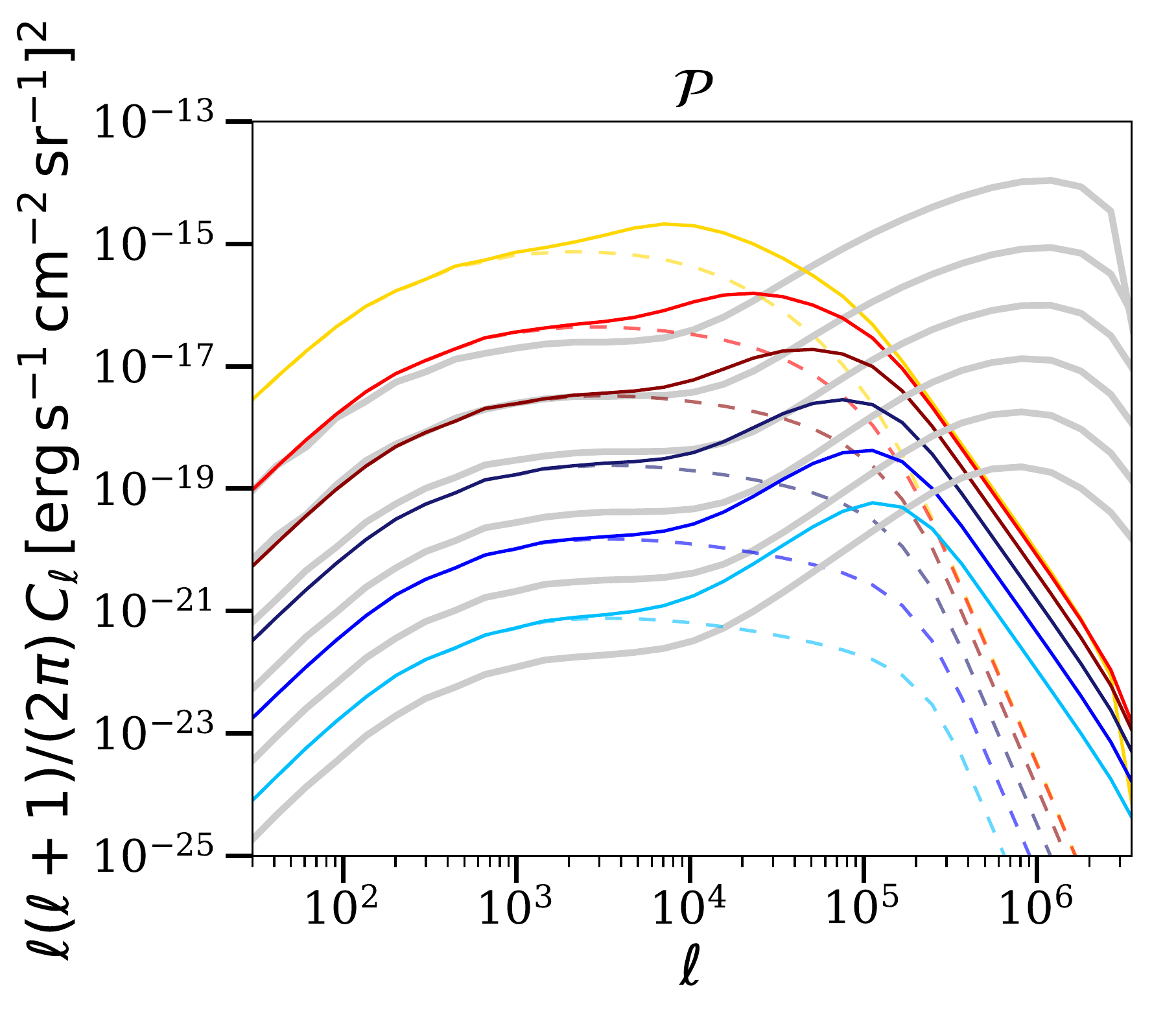}
\includegraphics[width=0.51\textwidth]{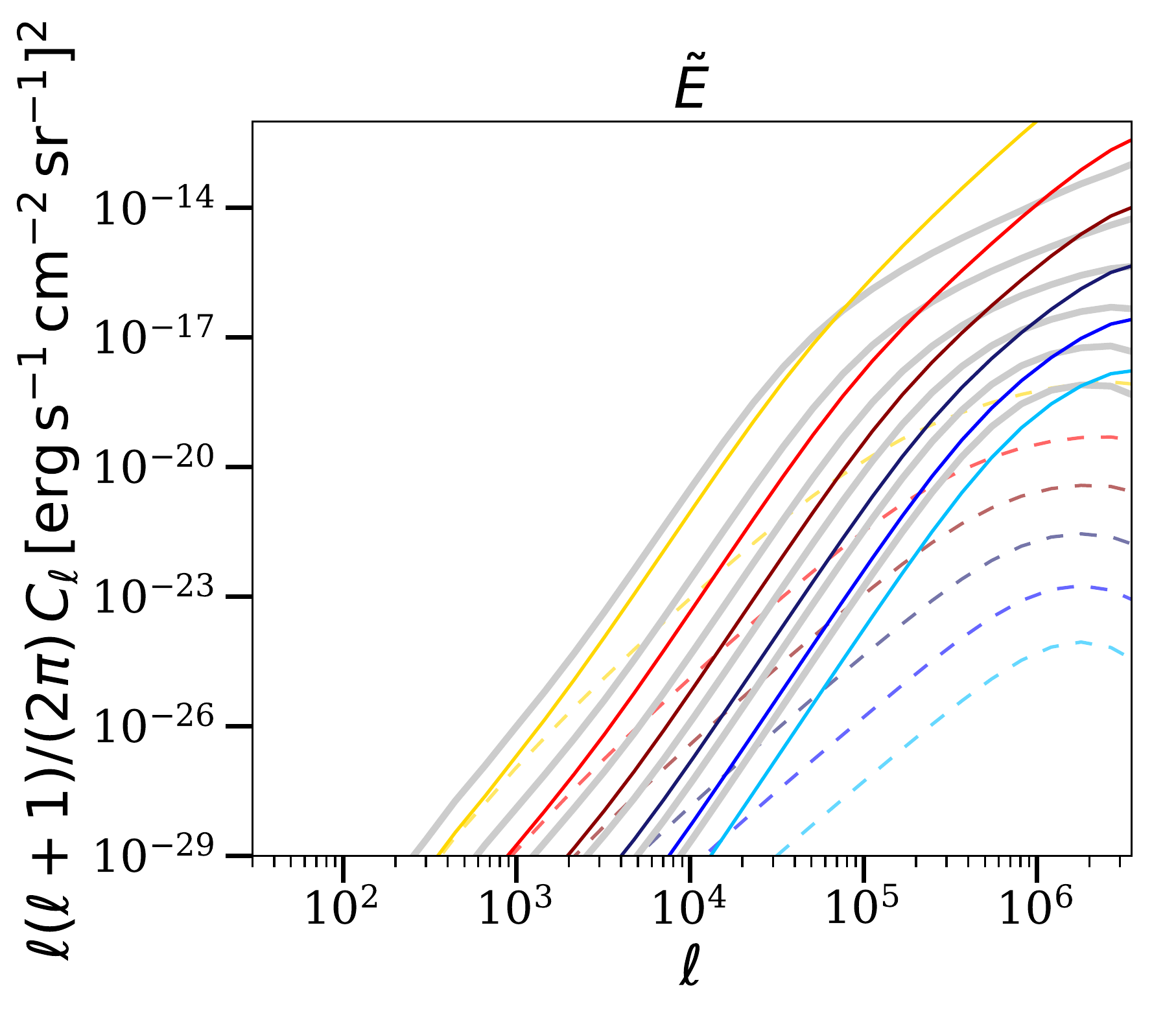}\includegraphics[width=0.51\textwidth]{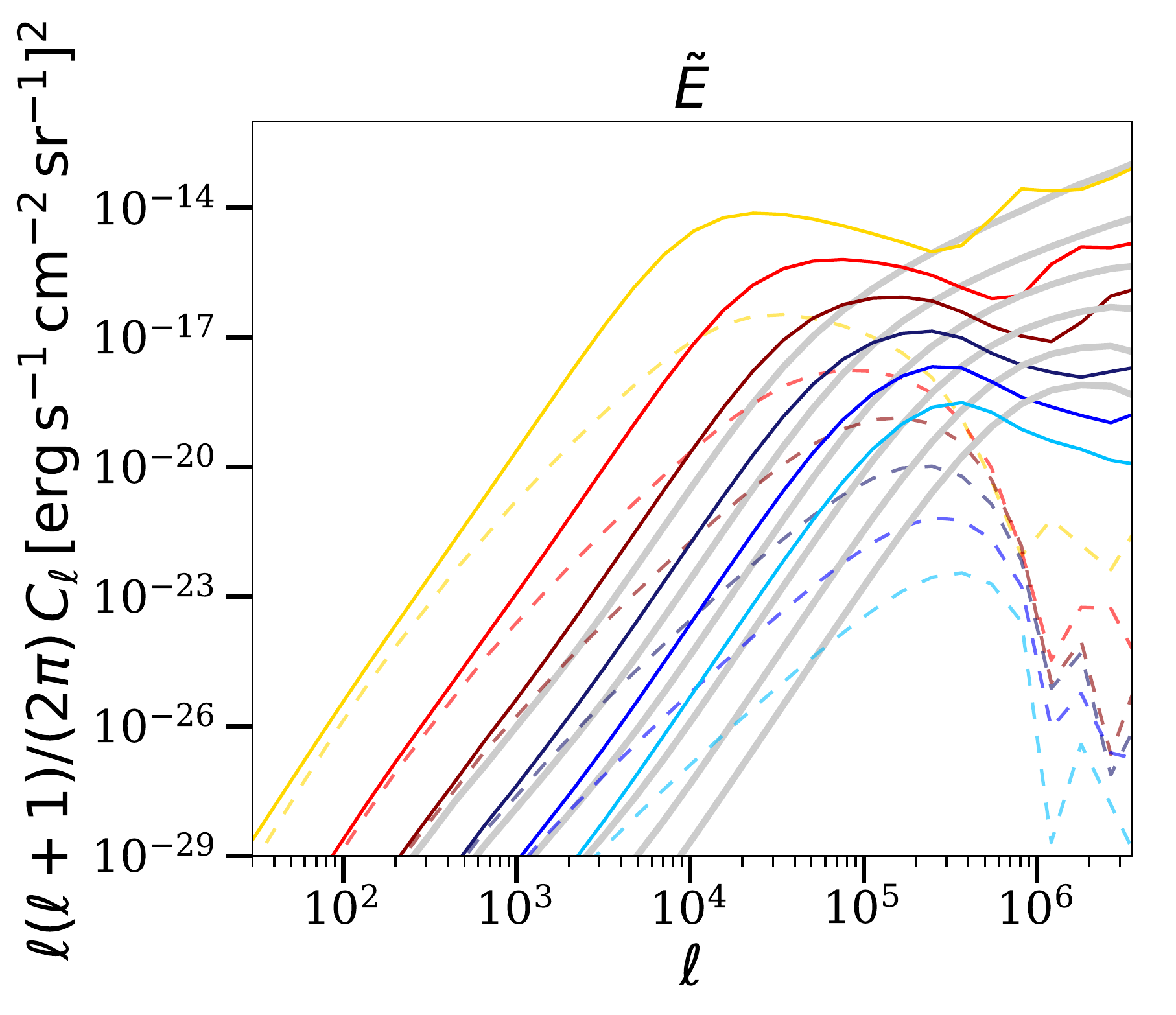}
\caption{{\it Left column:} Power spectra considering a constant polarization fraction profile with a 
value of $10\,\%$, out to one virial radius ({\it colored lines}). {\it Right column:} Power spectra for 
the case of a flat surface brightness profile extending out to three virial radii ({\it colored lines}). For 
comparison, the fiducial power spectra of Figure \ref{fig:psplot} are shown as {\it gray lines}.} 
\label{fig:flat}
\end{figure}

\item{Surface brightness and polarization fraction profile shape}. 
The {\it right panels} in Figure \ref{fig:flat} show the 
power spectra adopting a flat surface brightness profile extending out to three virial radii of the 
halos  ({\it color lines}). The power spectra for intensity ({\it top right panel}) show a 
one-halo term peak at $\ell \sim4-5 \times 10^5$. 
For $\po$ ({\it middle right panel}), the power previously in the one-halo terms is transferred to lower 
multipoles, especially at low redshifts. The power spectra of $\tilde E$ ({\it bottom right 
panel}), in general, are shifted toward lower $\ell$ compared to the fiducial case, and the knees 
become sharp peaks easier to identify, and whose position depends on redshift similarly to $\Pi$.

      The {\it left panels} in Figure \ref{fig:flat} display 
the comparison between the fiducial power spectra ({\it gray lines}), and,   
from top to bottom, those for $\Pi$, $\po$, and $\tilde E$, resulting from considering a 
constant polarization fraction value of $10\,\%$, from the center of the halo out to one virial radius 
({\it color lines}). The power spectra of $\Pi$ present lower amplitudes than the fiducial calculation, while  
the power spectra of $\po$ and $\tilde E$ show milder variations, mostly resulting in steeper one-halo 
terms at the lowest redshifts.

\begin{figure}[h]\center 
\includegraphics[width=0.51\textwidth]{psI.pdf}\includegraphics[width=0.51\textwidth]{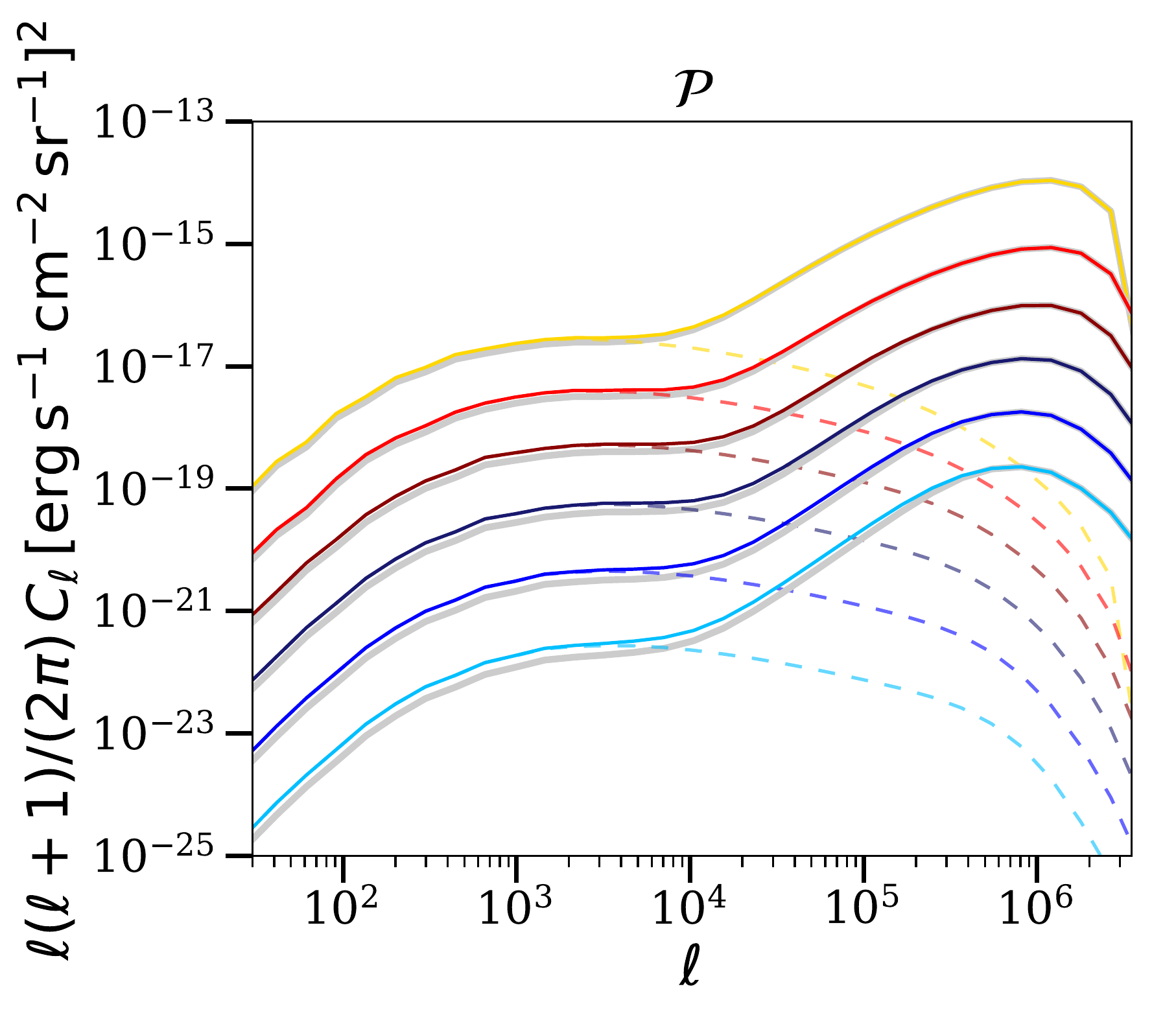}
\includegraphics[width=0.51\textwidth]{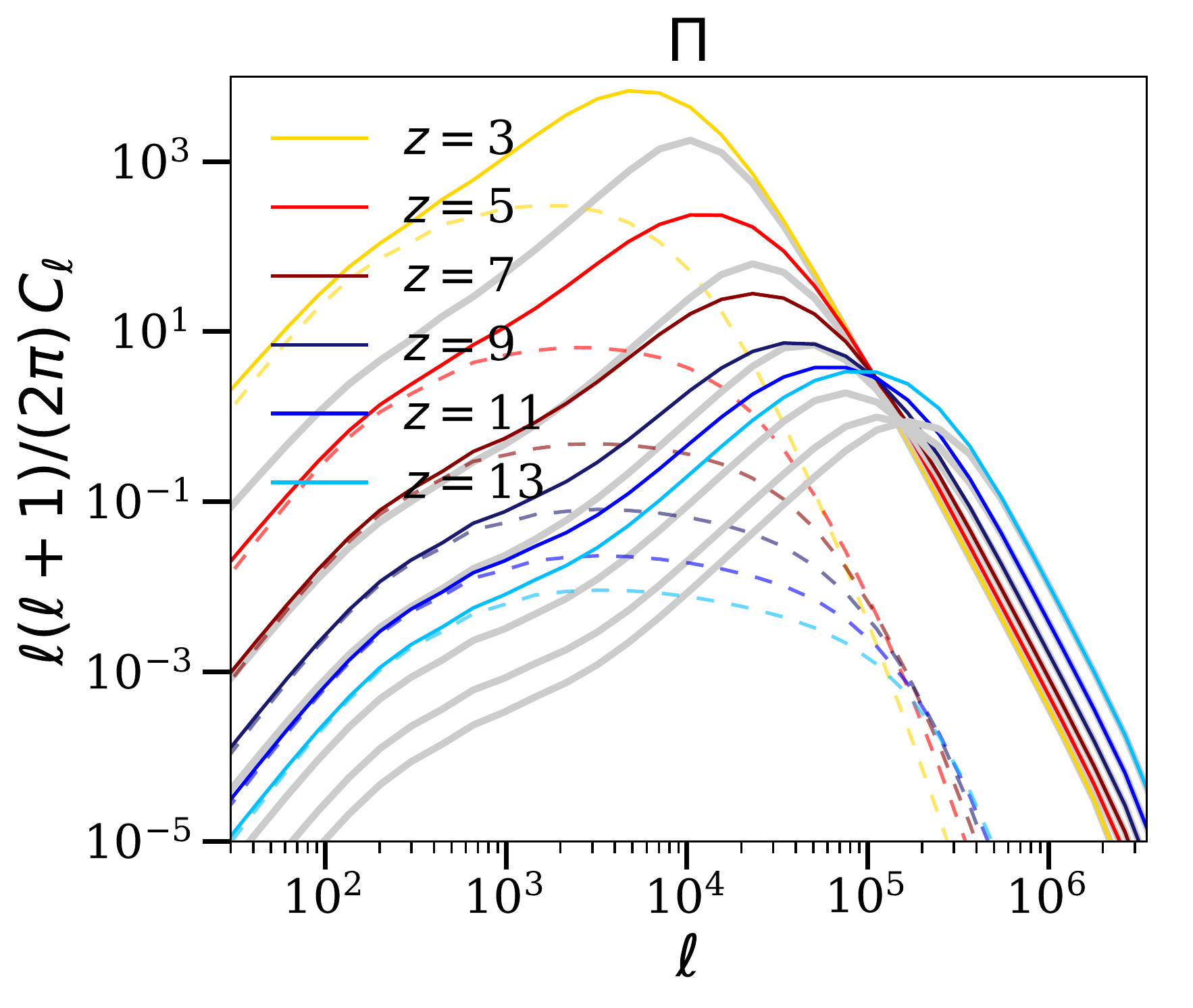}\includegraphics[width=0.51\textwidth]{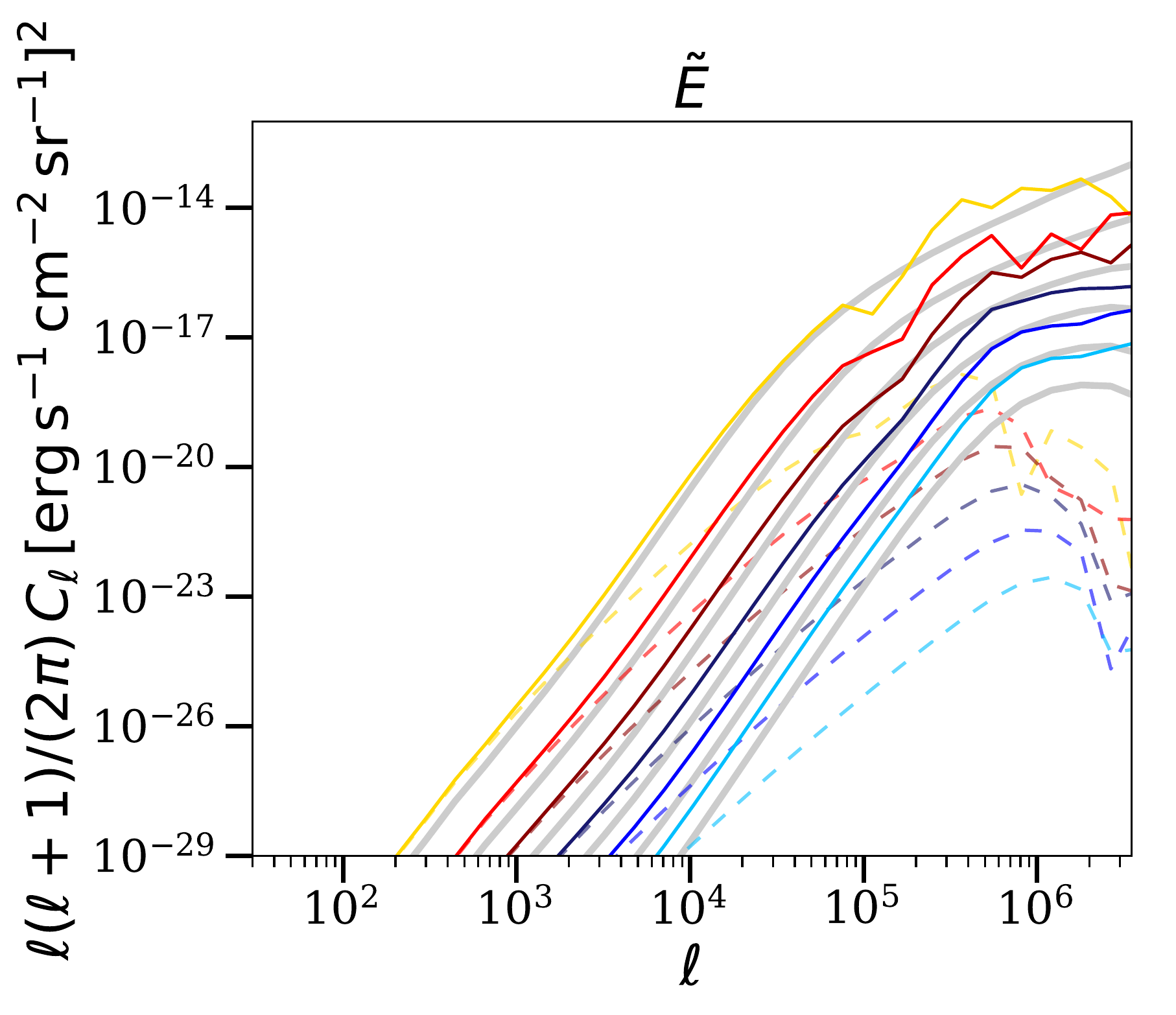}
\caption{Power spectra considering a slow decay, proportional to $\exp{1 - r_{\perp}/r_{\rm vir}}$, 
of the polarization fraction beyond one virial radius ({\it color lines}). For comparison, the {\it gray 
lines} denote the fiducial power spectra of Figure \ref{fig:psplot}, with a decay proportional to 
${\rm exp}[1 - (r_{\perp}/r_{\rm vir})^5]$.} 
\label{fig:slowdecay}
\end{figure}

    We have also tested the impact of a polarization fraction profile that decays slowly after the 
virial radius, proportional to $\exp{1 - r_{\perp}/r_{\rm vir}}$ instead of 
${\rm exp}[1 - (r_{\perp}/r_{\rm vir})^5]$. Figure \ref{fig:slowdecay} shows that only the power 
spectra of $\Pi$ are impacted by this variation, resulting in smoother shapes for the profiles, and 
enhanced amplitudes compared to the fiducial case. The steep slope of the surface brightness 
profile is the reason why varying the decay of the polarization fraction has little impact on the other 
quantities. Regardless of the polarization fraction value, at large physical distances the number of 
photons is very small compared to the center, and their contribution to the shape of the power is 
therefore also small.

\end{enumerate}

\end{document}